# FREE-SPACE NPR MODE LOCKED ERBRIUM DOPED FIBER LASER BASED FREQUENCY COMB FOR OPTICAL FREQUENCY MEASUREMENT

by

TURGHUN MATNIYAZ

B.Eng., Harbin Institute of Technology, 2011

A THESIS

submitted in partial fulfillment of the requirements for the degree

MASTER OF SCIENCE

Department of Physics
College of Arts and Sciences

KANSAS STATE UNIVERSITY
Manhattan, Kansas

2014

Approved by:

Major Professor
Brian R. Washburn

# Abstract

This thesis reports our attempt towards achieving a phase stabilized free-space nonlinear polarization rotation (NPR) mode locked erbium doped fiber laser frequency comb system. Optical frequency combs generated by mode-locked femtosecond fiber lasers are vital tools for ultra-precision frequency metrology and molecular spectroscopy. However, the comb bandwidth and average output power become the two main limiting elements in the application of femtosecond optical frequency combs.

We have specifically investigated the free-space mode locking dynamics of erbium-doped fiber (EDF) mode-locked ultrafast lasers via nonlinear polarization rotation (NPR) in the normal dispersion regime. To do so, we built a passively mode-locked fiber laser based on NPR with a repetition rate of 89 MHz producing an octave-spanning spectrum due to supercontinuum (SC) generation in highly nonlinear fiber (HNLF). Most significantly, we have achieved highly stable self-starting NPR mode-locked femtosecond fiber laser based frequency comb which has been running mode locked for the past one year without any need to redo the mode locking.

By using the free-space NPR comb scheme, we have not only shortened the cavity length, but also have obtained 5 to 10 times higher output power (more than 30 mW at central wavelength of 1570 nm) and much broader spectral comb bandwidth (about 54 nm) compared to conventional all-fiber cavity structure with less than 1 mW average output power and only 10 nm spectral bandwidth.

The pulse output from the NPR comb is amplified through a 1 m long EDF, then compressed by a length of anomalous dispersion fiber to a near transform limited pulse duration. The amplified transform limited pulse, with an average power of 180 mW and pulse duration of 70 fs, is used to generate a supercontinuum of 140 mW. SC generation via propagation in HNLF is optimized for specific polling period and heating temperature of PPLN crystal for SHG around 1030 nm.

At last, we will also discuss the attempt of second harmonic generation (SHG) by quasi phase matching in the periodically polled lithium niobate (PPLN) crystal due to nonlinear effect corresponding to different polling period and heating temperature.

# Table of Contents

# List of Figures

# List of Tables

# List of Equations

# Acknowledgements

First of all, I would like to thank Utuq Ablikim for his help and encouragements for all those years from college to graduate school at K-state. He is the one who told me about this fine university and the excellent graduate program in Physics. That is being said, Dr. Carlos Trallero was the first faculty of physics who has introduced me to the research in AMO and welcomed me to his group even before I came to K-state. It was very nice of him to give me a JRM lab tour on the second day I arrived at Manhattan. I am especially thankful to his generous advice of telling me to feel not bonded to search for my interest among the different groups in Physics. Of course, I would never forget the first day that entered the Corwin/Washburn lab in JRM when Shun Wu gave me a detailed and amazing lab tour. To me, she is not only a good colleague but also an excellent mentor who helped every time when I get stuck on my lab work. Xiaohong Hu, a visiting scholar from China, had taught me lots of first hand experimental skills about the mode-locked lasers and frequency combs. Frankly, without Hu's help this research project would have not been done this far. He was not only an awesome teacher, but also a good lab mate of mine during those tough times. Chenchen Wang and Rajesh Kadel had also never hesitated to help me out in the lab and explained the essential aspects of frequency comb technique. To all of them, I owe sincere gratitude.

The support and understanding from friends around is certainly one of the most precious things one needs, especially when you are far away from your family and country. Among them, the Muslim brothers at the Islamic Center of Manhattan are of huge help to me. Most of all, as a Physics graduate student, I would never forget the fellow graduate students and my classmates in Physics. Hui Wei, a smart and hardworking guy, had helped me to go over the puzzles of many physics graduate classes, especially Quantum Mechanics. Pratap Timilsina, a funny and optimistic guy, has been a good friend and an awesome study mate. I will not forget those times we spent together on preparing for departmental exams and having homework discussions in the map room of Jardine apartments. Neda Dadashzadeh and Mary Harner have definitely been very nice lab mates to work together in the UNFO/LUMOS lab. To all friends including many other nice persons in Physics Department, I owe deep gratefulness.

There are many excellent professors in the Department of Physics. I would like to express my special gratitude to Dr. Larry Weaver, Dr. Chii-Dong Lin, Dr. Uwe Thum, Dr. Andrew

Ivanov and Dr. Artem Rudenko for their terrific teaching of all those essential graduate courses. Without their tremendous work in the classroom and kind extra help outside the classroom, I would not been able to understand the basics of physics this far. I am also very thankful to all of the JRM machine-shop staff that has been very helpful to setup my experiments in JRM lab. Last but never the least, I express my utmost gratitude to my advisor Dr. Brian Washburn for taking me into his research group and giving me the chance to acquire some very interesting research skills about frequency comb technique. He is such a hard working person who enjoys his research work. Frankly, he is not the easiest person to try to win the argument with, but he is very intelligent so that he always provides very critical and useful suggestions to one's work. Not to mention, Dr. Kristan Corwin has also been of huge help by encouraging me on my lab work and giving precious advices during the twice weekly group meetings. Once again I would like to thank my thesis supervisory committee members Dr. Washburn, Dr. Corwin and Dr. Rudenko for approving the final thesis defense exam in spite of their very busy schedule at the beginning of this new semester.

Finally, I would like to express my gratefulness to my parents for giving me continuous encouragement and support in my pursuit of academic life. I wish to express thanks deep from my heart to my siblings:  Aysakhan Matniyaz, Turnisakiz Matniyaz, Gulkiz Matniyaz, Tursunjan Matniyaz and Alimjan Matniyaz. Very fortunate as I am, my wife has been standing by my side during the toughest times at K-state. Most exciting of all, we just had our baby son Otkur on the 8th of October. Thanks to you, both my beloved wife and son, for standing by me during not only the fun times but also the hard times and being the love of my life. In the end, thank you Allah, the almighty God, for all those blessings. Oh Lord, please guide me to the straight way; the way of those on Thou has bestowed Thy grace, those who have not incurred Thy displeasure and those who have gone astray (Al-Fatiha, The Quran).

# Dedication

This work is dedicated to my family. To my father **Matniyaz Amet** and to my mother **Hörnisakhan Meshuk** who have always been very supportive during all those years. To my beloved wife **Khalide Imir** who has been giving encouragement, support, happiness and love to me. To my baby son **Otkur Turghun**.

# Chapter 1 - Introduction and Background

Over the past two decades or so, the optical frequency comb technology witnessed major accomplishments due to the fast development of ultrashort pulse mode-locked laser technology. The importance of these accomplishments has been recognized by the 2005 Nobel Prize in Physics, shared by Theodor Hänsch and John Hall for their contributions to the mode-locked laser based optical frequency comb technology [1, 2]. Soon after the first invention of the Ti:Sapphire laser [3] based frequency combs, the demonstration of fiber based frequency combs appeared as a strong alternative. Compared to Ti:Sapphire lasers, fiber lasers have advantages in terms of cost effectiveness, compactness, passive cooling, etc., which make them an attractive technology. In addition, conventional Ti:Sapphire mode-locked lasers are pumped by bulky, expensive and complex pump sources, such as, argon-ion laser, Nd-YAG laser or dye laser, with as high as 10 W pump power [3-8]; while most of the fiber based mode-locked lasers are pumped by compact and cheap diode laser at 100 mW~500 mW level pump power [9-20]. Only until 2009, several attempts have been made by direct diode laser pumping of Ti:Sapphire laser to overcome the challenging pumping requirement for Ti:Sapphire laser [21-25] . Nonetheless, due to less readily availability of commercial green diode laser and the Ti:Sapphire crystal's favoring of high brightness pump sources, the directly diode pumped Ti:Sapphire mode-locked lasers are still inferior to fiber based mode-locked lasers in terms of compactness, robustness, and inexpensiveness. Therefore, the $Er^{3+}$ doped mode-locked fiber laser based optical frequency combs are preferable in many out-of-the-lab applications. In 2013, Menlo Systems has developed RF-stabilized fiber frequency comb for the application of precision spectroscopy in space [26]. In the same year [27], a group of South Korean (from Korea Advanced Institute of Science and Technology,  KAIST) scientists conducted research that involved the outer space testing of frequency comb of mode-locked Er doped fiber femtosecond laser during which the erbium fiber oscillator was mode-locked over one year. Very recently, in 2014, Sinclair and Newbury *et al.* [28]reported the optically-stabilized 200-MHz fiber laser frequency comb operated in the strong vibrational environments typical of terrestrial platforms. The potential benefits of fiber based frequency combs are quite clear with regard to expense, robustness and field operation. Thus, the fiber laser based comb was selected for the first outer space comb testing in the human history due to its unique characteristics, such as, high alignment stability, small footprint, light weight, optical robustness and high power conversion efficiency.

In this chapter, we begin with the overview of femtosecond fiber laser based frequency comb technique development in the recent decades. First, in Section 1.1 we will review the erbium doped fiber laser comparing to all other rare earth doped fiber lasers and free-space cavity solid state lasers such as Ti:Sapphire laser. Moreover, we will discuss the optical frequency combs and its application in the field of optical metrology such as spectroscopy, astronomy, medical treatment and so forth. Finally, we will end this chapter with the outline of this thesis in Section 1.2.

## 1.1 Femtosecond Fiber Laser Frequency Combs

A frequency comb, usually generated by a mode-locked femtosecond laser oscillator, has contributed to the revolutionary development of precision measurement in the field of metrology and spectroscopy in the past 20 years or so. The idea of using a frequency comb for high-resolution spectroscopy came to be known theoretically as early as Hänsch and Eckstein had published their paper about two-photon high resolution spectroscopy in the 1970's [29]. However, it was not until in 1990 that the frequency comb technique was first realized experimentally in the laboratory [30]. It was the advancement of Ti:Sapphire mode-locked lasers [4, 31]that generate stable pulse trains with equidistant comb lines in the frequency domain (or longitudinal modes in the time domain), which made the frequency comb accomplishment possible. Although the Ti:Sapphire laser was very successful in terms of producing optical frequency combs, it encountered several limitations such as size, pump laser requirements and expense [32].Therefore, the implementation of rare-earth doped (especially $Er^{3+}$doped) mode-locked fiber lasers toward the optical frequency comb technique is very attractive, benefited from the development of mode-locked ultrafast fiber lasers [17, 33, 34].

### 1.1.1 Mode-locked Erbium Doped Fiber Lasers

#### *1.1.1.1 Brief History of Mode-locked Lasers*

The mode-locking technique of lasers had been achieved first using a He-Ne laser by Hargrove *et al.* in the 1960s [35]. In terms of the mechanism that used to mode-lock the laser, the mode-locking technique is classified into three categories including passive mode-locking, active mode-locking and hybrid mode-locking. Mode-locking (or phase-locking), in general, is a way to achieve fixed phase relationship between the longitudinal modes of the resonant cavity of the laser. In the time domain, the mode-locked laser generates stable pulse train due to the interference between the longitudinal modes. In case of active mode-locking, the laser is mode-locked by taking advantage of placing intracavity modulators, such as an acousto-optic modulator (AOM) for amplitude modulation and electro-optic modulator (EOM) for frequency modulation, which involves periodic modulation of resonator losses or round-trip phase change. In the case of passive mode-locking, however, the laser is mode-locked by taking advantage of self-phase or self-amplitude modulation induced by the passive mode-locking device, such as fast semiconductor saturable absorber mirror (SESAM). The hybrid mode-locking, as it implies, is a way to implement both of the above methods on a laser to obtain an ultrashort pulse train. Among the two, the passive mode-locking technique is much faster than active mode-locking because of the fast modulation of resonator losses due to short recovery time of saturable absorbers. Moreover, previously, shorter pulse have been generated using passive mode-locking techniques compared to active mode-locking technique.

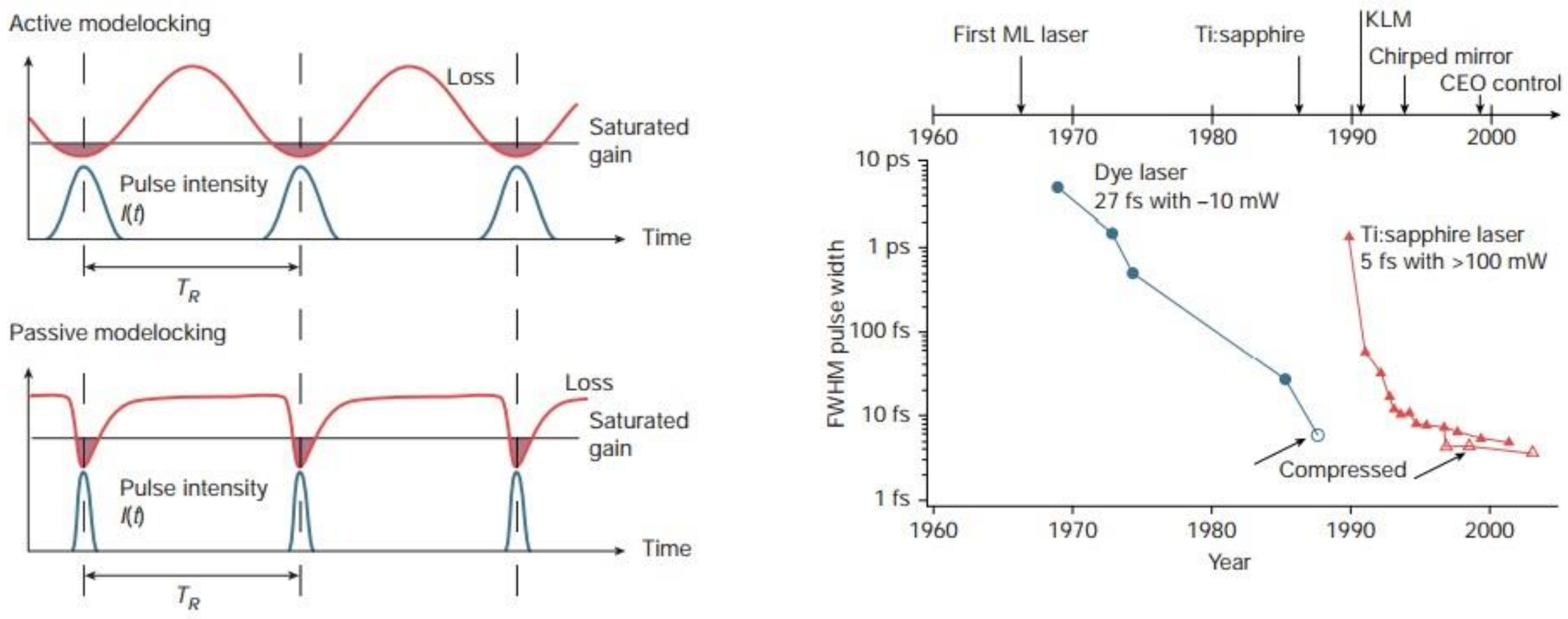

**Figure 1.1 Left: The Passive and active mode-locking in the time domain. Right: The progress in the ultrashort pulse generation in terms of pulse duration over the years. Figure reproduced from Ref. [36].**

In the 1970s, the passively mode-locked dye laser that generated as short as 1.5 picosecond pulse had been invented by Ippen *et al.* [37] at the Bell Telephone Laboratories.

Soon after, in 1980s, Moulton [3] had reported a new solid-state laser, the $Ti^{3+}$ doped sapphire (or Ti: Sapphire) laser.

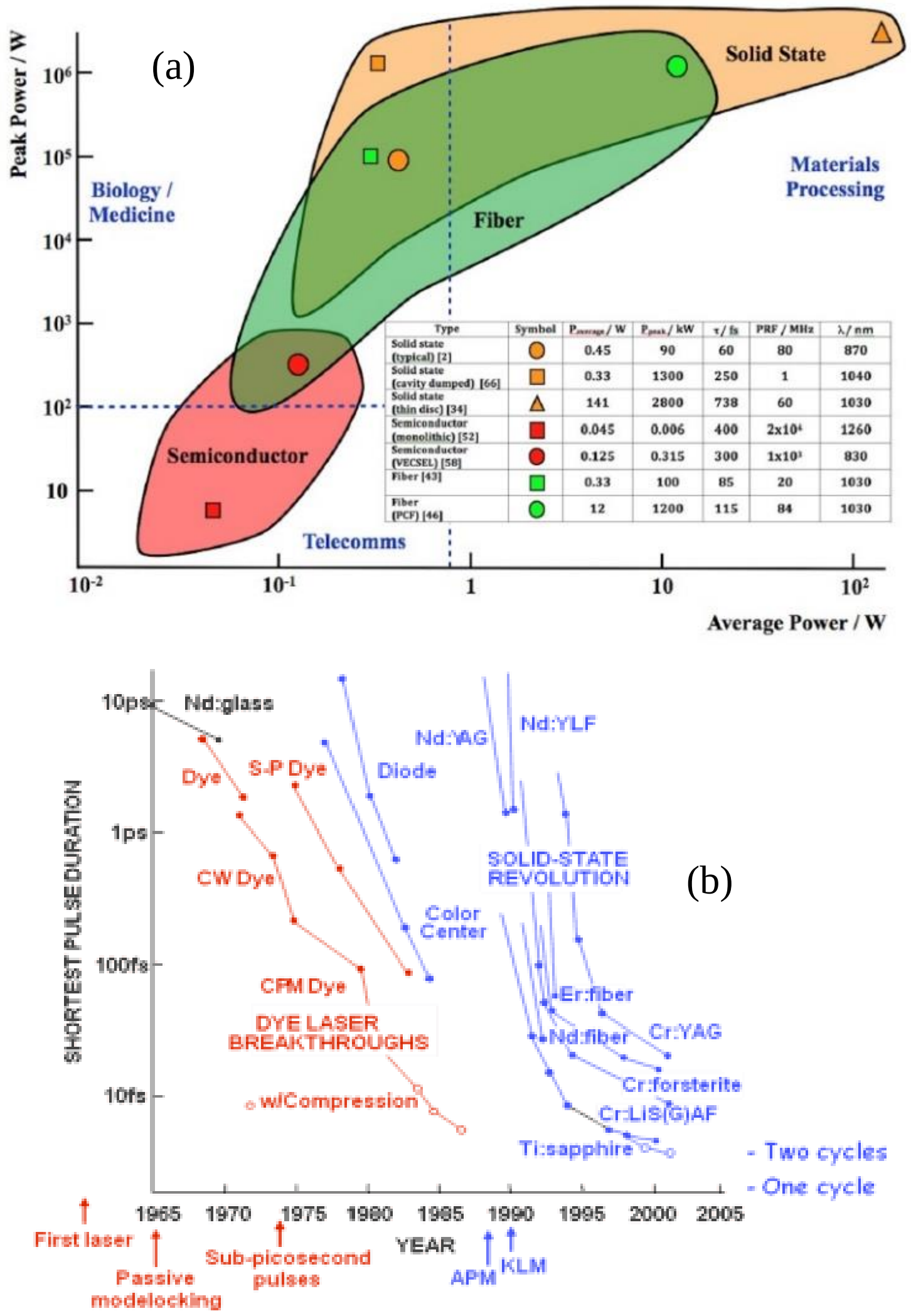

**Figure 1.2 (a) Peak power and average power for ultrafast laser oscillators for different sources. (b)The progress in generating short optical pulses from various laser materials. Figure reproduced from Ref. [17, 38]**

As the advance of solid state laser technology, the Ti:Sapphire ultrafast lasers replaced the previously dominant dye lasers due to the fact that the solid-state lasers are more stable in operation than dye lasers, even though the mode-locked ultrashort dye lasers were improved to generate pulses shorter than 100 fs, as showed in Figure 1.1. In the 1990s, the Kerr-lens mode-locked Ti:Sapphire laser technology had been developed by a group of researchers including T.W. Hänsch's group, J.L. Hall's group and W. Sibbett's group, which boosted the beginning of frequency comb technology [4, 30, 39]. At that point, the Ti:Sapphire solid-state mode-locked ultrafast laser technology has been fairly well established for the frequency comb [40, 41] application concerning the output power and pulse duration, as shown in Figure 1.2.

#### *1.1.1.2 Erbium Doped Fiber Laser*

Nevertheless, the rare earth doped solid-state laser (including $Er^{3+}$ doped fiber laser, $Yb^{3+}$ doped fiber laser, $Tm^{3+}$ doped fiber laser and etc.) technology has been competing with the Ti:Sapphire ultrafast laser technology whose application towards frequency comb is limited by long-term stability and operability. Fiber lasers have a number of practical advantages, including much more compact in size, lighter, more efficient power operation, lower cost, more robust and less alignment dependent [42, 43].In short, a fiber laser based frequency comb allows easier operation and longer measurement time compared to a Ti:Sapphire system [44]. Among these rare earth doped fiber lasers, the $Er^{3+}$ doped fiber laser has very attractive application in the telecommunication field owing to the near 1.5 $\mu m$ output wavelength. In 1989, Kafka and Baer [34] reported the first mode-locked erbium doped fiber laser that generated 4 ps pulse at 1530 nm. After that, in 1993, Tamura *et al.* [33] developed a stretched pulse mode locked erbium doped all fiber ring laser that generated 77 fs ultrashort pulse. It is worthwhile to mention that a new concept of mode locking, *i.e.* the nonlinear polarization rotation (NPR) mode-locking, had emerged at that time as a passive mode-locking technique. Usually, there are three types of NPR mode-locking schemes that exist in the mode-locking ultrafast solid-state lasers, including free-space NPR mode-locking, all-fiber NPR mode-locking and hybrid NPR mode-locking. The Ti:Sapphire lasers, in general, are mode-locked through a free-space mode-locking setup, while fiber lasers are mode-locked through either the all-fiber NPR setup or the hybrid NPR mode-locking setup. The schematics for both all-fiber (a) and hybrid (b) NPR lasers are shown in Figure 1.3. Passively mode-locked fiber lasers have been extensively investigated for the generation of ultrashort pulses. Various techniques, such as, NPR [45], nonlinear optical loop mirror (NOLM) [46], semiconductor saturable absorption mirrors(SESAM) [47], carbon-nanotube saturable absorbers (CNT-SA) [48, 49]and graphene saturable absorber [50], have been employed in ultrashort pulse fiber lasers for passively mode-locking [20].

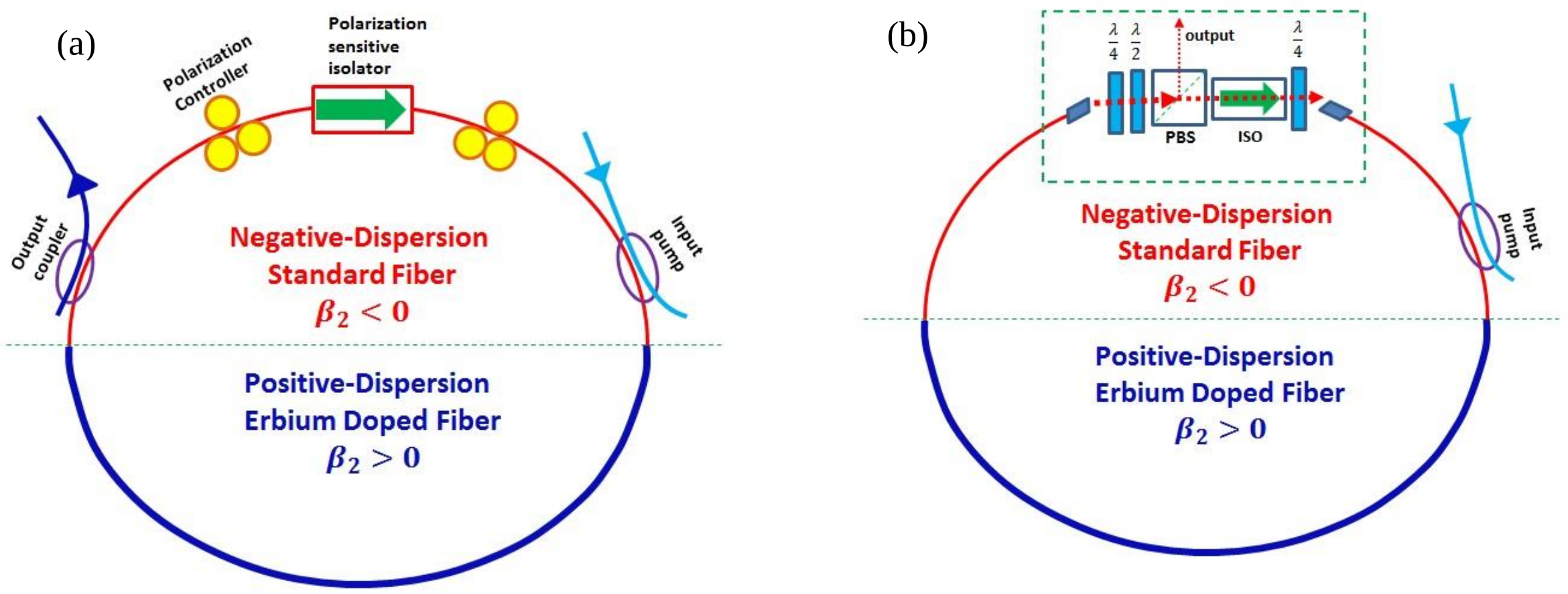

**Figure 1.3 The Two types of NPR mode-locking cavity schematics. (a) All-fiber NPR mode-locking $Er^{3+}$ doped laser oscillator [33]. (b) Hybrid NPR mode-locked $Er^{3+}$ doped fiber laser oscillator [51]. ISO: polarization sensitive free-space isolator, PBS: polarization beam splitter, $\lambda/4$ & $\lambda/2$: quarter and half wave plates (work as polarization controller).**

All-fiber NPR scheme output power is limited by the damage threshold of optical fiber and fiber components, therefore, currently the hybrid NPR mode-locking setup is more desirable due to the fact that hybrid mode-locking setup enables much higher average output power directly from the laser oscillator. In addition, the cavity length is readily tunable in the hybrid cavity scheme, which in turn provides more freedom to adjust the laser repetition rate by changing the cavity length. All of these advantages of the hybrid NPR erbium doped fiber laser are extremely important towards the application of femtosecond fiber laser based frequency combs.

## 1.1.2 Optical Frequency Comb and its Application

### *1.1.2.1 Optical Frequency Combs*

An optical frequency comb is the spectral content of an infinite pulse train that is generated from a mode-locked laser. The most important part in a frequency comb is the laser oscillator, which is also called a comb generator. In short, the frequency comb is a system consisting of a mode-locked laser and the electronic feed-back loop that used for the stabilization of the laser. Therefore, sometimes people refer to the laser oscillator part as the comb in the literatures and the laser spectrum in the frequency domain are named the comb lines. The time domain and the frequency domain are mathematically connected by the Fourier transformation or its inverse.

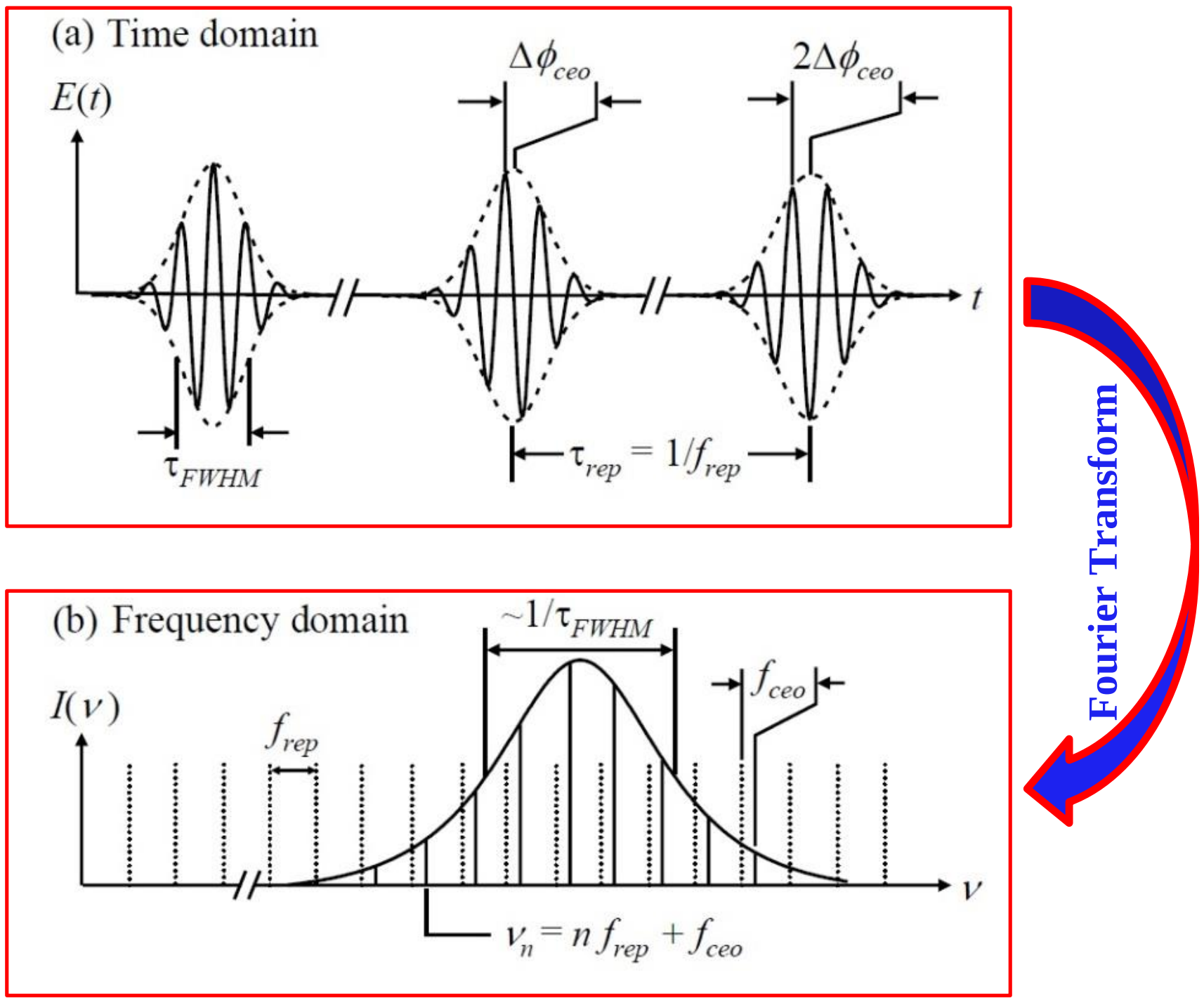

**Figure 1.4 The output of the mode-locked laser based optical frequency comb in time domain (a) and frequency domain (b). The laser produces a periodic pulse train of pulse duration $\tau_{FWHM}$ and pulse repetition frequency $f_{rep}$. The total comb offset from the zero frequency by $f_{ceo}$ (*i.e.* the $f_0$) and $f_{ceo} = f_{rep}\frac{\Delta\phi_{ceo}}{2\pi}$. Figure reproduced from Ref. [52].**

The Ti:Sapphire laser based frequency combs has been proven to be the very stable tool for ultraprecise measurement among the solidstate-laser based frequency combs. However, the there are several drawbacks of a Ti:Sapphire based comb system, such as large size and operation at a wavelength of 800 nm, which is not desirable for the application of combs in the telecommunication field that requires compact size and operating wavelength near 1550 nm. On the other hand, fiber laser based frequency combs are very compact, portable, less alignment dependent and less expensive. Especially, the $Er^{3+}$ doped mode-locked fiber laser based combs, of which output wavelength is near the telecom window at 1550 nm, are very attractive sources of compact, self-starting, high output power optical frequency combs.

The goal of this work is to build an NPR mode-locked $Er^{3+}$ doped femtosecond fiber laser based frequency comb and phase stabilize the comb system by self-referencing method using a phase-locking feedback loop related electronic devices that are reference to a GPS disciplined Rb clock in our lab. The most desirable feature of this frequency comb for our purpose is to generate high average output power and wide bandwidth ultrashort pulse directly from the oscillator.

#### *1.1.2.2 Application of Combs*

The first practical application of an optical frequency comb in the past was the measurement of absolute frequency of the ultraviolet hydrogen 1S-2S two-photon resonance by Hänsch's group in 1999 [53, 54]. High resolution optical spectroscopy is of vital importance in conducting several precision measurements, such as molecular fingerprinting of gas molecules, identification of drugs and biological species and characterization of optical devices. The frequency comb is also a potential tool to generate attosecond pulses by phase-locking the equidistantly spacing high harmonics. In 2000, Holzwarth *et al.* has managed to build an optical frequency synthesizer for precision spectroscopy. This synthesizer were based on a frequency comb generated by a femtosecond mode-locked laser which is broadened to more than an optical octave in a photonic crystal fiber [55]. In 2001, the research teams from the National Institute of Standards and Technology (NIST) in Boulder, USA and the Max-Planck-Institute of Quantum (MPQ) in Garching, Germany developed the all-optical atomic clock based on single trapped $^{199}Hg^{+}$ ion, which proved to be substantially better than world's best microwave atomic clocks based on cesium [56, 57]. The stable optical clocks are potentially useful for tracking and communication between satellites and spacecrafts beyond our planet [58]. In 2006, Joo *et al.* measured absolute distance using Ti:Sapphire laser based frequency combs by the way of dispersive interferometry [59]. In 2009, Hyun *et al.* carried out absolute length measurements using a frequency comb, which is directly traceable to the atomic clock that used for time standard [60]. In the field of astronomical observation, optical frequency combs can also be used to conduct Doppler shift measurement, which is an attractive method to detect Earth-sized planets in the future [61, 62]. As far as the commercial availability of frequency combs is concerned, the Menlo Systems [63]has launched its first commercial fiber laser based optical frequency combs in 2005.

In conclusion, the applications of optical frequency combs can be found in many fields, including optical clockworks, precision metrology, laser calibration, laser cooling, high resolution spectroscopy, satellite navigation, telecommunication, astronomy and fundamental research in physics [64, 65].

## 1.2 Thesis Outline

In this thesis, we will present our work towards the building of an NPR mode-locked $Er^{3+}$ doped fiber laser based frequency comb system. At the beginning of this project, we aimed to use this frequency comb for frequency measurement by beating with another existed comb in our lab, once it is completely phase-stabilized. Therefore, we have set the repetition frequency of the comb to about 89 MHz.

Chapter 1 is the introduction and background of femtosecond frequency comb; including the femtosecond ultrafast laser by focusing on the $Er^{3+}$ doped mode-locked fiber laser and fiber laser frequency comb. Chapter 2 focuses on the free-space hybrid cavity NPR mode-locked erbium doped femtosecond fiber laser oscillator setup design and output pulse characteristics. The solitonic regime and stretched pulse regime of the mode-locked output has been especially analyzed. In addition, the experimental result of NPR mode-locked fiber laser has been presented with 56 nm full width half maximum (FWHM) pulse duration and 30 mW average power at 1575 nm.

In Chapter 3, we will put forward the experimental layout of ultrashort pulse compression and amplification. First, we are going to discuss the NPR mode-locked laser output pulse amplification through the erbium doped fiber amplifier (EDFA), focusing on the EDF length optimization and EDFA output pulse characteristics. Second, we will provide the experimental results of pre-compressor of the NPRL comb output and post-compressor of the EDFA output. Moreover, there will be given related theoretical explanation on the final amplified and compressed pulse of 181 mW average output power and 70 fs FWHM pulse duration.

In Chapter 4, we will start with the basic principles of supercontinuum (SC) and the experimental result of SC generation in dispersion flattened highly nonlinear fiber (HNLF) will be presented. The core task at this point is to optimize the HNLF length for best broadband spectra that will be used for *f to2f* self-referencing.

In Chapter 5, we will present the work of our attempt towards the phase-stabilization of the comb through *f to 2f* free space interferometer setup. The theoretical background of SHG will be presented and there will be given some explanation of the final experimental result. Also, we will discuss the current status of the SHG result and summarize the existing problems and potential methods to improve the result. Finally, Chapter 6 summarizes the whole thesis and gives some future work outlook to accomplish the phase-stabilization of the NPR mode-locked $Er^{3+}$ fiber laser comb system.

# Chapter 2 - NPR Mode Locked Erbium Doped Femtosecond Laser Frequency Comb

This chapter is about the cavity design and optimization of the nonlinear polarization rotation (NPR) based mode locked $Er^{3+}$ doped fiber laser, which is the first part of the phase stabilized femtosecond laser frequency comb system. At the same time, we will discuss briefly about the passive mode locking principle in this experimental setup. In addition to that, the spectral characteristics of the erbium doped fiber (EDF) including gain spectrum and nonlinear spectral broadening effect under the diode pumping will be discussed theoretically and experimentally. At last, we will present the experimental measurement of the ultra-short pulse duration using autocorrelation.

## 2.1 Design of the Hybrid Frequency Comb Cavity

In recent years, researchers have been able to find several different ways to realize the mode locking of ultrafast lasers. As the first mode locked ultrafast laser, the passively mode locked dye laser was built using a free-space cavity, as well as the later achieved Titanium Sapphire ($Ti:Al_2O_3$) mode locked laser in 1986 [3]. However, the free-space mode locking configuration is not a good choice to pursue a compact and environmentally-stable ultrashort laser output due to its need of a highly stable optical table in the laboratory environment and high power consumption. The potential of making compact and rugged laser systems with low power consumption at relative low price make all fiber cavity based mode locked lasers a very promising alternative to classical solid state lasers [66]. Although fibers have a number of beneficial properties for short pulse generation, the dispersion and particularly the high nonlinearity of fibers severely limits the performance of mode-locked fiber lasers particularly in terms of pulse energy, pulse duration, and often also pulse quality [67].There are several difficulties in achieving high energy ultrashort pulses from all-fiber passively mode locked erbium doped fiber lasers [33].

To improve the pulse energy and output power from the oscillator, Tamura *et al.* reported a new technique using free-space NPR scheme with EDF as gain medium, which enables extracting high average power from the cavity [68].

### 2.1.1 Free-space NPR Mode-locked Frequency Comb

In our experiment, we have used a hybrid design of all-fiber cavity and free-space cavity, which is a free-space NPR mode-locked erbium doped fiber laser. The experimental setup configuration is shown in Figure 2.1.

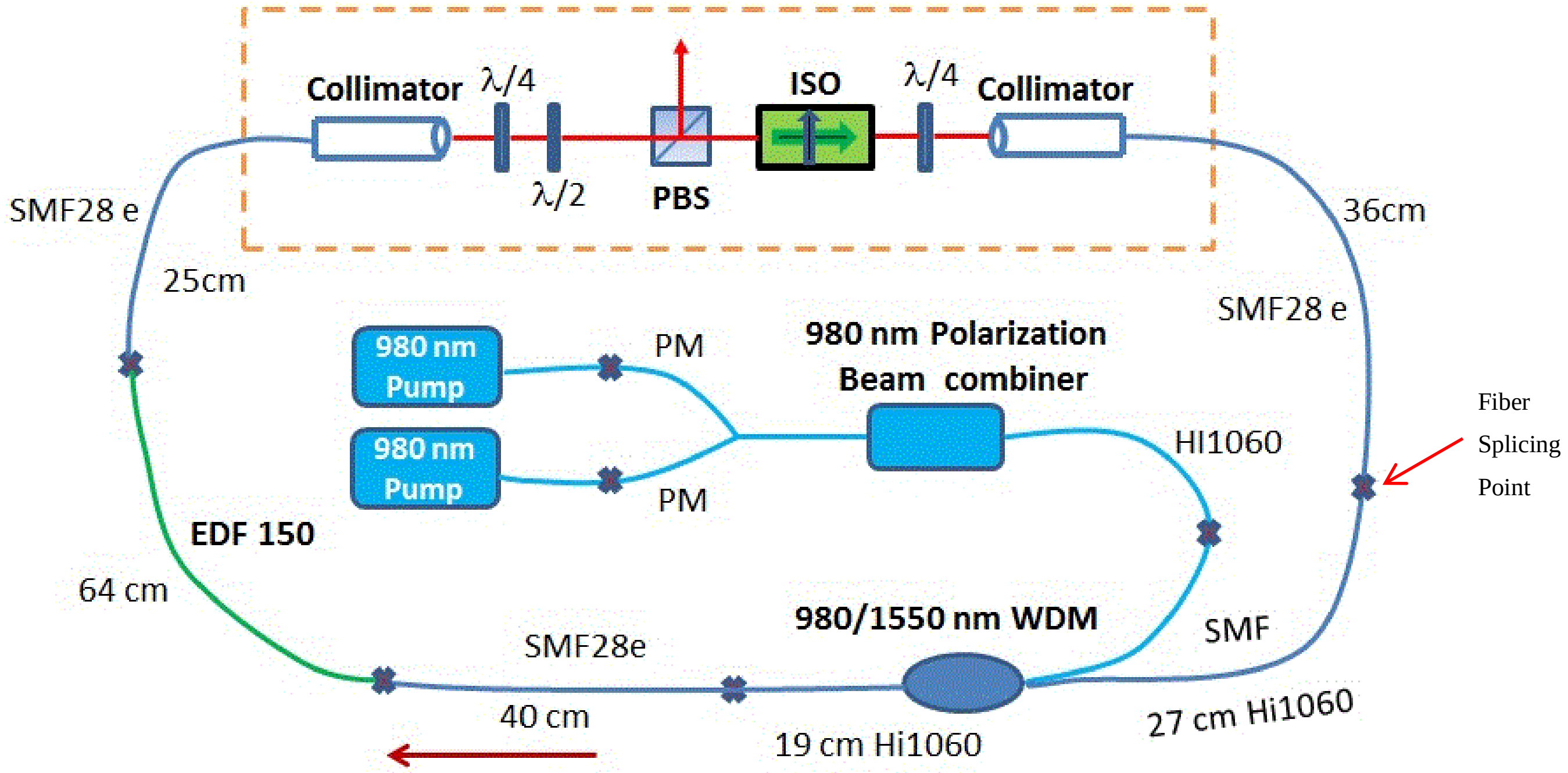

**Figure 2.1 Schematic of the free-space NPR mode-locked $Er^{3+}$ doped fiber laser oscillator. PBS: polarization beam splitter, ISO: polarization-sensitive isolator, WDM: wavelength-division multiplexing, SMF: single mode fiber, EDF: erbium doped fiber, HI1060: HI1060 specialty flexcor fiber in 980/C band WDM, PM: polarization maintaining single mode optical fiber, λ/4&λ/2: quarter and half wave plate, Collimator: mounted aspheric lens with fiber connector.**

In the above setup, we have implemented double-pumping scheme using two laser diodes of central wavelength at around 980 nm. The double-pumping is used so that we could have enough pump power to obtain high enough output power. Besides, we have chosen the forward pumping method instead of backward pumping so that we could decrease the soliton effect by placing the pump closer to the gain fiber. Then, the two pump lasers were spliced to panda PM fiber input port of a AC Photonics polarization beam combiner (PBC), with estimated splicing loss about 0.03 dB using a Ericsson FSU-995 Arc fusion splicer. Thus, the two pumping sources were combined through the PBC. The output port of the PBC then spliced to the pump port of the SIFAM 980/C band WDM, with estimated splicing loss of 0.02 dB. A segment of 40cm SMF is used between the HI1060 flexcor fiber on the common port of the SIFAM 980/C band WDM, as a bridge between EDF and WDM, to decrease the splicing loss. The splicing loss between SMF and HI1060 fiber was estimated to be 0.02 dB, while the estimated loss between the SMF and EDF was 0.01 dB. After that, the signal light pumped from the EDF is collimated through an aspheric lens of 7.5 mm focal length. The quarter wave plate converts the elliptically polarized light into linearly polarized output, which in turn rotated in the polarizing direction so that the components that polarized perpendicular(s-polarized) to the plane (as seen from the top view) will be reflected by the PBS while the components that polarized parallel(p-polarized) into the plane will be transmitted through the PBS. Consequently, the p-polarized light passes through the ISO and will be converted back into the elliptical polarized state and coupled into the SMF fiber through the collimator with another aspheric lens of 7.5 mm focal length. And now, the signal can circulate in the feedback loop cavity which in turn extracts the s-polarized portion as reflected by the PBS.

### 2.1.2 Free-space NPR Mode-locking Principle

Nonlinear polarization can occur in a fiber only when the initial polarization state is elliptical [9]. There are two different ways of mode-locking technique, including active mode locking and passive mode locking. In the case of active mode locking, an optical modulator (*e.g.* EOM or AOM) is placed in the laser cavity, to modulate the intracavity losses or the round trip phase shifts which in turn enables the mod-locking between the various cavity modes. However, the method of actively mode locking the laser has its drawback that pulse duration of the generated mode-locked pulse is limited by the bandwidth of the optical modulator [69]. The passive mode locking, on the other hand, does not need optical modulators to achieve mode locking instead a semiconductor saturable absorber mirror (SESAM) is used in a free-space Fabry-Perot cavity mode locked laser. A saturable absorber is an optical device with an intensity dependent transmission property, which allows the transmission of high intensity pulse and absorbs low intensity pulse. Spence *et al.* generated 60 fs ultrashort pulse using a passively mode-locked Ti:Sapphire laser with a saturable absorber [4]. Later, Ell *et al.* has succeeded to obtain 5 fs ultrashort pulse [6] without using a saturable absorber in the cavity but taking advantage of the Kerr Lens Mode-locking (KLM) which is based on the optical Kerr lens effect in a Kerr medium. The optical Kerr effect is third order nonlinear effect which induce change in the refractive index of the medium. The change in the total refractive index under intense laser pulse is intensity dependent and defined as:

$$n(I) = n_0 + n_2 I \tag{2.1}$$

where $n_0$ is the linear refractive index of the medium and $n_2$ is the nonlinear refractive index off the medium, $I$ is the instantaneous pulse intensity.

For typical solid state materials, the change in the refractive index $n_2$ is on the order of several $10^{-16} cm^2/W$. Therefore, the total refractive index has significant change only when the pulse intensity is high enough. Thus, the Kerr lens effect only occurs when the intensity of the light is extremely high, such as the instantaneous intensity of a mode-locked pulse. In addition, low intensity part of a pulse will make no obvious change in the total refractive index of the Kerr medium that means there is no nonlinear effect for the low peak intensity portion of a pulse.

As one of the many mode-locking mechanisms that have been developed over the past decades, the Kerr lens mode-locking (KLM) is used widely in the area of free-space cavity based solid state ultrashort pulse laser and contributed to the generation of the shortest pulse. In contrast with free space cavity based lasers, the fiber ring cavity based lasers are known for its compactness, lack of optical alignment, high efficiency and immunity against thermo-optical problems [70].

In 1993, Matsas *et al.* [71] reported a self-starting passively mode-locked all-fiber cavity erbium doped fiber laser, which is based on nonlinear polarization evolution (NLPE). In fact, the laser cavity is constructed using low birefringence fiber, fiber based optical isolator and fiber based optical polarization controller (PC), so that the combination of fiber and PCs form an artificial saturable absorber. Generally, the shortest pulses can be obtained with passive mode-locking techniques, and over the years, a large number of cavity designs have been investigated. The first all-fiber mode-locked lasers were based on the use of nonlinear Sagnac interferometers inside the fiber ring cavity. The Kerr-type mode locked fiber lasers were demonstrated in ring cavities by taking advantage of ring cavity that has less susceptible to back reflection (which favor CW over mode-locked operation) compared to a Fabry-Perot cavity. Another advantage of the ring cavity is possible extraction of high pulse energy [72].

As mentioned above, the all-fiber setup has its limits in terms of high power output due to the inherent properties of optical fiber. Therefore, a hybrid design of free-space and all-fiber cavity is necessary to get high output power and high pulse energy self-starting NPR mode-locked fiber laser [73].

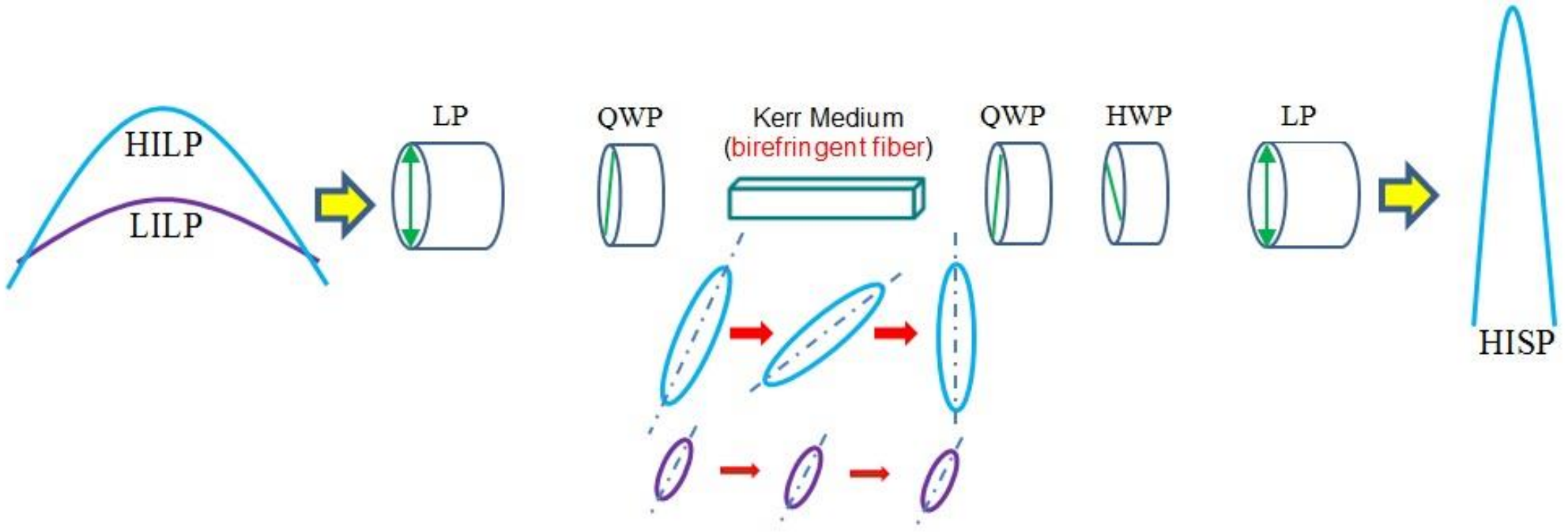

**Figure 2.2 The basic structure of the NPR output polarization controlled free-space cavity. Also, the process of pulse shortening in NPR cavity and nonlinear elliptical polarization rotation in fiber is modeled. High intensity pulse is sensitive to the NPR, while the low intensity pulse is not and blocked by the polarizer since the polarization direction is not adjusted. LP: linear polarizer, QWP: quarter waveplate, HWP: half waveplate, HILP: high intensity long pulse, LILP: low intensity long pulse, HISP: higher intensity short pulse [9, 73].**

In this free-space NPR mode-locking cavity, waveplates and linear polarizers are used in specific sequence, in contrast with fiber based polarizer controllers (PC) in an all-fiber NPR laser cavity [74]. By comparing the Figure 2.1and Figure 2.2, we can see that the cavity fiber section is corresponding to the Kerr medium, while the PBS and ISO are corresponding to the linear polarizer in the free-space. Previously, the NPR mode-locking technique is also called additive pulse mode-locking (APM), since right and left hand circular polarization components lead to a differential nonlinear phase shift, that employs a nonlinear interferometer to achieve pulse shortening. From the conceptual point of view, the mode-locking mechanism of NPR is identical to APM except that in NPR the orthogonally polarized components (in a birefringent fiber) of the same pulse are used instead of counter-propagating waves as in APM. One advantage of NPR mode-locking is that: it is extremely fast because it is based on self-phase modulation (SPM) from Kerr effect in fused silica glass made fiber. The NPR technique has been widely used in fiber lasers, where pulse shortening is achieved through SPM and polarization control.

### *2.1.2.1 Unidirectional Ring Laser Cavity*

Tamura *et al.* [75] reported that the choice of using ring cavity or linear cavity will affect how easily you can mode-lock the laser and self-starting of the mode-locking. The spurious reflections will create injection signals that pull the mode frequencies away from the desired even spacing, which will make the mode-locking harder to happen. The ring cavity geometry reduces the self-starting threshold by decreasing the mode-pulling due to spurious reflections and spatial hole burning. In our experimental setup of the NPR mode-locked laser, the ISO (shown in Figure 2.1 ) was used to ensure unidirectional laser operation.

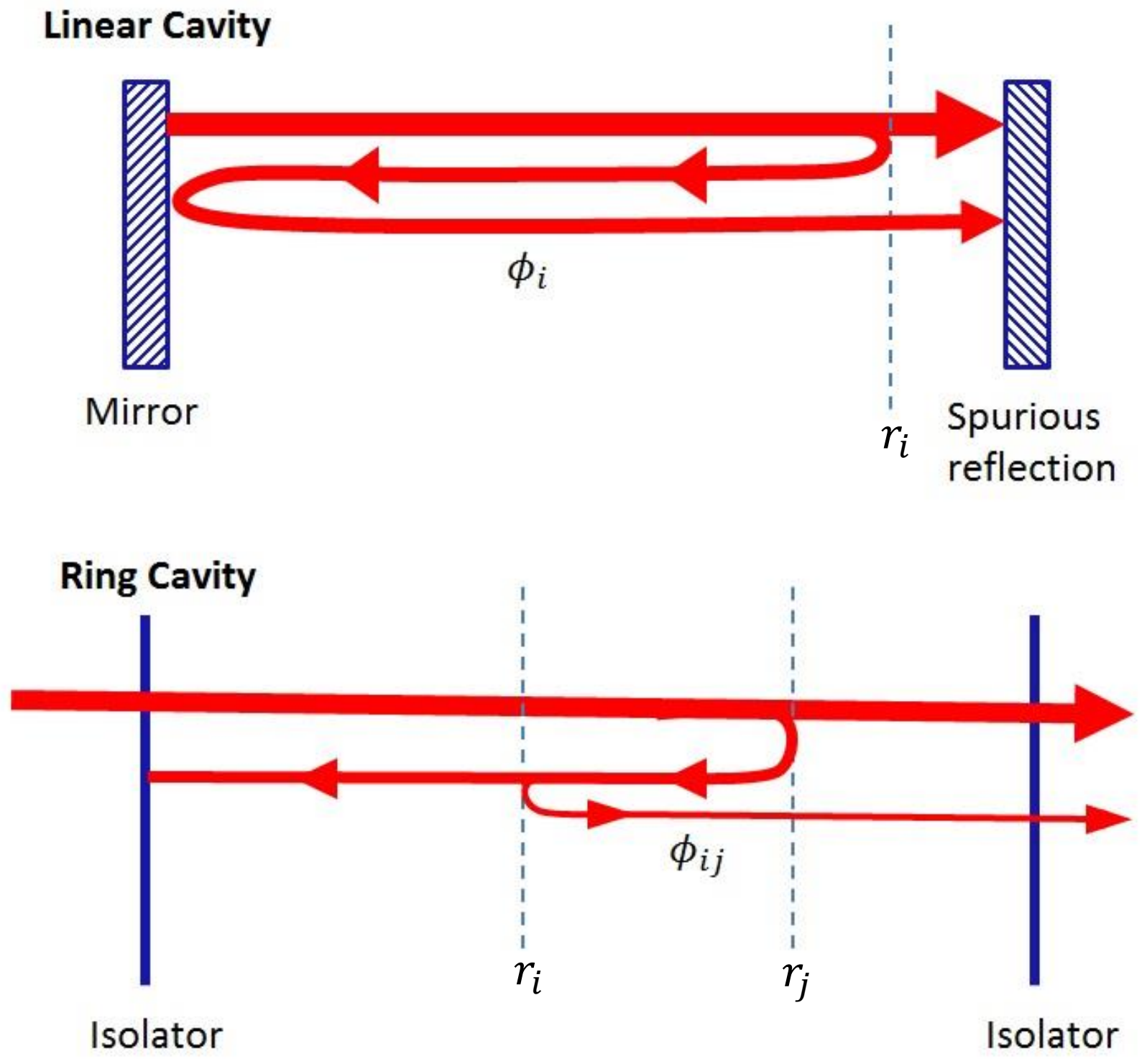

**Figure 2.3 Schematic of the etalon effect in a linear cavity laser (top) and a unidirectional ring cavity (bottom) laser reported by Tamura. Figure redrawn from Ref. [73].**

### 2.1.3 Dispersion Compensation in the Cavity

In the previous sections, we have mentioned the cavity structure of hybrid design. To get the laser mode-locked easily and obtain different mode-locking regime, we need to have the net cavity dispersion compensated. Basically, we will have the NPR oscillator operate in the soliton mode-locking regime if net cavity dispersion is negative (*i.e.* anomalous dispersion). On the other hand, to get stretched pulse mode-locking we need to have positive net cavity dispersion. Nevertheless, the net cavity dispersion should be managed less positive since too much positive cavity dispersion makes the laser mode-locking hard to obtain.  In order to calculate net cavity dispersion, we need cavity length and group velocity dispersion (GVD) of each section and parts in the cavity. First of all, have to calculate the refractive index of different optical materials using Sellmeir formula [76] which is proposed by Sellmeier in 1871.

$$n^2(\lambda) - 1 = \sum_{i=1} \frac{A_i \cdot \lambda^2}{\lambda^2 - \lambda_i^2} \tag{2.2}$$

As shown in Equation 2.2, this Sellmeier formula is given to calculate the refractive index as a function of wavelength rather than wave number, frequency or energy. In practice, we only need the three term Sellmeier formula (*i.e.* $i = 1,2,3$) in order to get the refractive index for fibers. A generalized form of the short-wavelength approximation to the Sellmeier formula is the Cauchy formula developed in 1836.

$$n(\lambda) = A_0 + \sum_{i=1} \frac{A_i}{\lambda^{2i}} \tag{2.3}$$

We can do the square and power series expansion of the Equation 2.3 to get the general form of Sellmeier formula for glasses known as Schott glass.

$$n^2(\lambda) = A_0 + A_1\lambda^2 + A_2\lambda^{-2} + A_3\lambda^{-4} + A_4\lambda^{-6} + A_5\lambda^{-8} \tag{2.4}$$

#### *2.1.3.1 Free-space Section*

First, we will consider the free-space optical bulk devices. Table 2.1 lists the material, Sellmeier coefficients and calculated GVD at 1560 nm for each free-space bulk optical devices (*e.g.* waveplate and PBS) [77].

**Table 2.1 NPR hybrid cavity free-space section bulk devices property**

| Devices | PBS | Quarter Wave Plate | Half Wave Plate |
|---|---|---|---|
| Material | SF2 | Crystalline Quartz | Crystalline Quartz |
| Sellmier Formula | Equation 2.2 | Equation 2.4 | Equation 2.4 |
| Sellmeier Coefficients | $A_1 = 1.40301821$<br>$A_2 = 0.231767504$<br>$A_3 = 0.939056586$<br>$\lambda_1^2 = 0.0105795466$<br>$\lambda_2^2 = 0.0493226978$<br>$\lambda_3^2 = 112.405955$ | $A_0 = 2.35728$<br>$A_1 = -1.17 \times 10^{-2}$<br>$A_2 = 1.054 \times 10^{-2}$<br>$A_3 = 1.34143$<br>$A_4 = -4.454 \times 10^{-7}$<br>$A_5 = 5.926 \times 10^{-8}$ | $A_0 = 2.35728$<br>$A_1 = -1.17 \times 10^{-2}$<br>$A_2 = 1.054 \times 10^{-2}$<br>$A_3 = 1.34143$<br>$A_4 = -4.454 \times 10^{-7}$<br>$A_5 = 5.926 \times 10^{-8}$ |
| Effective Length | 20 mm | 10 mm | 5 mm |
| GVD ($fs^2/nm$) | $1.90221 \times 10^{-7}$ | 0.00322784 | 0.00322784 |

In Table 2.1, the GVD values are calculated based on the relationship as below [78]:

$$\beta(\omega) = \frac{\omega}{c} n(\omega) \tag{2.5}$$

$$\beta_2(\omega) = \frac{d^2\beta(\omega)}{d\omega^2} \tag{2.6}$$

There are the simulation results for the refractive index and the GVD for the waveplates and the PBS in Figure 2.4. Furthermore, the refractive index and the GVD for waveplates have been calculated by Mathemetica as shown in Figure 2.5.

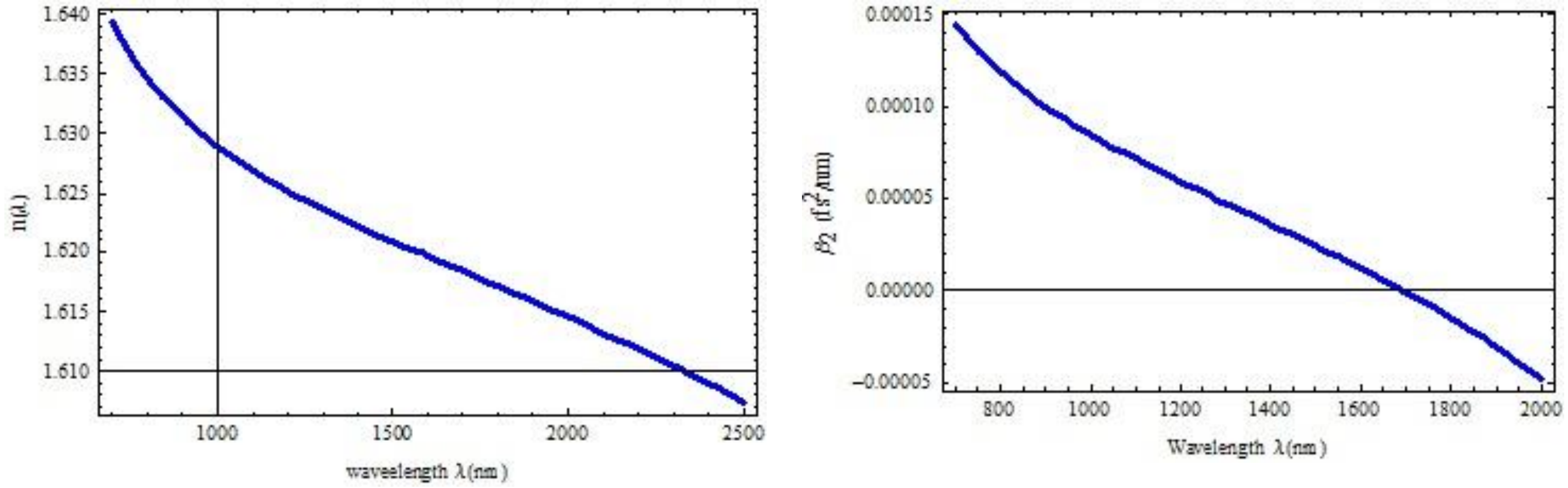

**Figure 2.4 The simulation of the calculated refractive index and the GVD for the PBS according to the Sellmier formula in Equation 2.2.**

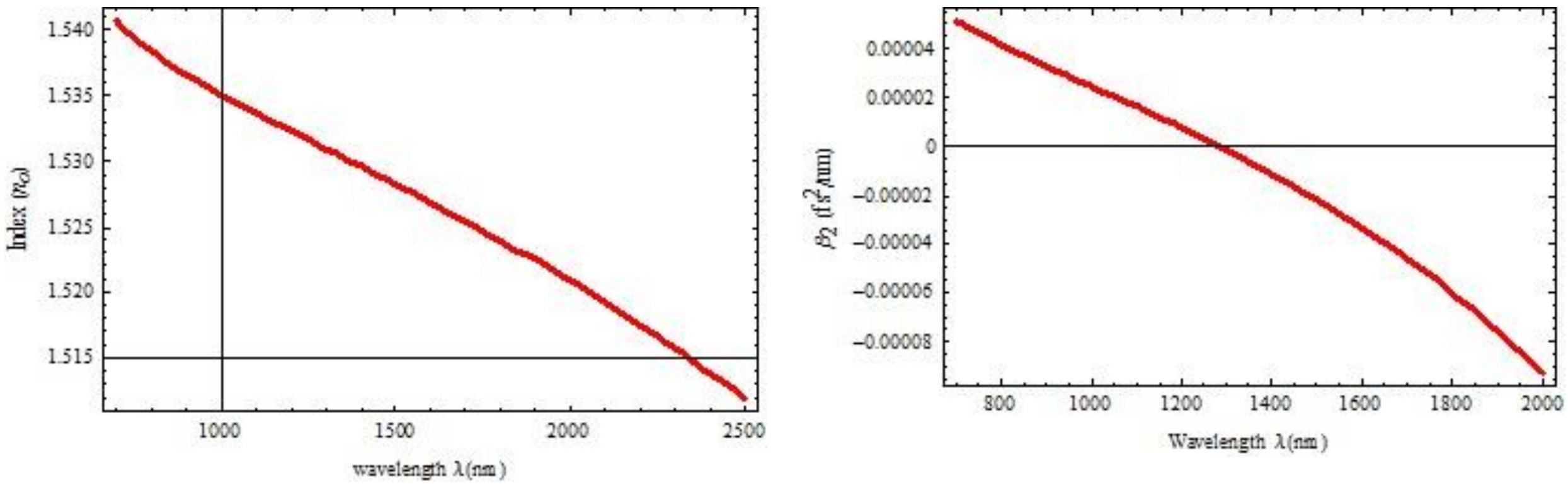

**Figure 2.5 The simulation of the calculated refractive index and the GVD for the waveplates (*e.g.* HWP and QWP) according to the Sellmier formula in Equation 2.2.**

### *2.1.3.2 Fiber Section*

We have used OFS EDF-150 as the gain fiber and Corning SMF-28e as the anomalous dispersion fiber and pigtail of the collimator. In addition, there is certain length of HI1060 flexcor fiber on the common port and signal port of the SIFAM 980/C-band WDM. Below are the essential parameters and properties for those fibers used in the cavity.

**Table 2.2 NPR cavity fiber section essential fiber parameters**

| Fibers | EDF-150 | SMF-28e | HI1060 |
|---|---|---|---|
| Cutoff Wavelength ($\lambda_c$) | 925 nm | 1260 nm | 920±50 nm |
| Numerical Aperture (NA) | 0.29 | 0.14 | 0.14 |
| Mode Field Diameter (MFD) | 4.3 $\mu m$@1550 nm | 10.4 $\mu m$@1550 nm | 6.2 $\mu m$@1060 nm |
| Dispersion Parameter (D) | -48 ps/nm/km | 17.929 ps/nm/km | 3.4 ps/nm/km |
| GVD Parameter ($\beta_2$) | 61.2215 $ps^2/km$ | -22.868 $ps^2/km$ | -4.3365 $ps^2/km$ |
| TOD Parameter ($\beta_3$) | -0.10006 $ps^3/km$ | 0.18704 $ps^3/km$ | 0.09398 $ps^3/km$ |
| Nonlinear Coefficient | 8.3741 $W^{-1}km^{-1}$ | 1.4315 $W^{-1}km^{-1}$ | 4.0280 $W^{-1}km^{-1}$ |
| (γ) | | | |

## 2.2 $Er^{3+}$doped Mode-locked Ultrashort Fiber Laser

In this section, we will present general description of the erbium doped mode locked fiber lasers. After that EDF gain and emission spectrum will be discussed theoretically and the experimental result measurement of ASE spectrum in EDF will be reported. In addition, we will review the pumping efficiency and gain saturation effect in EDF corresponding to the laser diode.

### 2.2.1 EDF Gain and Emission Spectrum

Rare earth doped fibers have been very popular in the area of fiber lasers and fiber amplifiers. In order to make laser comb for the application of optical metrology at central wavelength of 1.55 $\mu m$, we would choose the erbium doped fiber as gain fiber since it has absorption spectrum at 980 nm and 1550 nm and has emission spectrum at around 1550 nm.

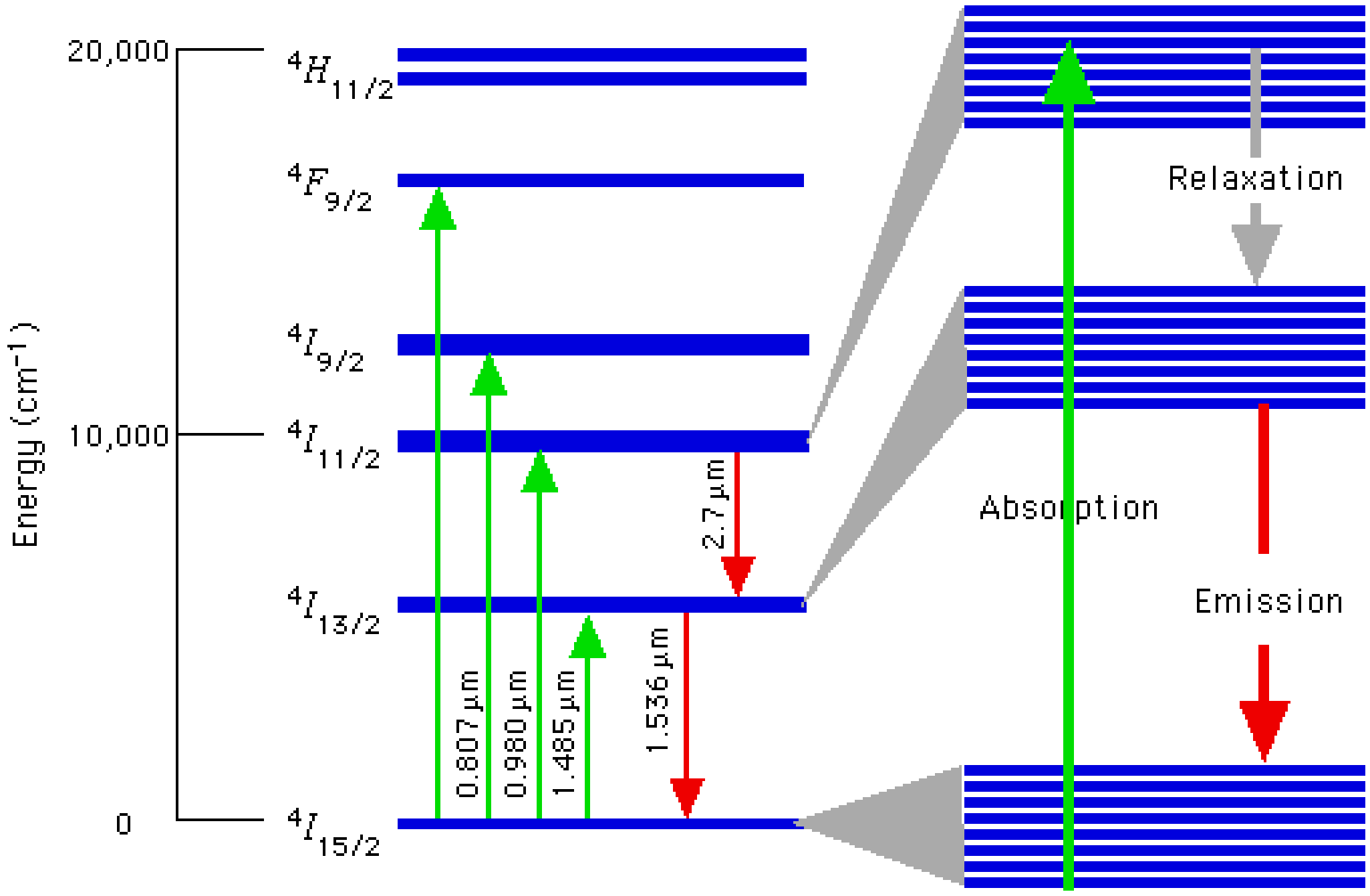

**Figure 2.6 EDF energy level diagram showing transition of absorption and emission spectrum of $Er^{3+}$ ion. In the laser cavity, EDF could be pumped by $0.98\ \mu m$ and $1.485\ \mu m$ and lase respectively at $1.536\ \mu m$ and $2.7\ \mu m$. Figure reproduced from Ref. [79].**

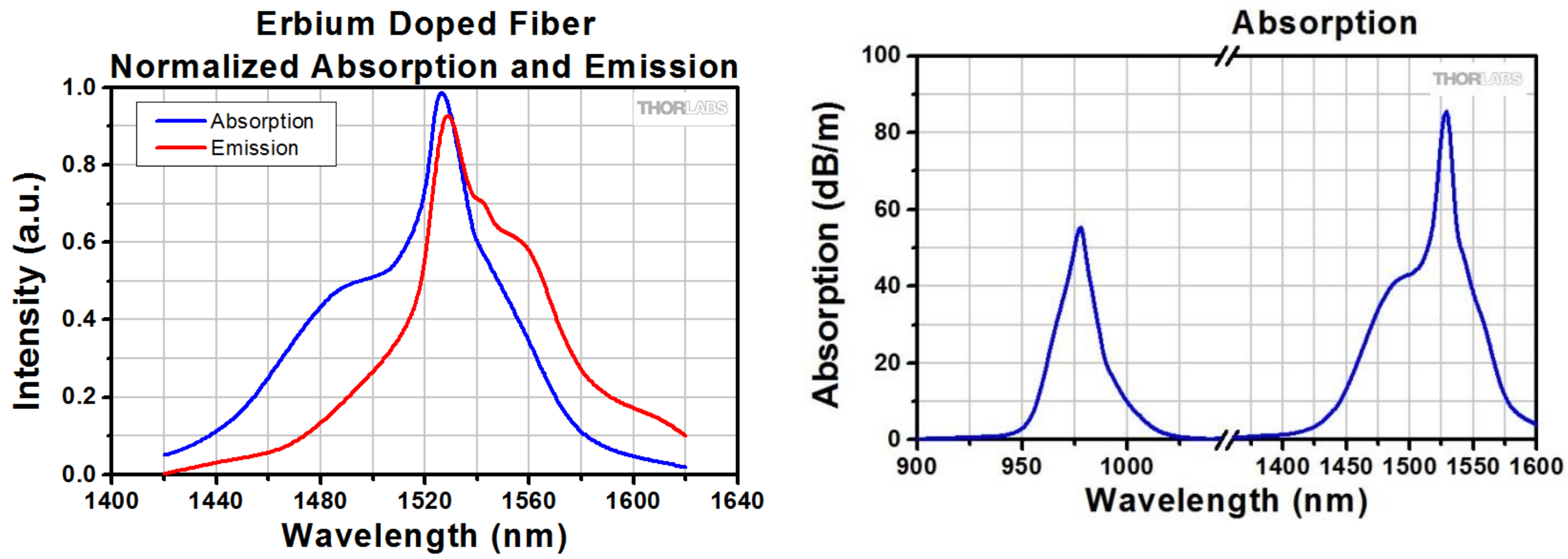

**Figure 2.7 Absorption and emission spectra for EDF from Thorlab. Figure reproduced from Ref. [80].**

From Figure 2.7, we can see that EDF has strong absorption characteristics at the wavelength of 980 nm and 1550 nm which makes it available for diode laser pumping sources.

### 2.2.2 ASE Spectrum Corresponding to Different Pump Power

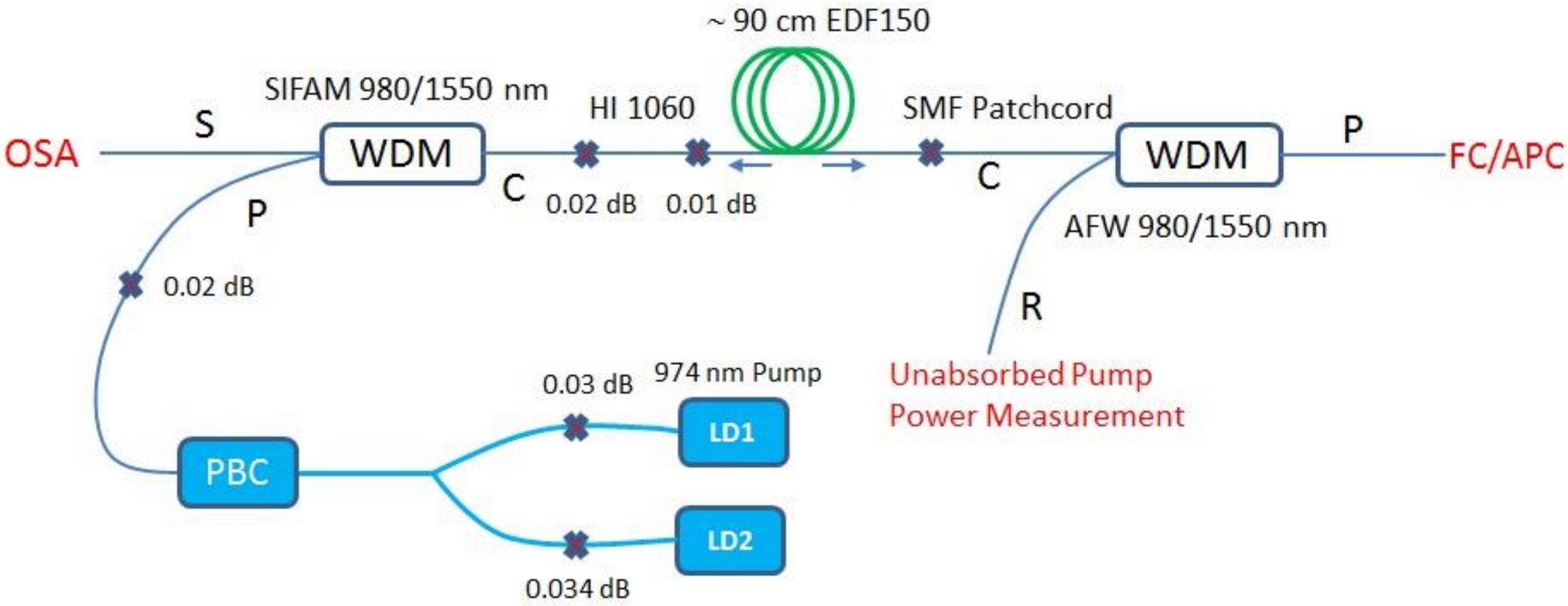

**Figure 2.8 Experimental setup schematic for forward (at pump port of AFW WDM) and backward (at signal port of SIFAM WDM) ASE measurement.**

Amplified spontaneous emission (ASE) in NPR mode locked erbium doped fiber laser operating at 1.5$\mu m$ is measured in both forward and backward pumping cavity.

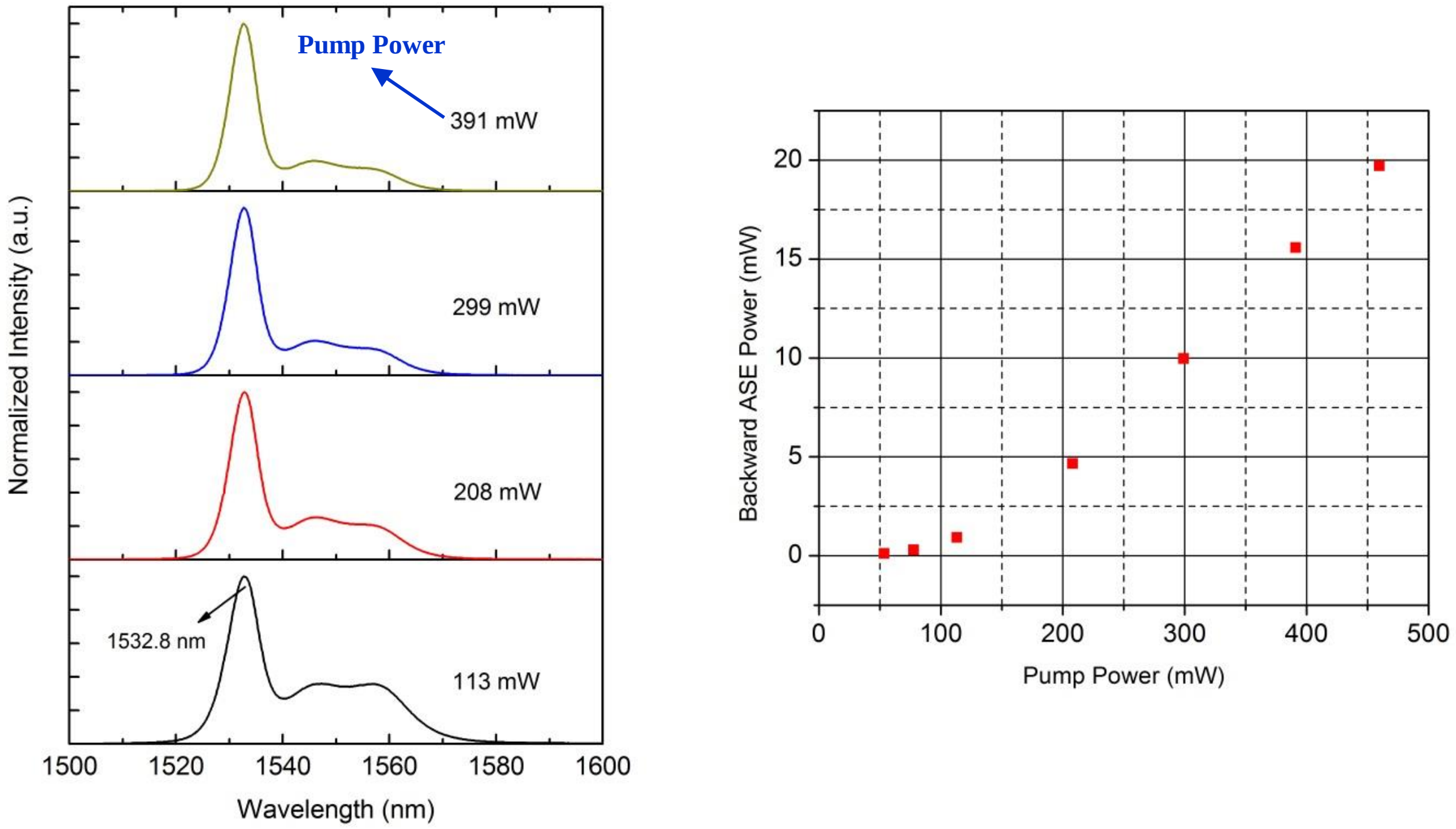

**Figure 2.9 Backward ASE spectrum (left) and ASE output power (right) corresponding to different pump power.**

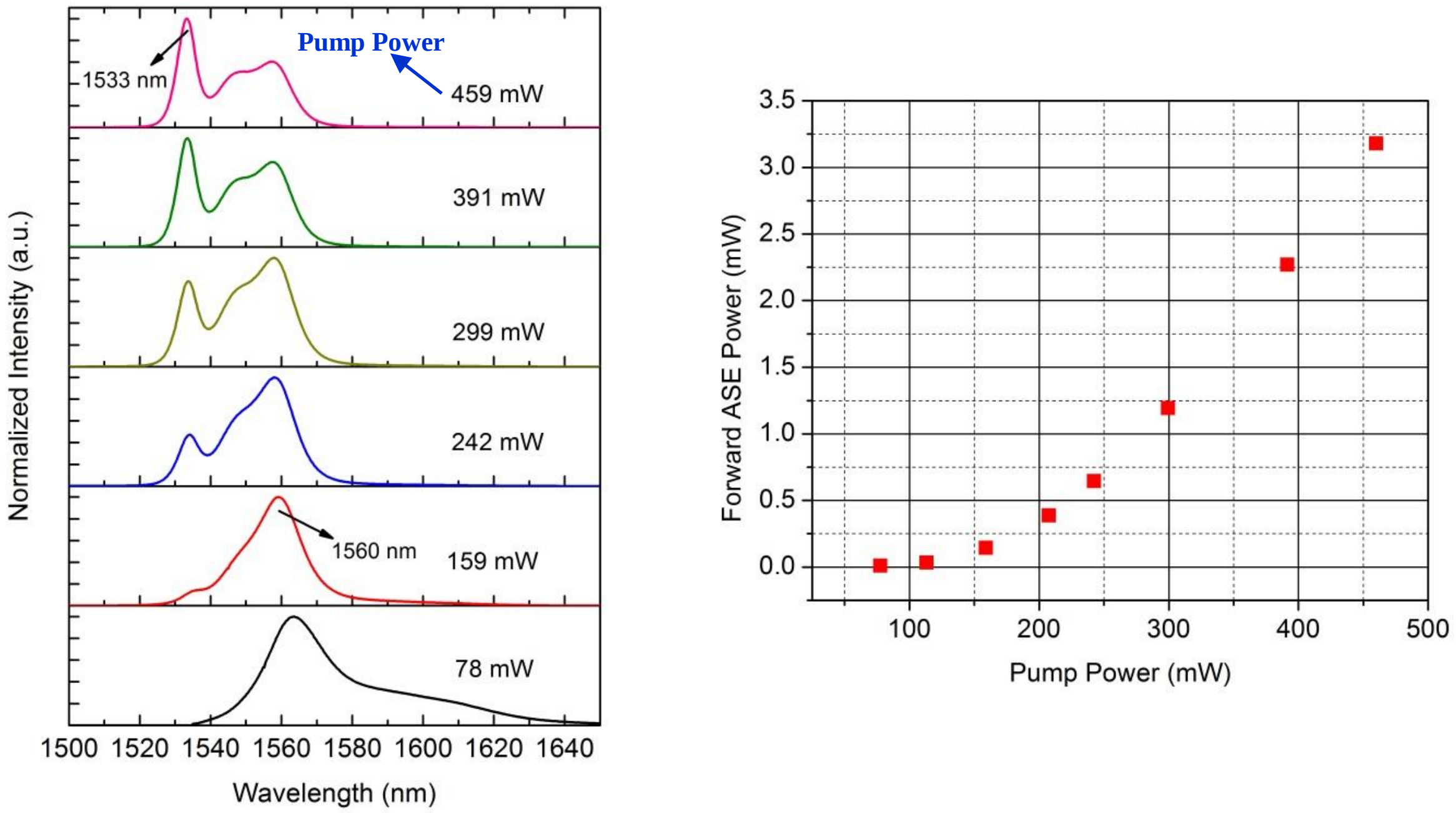

**Figure 2.10 Forward ASE spectrum (left) and ASE output power (right) corresponding to different pump power.**

We have measured the forward and backward ASE by Optical Spectrum Analyzer (OSA) with the gain fiber of 90 cm EDF-150. After the measurement, we found that there is less ASE noise in the forward pumping scheme compared to backward pumping scheme. That is the other reason why we choose forward pumping cavity rather than backward pumping cavity since less ASE noise means more stable laser operation.

### 2.2.3 Pumping Efficiency

As the pump power reaches the threshold of EDF the lasing will occur and will be increased correspondingly until the gain fiber is saturated and there is no more increase in the output power. In Figure 2.11, we can see that for a length of 39 cm EDF the pump light is not fully absorbed by the gain fiber, while the pump is efficiently absorbed for the length of 90 cm EDF.

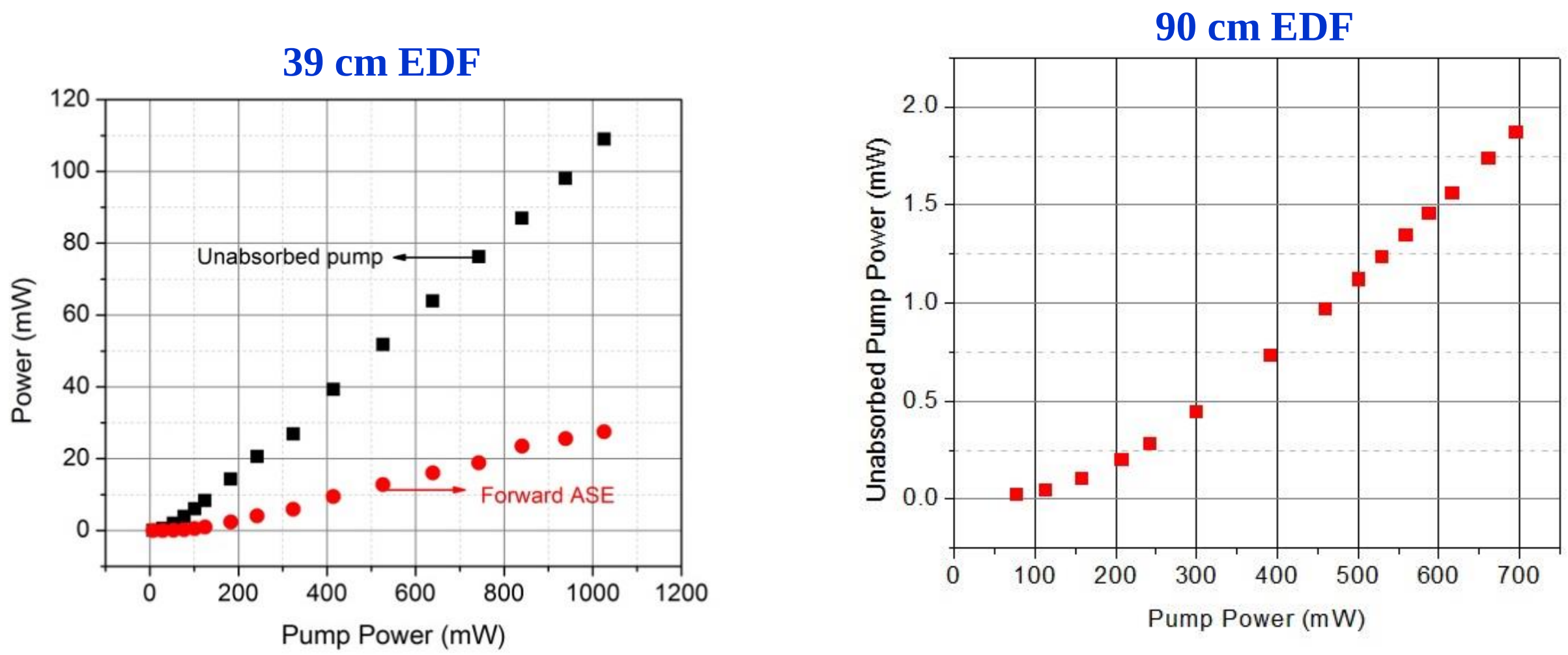

**Figure 2.11 The measured unabsorbed power measured after different length of EDF through reflective port of AFW WDM as in Figure 2.8.**

To describe the light propagation in the weakly birefringent fibers, we used coupled complex nonlinear Schrödinger equations. For erbium doped fiber, the gain saturation can be written as [81]:

$$g = G\,exp\left(-\frac{\int(|u|^2 + |v|^2)dt}{P_{sat}}\right) \tag{2.7}$$

where $G$ is small signal gain coefficient, $P_{sat}$ is the normalized saturation energy, and $u$ and $v$ are the normalized envelopes of the optical pulses along the two orthogonal polarization axes of the fiber.

## 2.3 Time and Frequency Characteristics of the Pulse Generated by the Comb

In this section, we will discuss the nonlinear effects in optical fiber. Cavity mode competition and multi-pulse suppression in the erbium doped fiber laser oscillator will be briefly discussed. Then, the single-pulse operation of the oscillator will be presented.

### 2.3.1 Cavity Mode Competition

#### *2.3.1.1 Cavity Modes*

When a laser is lasing there exist longitudinal mode in the cavity. In a linear cavity, the oscillation frequency condition is:

$$\omega = \omega_q = q \times 2\pi \times (\frac{2L}{c}) \tag{2.8}$$

The set of frequencies $\omega_q$ are called longitudinal cavity modes (sometimes also called axial modes) since they represent the resonant frequencies at which there exist exactly q half-wavelengths along the resonant cavity. The number of longitudinal modes is usually very large, *e.g.* for the standing wave case [82]:

$$q = \frac{\omega_q L}{\pi c} = \frac{2L}{\lambda_q} \approx 10^5 \sim 10^6 \tag{2.9}$$

Then, the laser gain bandwidth and the laser output spectrum can be related as shown below.

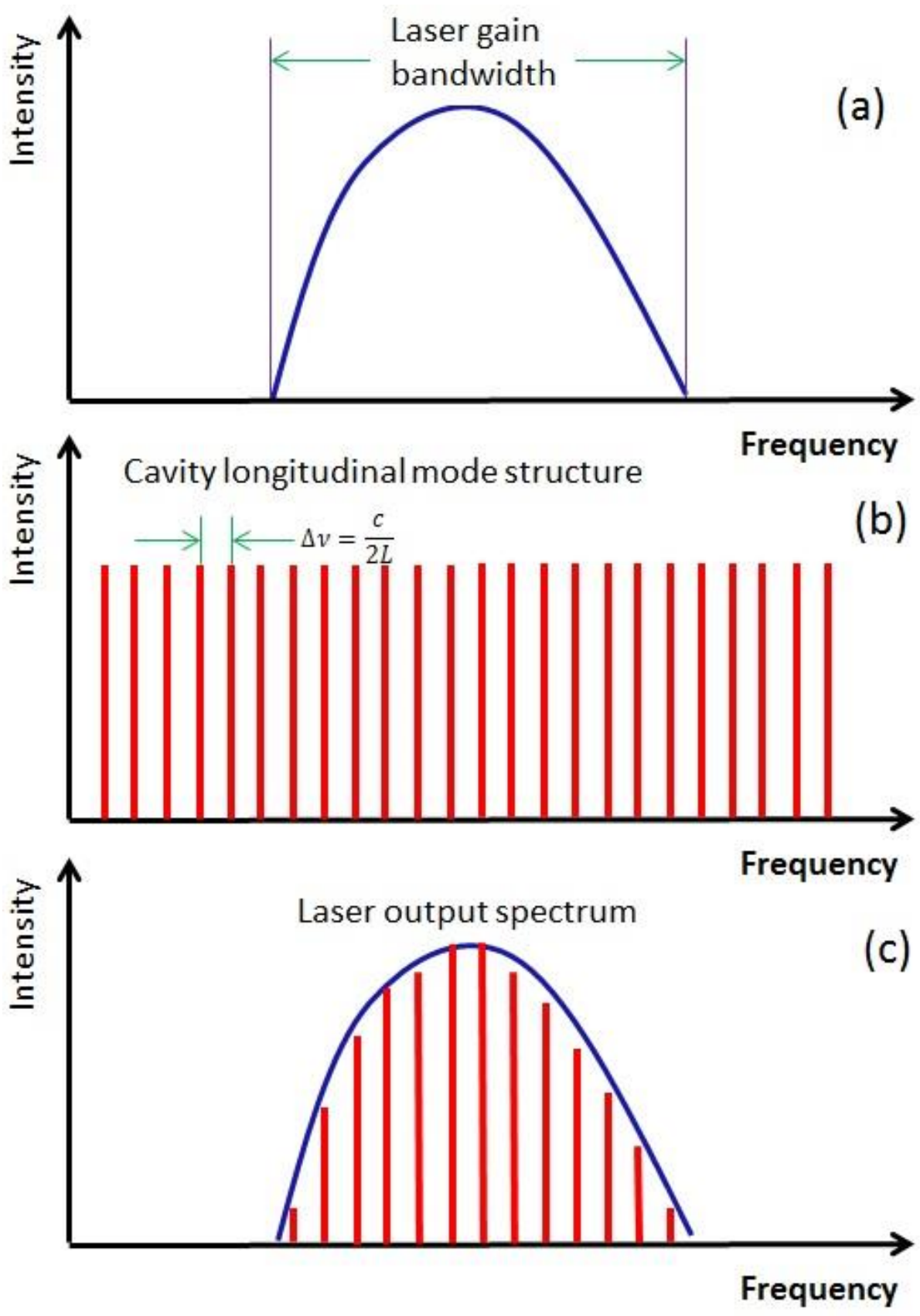

**Figure 2.12 A schematic of the longitudinal mode structure in a laser. (a) The gain medium of the laser will only amplify light over certain range of frequencies. This is corresponding to a laser's gain bandwidth. (b) The longitudinal modes are equally spaced by $\frac{c}{2L}$ (c) Only the modes whose corresponding frequencies fall into the laser's gain bandwidth will be amplified and lased out. Figure redrawn from Ref. [83].**

In a fiber, all the modes (including TE, TM, and HE modes) can be classified into linearly polarized ($LP_{lm}$) modes in approximation of weak guidance, due to the refractive index difference between core and cladding. The mode subscripts $l$ and $m$ describe the electric field intensity profile. There are $l$ field maxima around the fiber core circumference and $m$ field maxima along the fiber core radial direction [84].

In terms of fundamental mode ($LP_{01}$) or higher order modes in a step-index fiber, we have the standard parameter called V number:

$$V = \frac{2a\pi}{\lambda}\sqrt{n_{core}^2 - n_{cladding}^2} \tag{2.10}$$

where $a$ is the radius of the fiber core, $\lambda$ is the vacuum wavelength. For a V value that is below $V \approx 2.405$, a fiber supports only one mode per polarization direction which is a single-mode fiber. On the other hand, for a V value that is much higher than 2.405 there are multiple modes (such as in multimode fibers). The number of modes in a step-index fiber with high V value can be calculated approximately as:

$$M \approx \frac{V^2}{2} \tag{2.11}$$

#### *2.3.1.2 Mode Competition*

The term mode competition refers to the differing gains of various transverse modes in the presence of saturation of the population inversion, often referred to as transverse hole burning. Spatial hole burning by a strong fundamental leaves population inversion that has a higher overlap with high order modes rather than fundamental modes. The cutoff wavelength of SMF28-e is 1280 nm, therefore, there will be multimode exist corresponding to the 980 nm laser diode pump for the NPR laser comb. In fact, all lasers become multi-moded when pumped hard enough (unless the gain medium is first destroyed by too high of a pumping rate) [85]. To attenuate the extra modes in the fiber, we have placed the fiber in small radius coiling. The differential bend loss between the fundamental and higher order modes may be sufficient to suppress higher order modes without excess loss of the fundamental [86].

In addition, mode competition caused by homogeneous gain broadening in EDF can be alleviated using inhomogeneous loss characteristics of NPR which induces intensity dependent loss for different frequencies. As a result, the balance between the inhomogeneous loss induced by NPR and the mode competition effect of EDF can lead to stable multi-wavelength oscillation at room temperature [87].

### 2.3.2 Relaxation Oscillation in EDF

The NPR laser is initially mode-locked with a lower pump power and operates in the pulsed lasing regime. However, we observed the phenomena of both CW lasing (shown on right in Figure 2.13) at 1564 nm and pulsed lasing, as we increase the pump power to 309 mW.

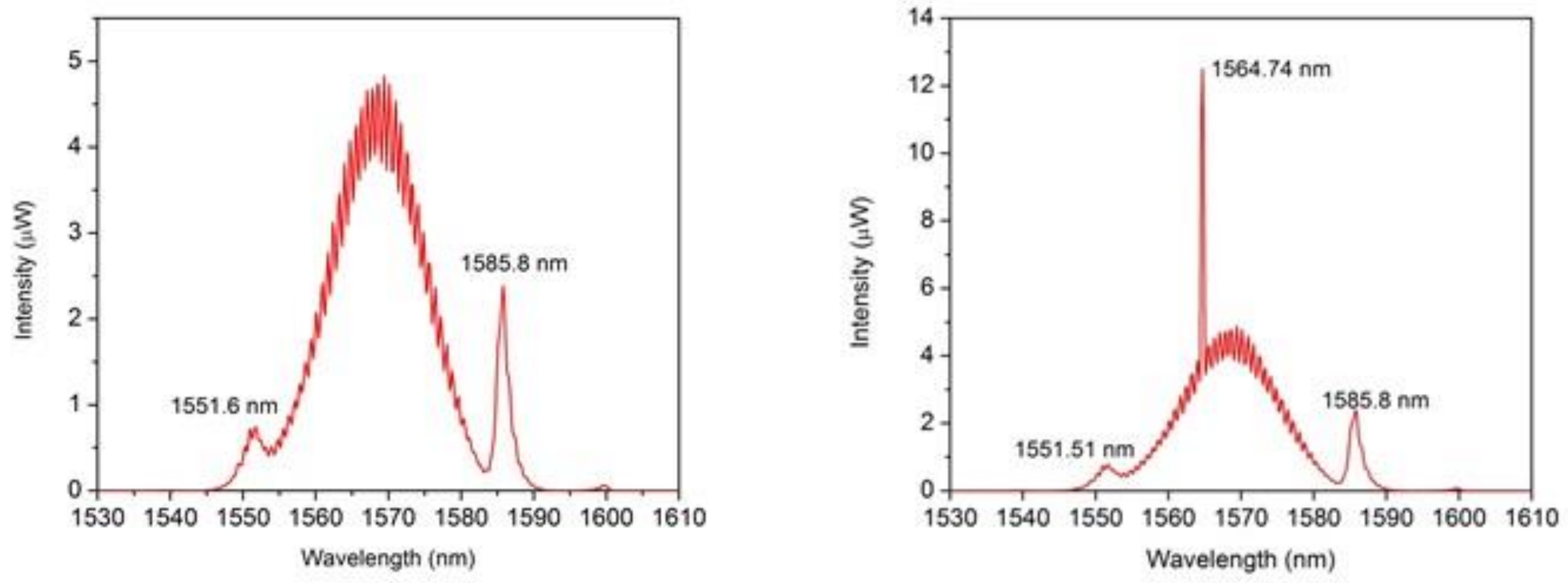

**Figure 2.13 Optical spectrum analyzer spectrum in linear scale for mode locked pulsed lasing (left) and simultaneous CW and pulsed lasing (right) operation of the NPR laser. The sidebands positioned symmetrically on the two sides of the central wavelength are called Kelly sidebands, which will be discussed next in Section 2.4.**

The CW lasing is due to the relaxation oscillations in EDF. We can describe the relaxation oscillations by the relaxation equation of the gain and the relaxation equation of the optical energy in the cavity. For the CW case, we assume [88]:

$$\frac{1}{\tau_{sp}} \ll g_N S_0 \ll \frac{1}{\tau_c} \tag{2.12}$$

where $\tau_{sp}$ is the lifetime of the excited state of $Er^{3+}$, $g_N S_0$ is the product of the differential gain $g_N$ and the photon number in the cavity, and $\tau_c$ is the lifetime of the photons in the cavity. The erbium ions have a very long excited state lifetime ($\sim 10\ ms$) while the lifetime of the photons in the cavity is of the order of 100 ns or less, which makes the approximation valid for this case. Then, the frequency of the relaxation oscillations can be written as:

$$\omega_r \approx \sqrt{\frac{g_N}{\tau_c} S_0} \tag{2.13}$$

It can be conclude that relaxation oscillations are those of the CW background, and the pulse train coexists without being affected by them and without affecting the relaxation oscillations in turn. In order to achieve low noise, we must eliminate CW radiation to avoid relaxation oscillations. In the experiment, we have managed to eliminate the CW radiation by decreasing the pump power or slightly adjusting the waveplates in the free-space section of the NPR laser cavity. The later method, which takes advantage of the fast saturable absorber like NPR effect to induce frequency dependent loss for the CW component, is more desirable when concerning about the high output power from the NPR laser comb.

## 2.4 Experimental Results

In this section, we will present the experimentally measured results and interpret theoretically the soliton mode locking and stretched pulse mode locking principle. We will also discuss the Kelly sideband generation in the soliton mode-locking regime. Most importantly, we will report the high output power stretched pulse generated by the positive net cavity dispersion scheme.

### 2.4.1 Solitonic Mode-locking

Prior to obtaining mode locking, we have considered all the elements of mode locking in the hybrid NPR mode locking laser cavity. First, the insertion loss of all optical devices have been measured, including WDM, PBC, PBS, ISO, waveplates and fiber splicing. Thus, we make sure that laser operation is not affected by insertion losses of the optical devices in the cavity.

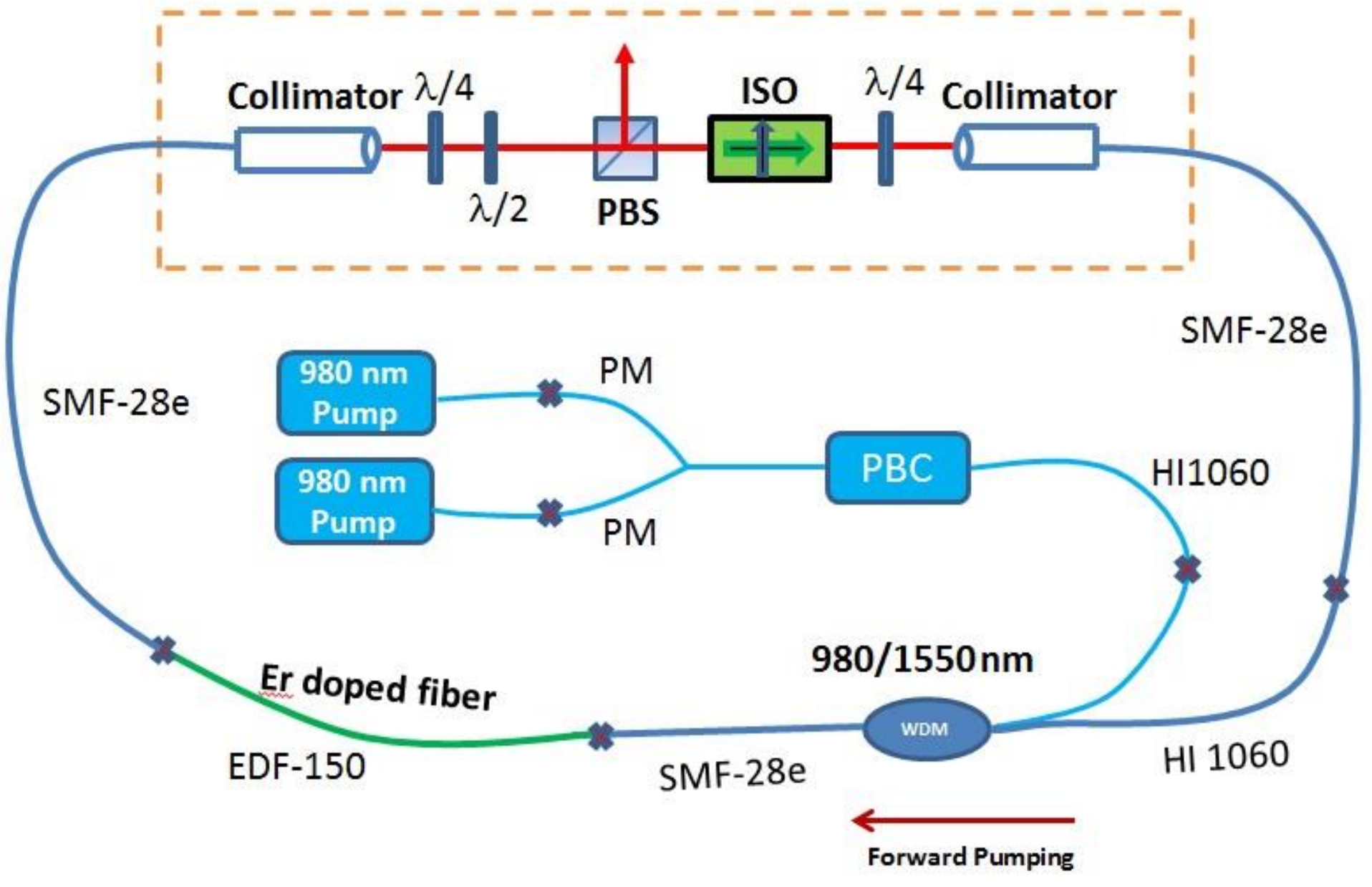

**Figure 2.14 The free-space NPR laser solitonic mode-locking experimental setup. All the abbreviations are already defined in previous figures.**

So, we calculate the GVD for the cavity including free-space section and fiber section. As we mentioned in Section 2.1 , we need to have anomalous net cavity dispersion in order to get solitonic mode-locked pulse generated from the oscillator.

It is well known that it is difficult or impossible to realize mode-locking if there are several reflecting surfaces in the cavity, therefore, we have changed the FC/PC fiber patchcord which is connected to the collimators as shown in Figure 2.14, to FC/APC. That is because usually there is about 4% reflection from a flat cleaved (*i.e.* in FC/PC case) fiber end, but the reflection can be eliminated greatly by angle cleaving (*i.e.* in FC/APC case) the fiber end.

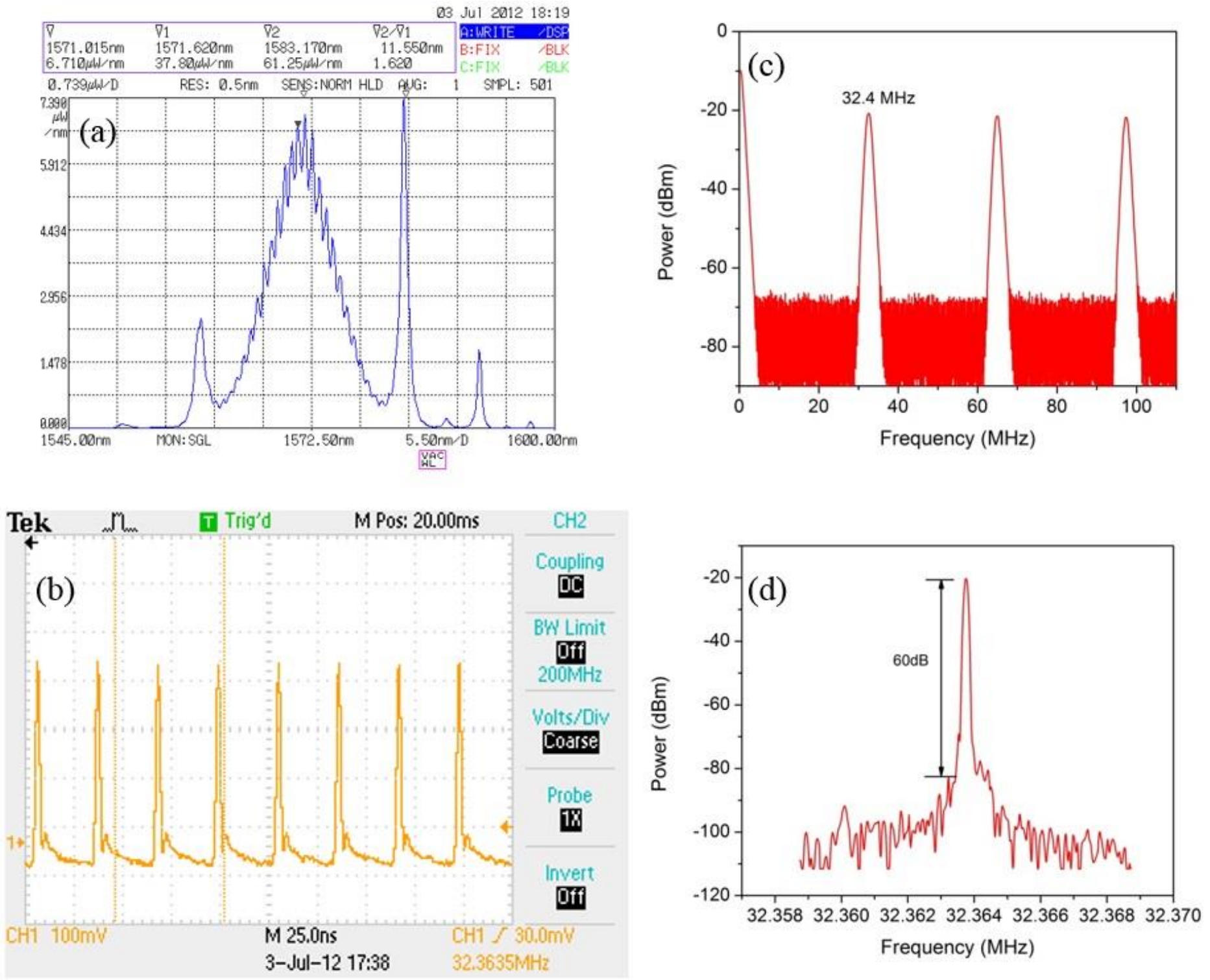

**Figure 2.15 (a) Optical spectrum measured with OSA in linear scale. (b) The oscilloscope trace of the pulse train. (c) RF spectrum of 32.4 MHz pulse measured with ESA. (d) Zoom-in figure of the fundamental mode.**

This solitonic pulse is obtained after we have adjusted the waveplates back and forth so many times. In addition, changing the pump power a little every time after we have tilted one set of angles for the three waveplates which corresponds to different polarization state of the NPR comb. At this point, we have 80 cm length of EDF-150, 3.3 m length of SMF-28e and 55 cm of HI1060 flexcore fiber in the fiber section of the cavity. To make sure it is in the solitonic mode-locking regime, we have calculated the net cavity dispersion of $-0.11625\ ps^2$, which is obviously anomalous. The pump power from the double laser diode pump is about 540 mW at 980 nm.

Our goal has been to find the mode locking state where the net cavity dispersion is positive and the average output power of the comb is high enough about tens of mW level. To do that, we need to manage the cavity GVD by optimizing the length of EDF-150 and the SMF-28e, which respectively have positive and anomalous dispersion.

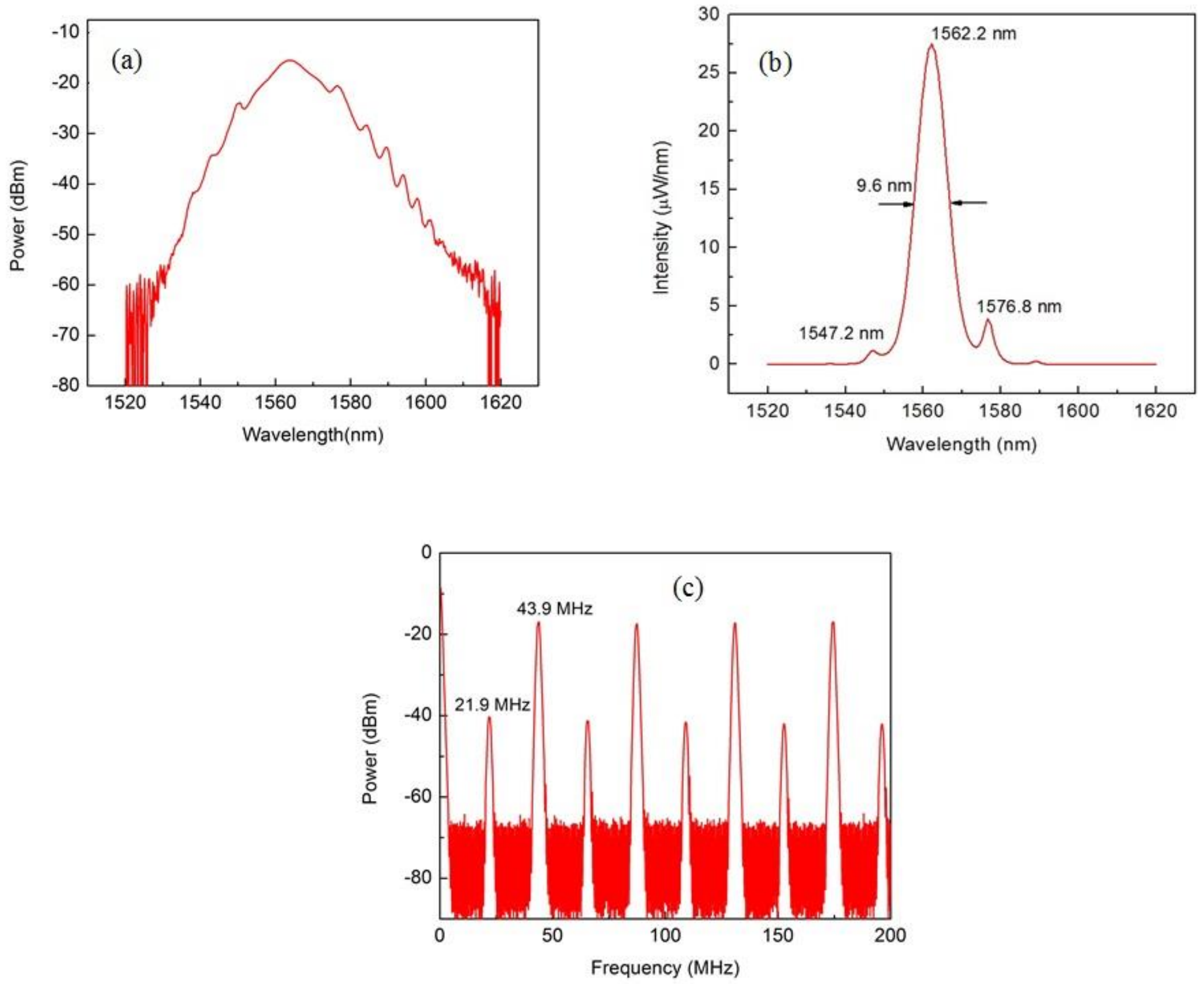

**Figure 2.16 (a) Optical spectrum of solitonic mode-locking of multi-pulsing in logarithmic scale. (b) Optical spectrum of solitonic mode-locking of multi-pulsing in linear scale. (c) RF frequency of two different frequencies from multi-pulsing.**

At this point, we have the net cavity dispersion of $-0.0343ps^2$ for a total length of 3.26 m SMF-28e, 46 cm of HI1060 flexcore fiber and 69 cm of EDF-150. The pump power from double laser diode is about 240 mW, while the average output power from the comb is 15.23 mW at 1560 nm. To get rid of the multi-pulsing, we have decreased the pump power to about 200 mW. As a result, the laser is operating at a single repetition frequency of 43.9 MHz.

### 2.4.2 Sideband Generation

As we can see from the result in Figure 2.15(a), there exist symmetric positioned sidebands in the soliton spectrum, which are called Kelly sidebands. This sideband presents itself when the soliton pulse is periodically perturbed by the linear effect of anomalous GVD and the nonlinear effect of self-phase modulation (SPM) [89]. Kelly sidebands are usually not observed in mode-locked bulk lasers (*e.g.* Ti:Sapphire laser), but it is common in soliton fiber lasers. That is because the intracavity nonlinearity and dispersion is much stronger especially in mode locked fiber lasers. Dennis *et al.* argued that soliton is generated due to the resonant enhancement after many cavity round trips of dispersive wave components shed from the soliton-like cavity pulse when perturbed by the discrete loss in the cavity [90]. The pulse energy experiences discreet loss due to the cavity components including output coupler (PBS in our case), fiber splices, waveplates, isolator and etc. In order to obtain the expression for Kelly sideband, we can write down the dispersive wave propagation constant as:

$$\beta_d(\omega) = \beta_0 + \beta_1 \Delta\omega + \frac{1}{2}\beta_2 \Delta\omega^2 + \frac{1}{6}\beta_3 \Delta\omega^3 + \cdots \tag{2.14}$$

Then the wavelength shift from the soliton central wavelength can be expressed[90].

$$\Delta\lambda_N = \pm N \cdot \lambda_0 \sqrt{\frac{2N}{cDL} - 0.0787 \frac{\lambda_0^2}{(c\tau)^2}} \tag{2.15}$$

where *N* is the sideband order, $\lambda_0$ is the central wavelength of the soliton, *c* is the velocity of light in vacuum, *D* is the fiber dispersion parameter and $\tau$ is the experimentally measured FWHM pulse duration.

As Kelly sidebands are formed from the constructive interference of the dispersive waves generated from the successive round trips in the laser, we can use it to accurately measure phase velocities of solitons. Most importantly, we could find the total cavity dispersion taking advantage of these sidebands.

### 2.4.3 Stretched Pulse Mode-locking

When we have cut-back the SMF-28e to the length of 1.96 m while keeping the length of HI060 and EDF-150 respectively at 0.46 m and 0.69 m, we have net cavity dispersion of $-0.004573\ ps^2$. Then we have observed the mode-locked comb spectrum with no sidebands.

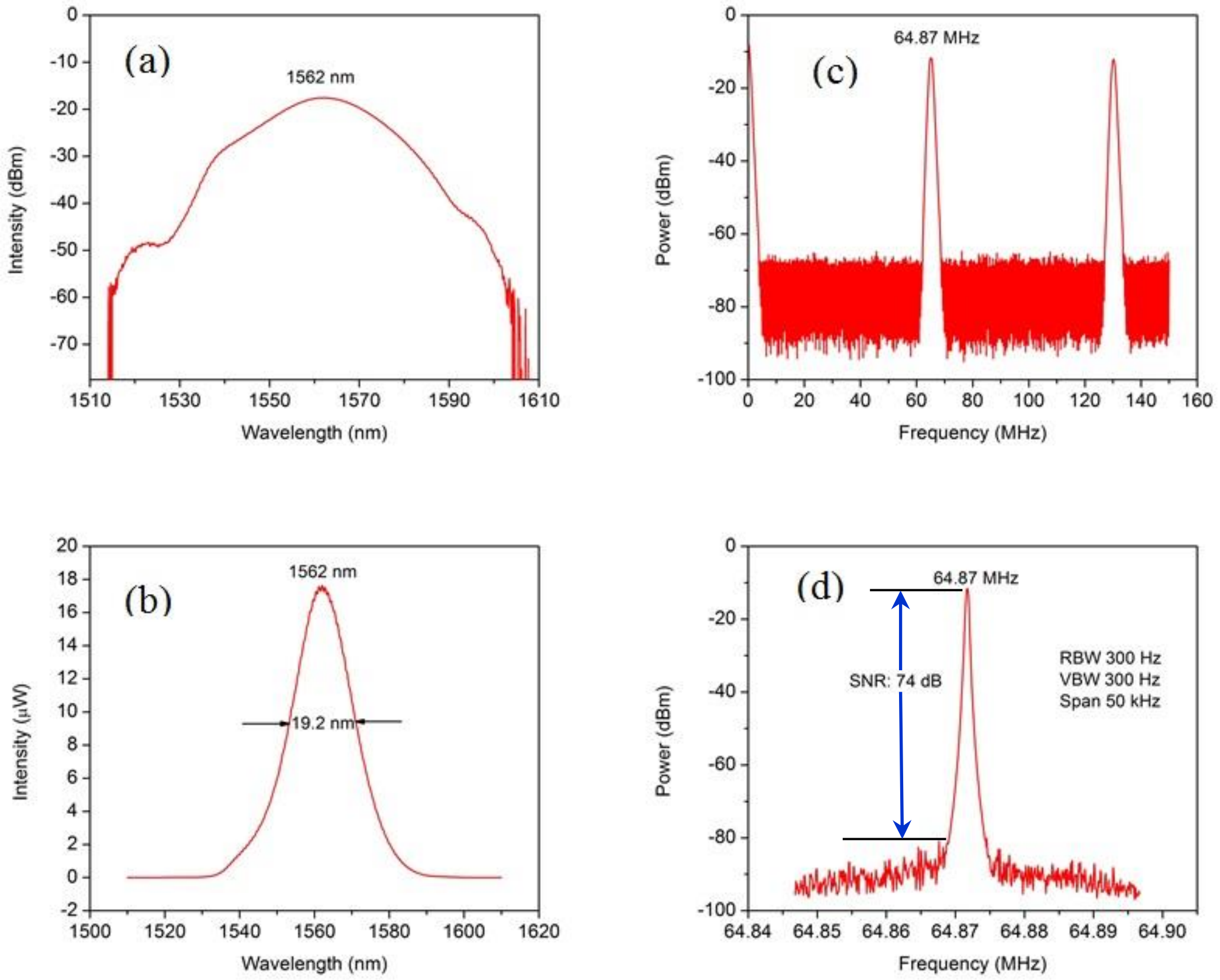

**Figure 2.17 The initial stretched pulse mode-locking spectrum experimental result. (a) Spectrum in logarithmic scale measured by OSA. (b) Spectrum in linear scale measured by OSA. (c) RF spectrum measured by ESA. (d) Zoom-in of the RF spectrum at 64.87 MHz.**

In order to use this comb in our lab, we need to tune the repetition frequency to 89.238 MHz. In addition, we need to obtain a stretched pulse mode-locking state that has output power and broader bandwidth to generate ultrashort high energy pulse. First, we have obtained a repetition frequency near 89 MHz by shortening the total length of fiber in the cavity, simultaneously, optimize the length of anomalous dispersion fiber and positive dispersion fiber so that net cavity dispersion is positive and near zero. To achieve the exact repetition frequency of 89.238 MHz for the NPR comb, we have taken advantage of the micrometer (which is placed under one of the collimator) to further adjust the total cavity length. Finally, we have managed to get the expected NPR mode-locked ultrafast laser pulse at the repetition rate of 89.238MHz and about 56 nm bandwidth.

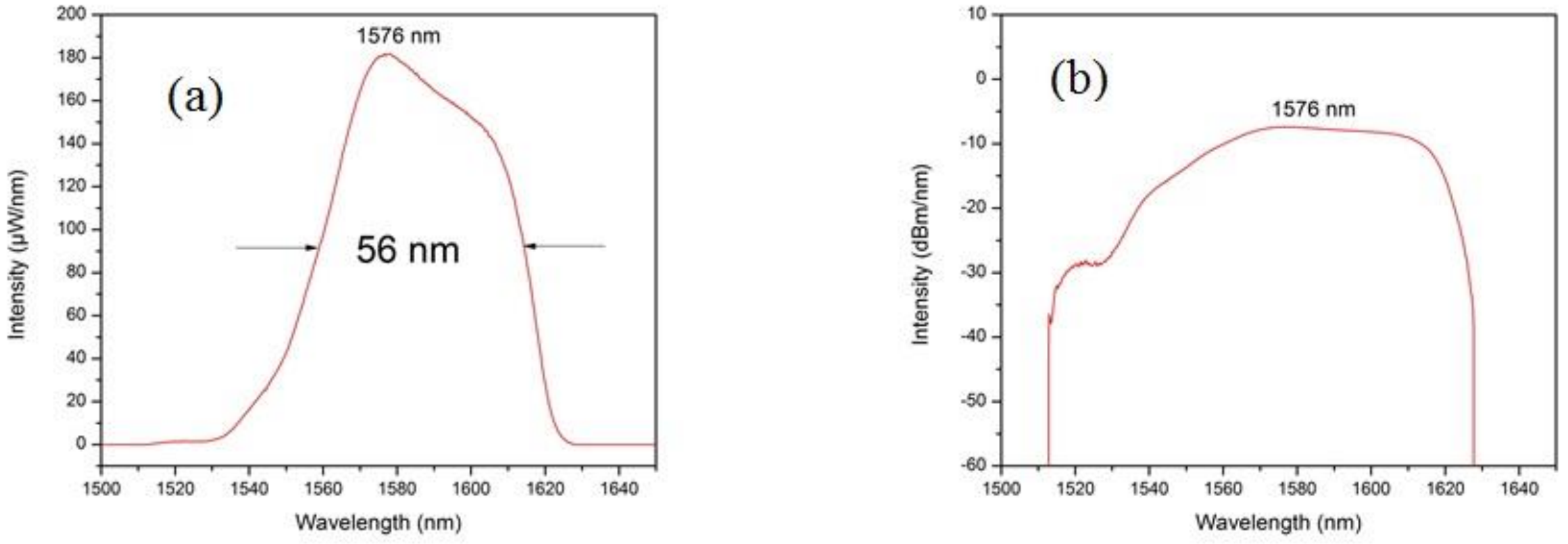

**Figure 2.18 Stretched pulse mode-locked spectrum of the NPR comb measured with OSA in linear scale (a) and logarithmic scale (b). The repetition frequency at this mode-locking state is 89.238 MHz.**

This mode locked laser oscillator has essential characteristics for the potential application in a frequency comb system, since it has more than 30 mW average output power and more than 50 nm frequency bandwidth, while conventional all-fiber based NPR mode locked lasers were merely able to generate less than 1 mW average output power and only around maximum 10 nm of frequency bandwidth [91].

## 2.5 Summary

In this chapter, we have first presented the hybrid cavity design of free-space and fiber based NPR mode locking laser. In addition, we have discussed the free-space NPR mode locking principle that implemented in our experimental setup. Solitonic pulse in the fiber is formed by common effect of SPM and GVD. When GVD is dominant in the cavity, we need to have the net cavity dispersion compensated so that we will have solitonic pulse in the anomalous dispersion regime, while one needs a net positive cavity dispersion in order to have stretched pulse generated.

We also have reviewed the NPR mode locked $Er^{3+}$ doped fiber laser. In the 1550 nm wavelength application area, we have taken advantage of the gain spectrum of the EDF. However, we have to avoid the noise effect of ASE emission which is due to inherent property of EDF. At the same time, the pumping efficiency and gain saturation of EDF is of vital important in erbium doped fiber lasers.

This chapter also discussed the cavity mode competition and the role of NPR mode locking to eliminate this phenomenon. In the experiment, we have found relaxation oscillations which lead to CW and pulse coexistence in the mode locked laser.

We have reported our experimental result in terms of achieving high output power, high pulse energy, broad frequency bandwidth and stable self-starting NPR mode-locked oscillator. Meanwhile, some of the typical nonlinear effects that affected our result of solitonic mode-locking pulses and stretched mode-locking pulse have been briefly discussed. To this point, we have presented our work to setup frequency comb oscillator setup. In the next chapter, we are going to discuss how to amplify and compress the pulse generated by this comb oscillator.

# Chapter 3 - Ultrashort Pulse Amplification and Compression

In this chapter, we will discuss the amplification of ultrafast pulses from the NPR mode-locked erbium doped fiber laser with 30 mW average output power and about 50 nm frequency bandwidth. In fact, prior to the erbium doped fiber amplifier (EDFA) the pulse is compressed to near transform limited (TL) pulse duration (*i.e.* time-bandwidth product for a $\text{sech}^2$ pulse $\approx$ 0.315) using certain length of anomalous dispersion (at 1550 nm) single mode fiber (SMF). After that, the unchirped pulse is injected as seed light into the EDFA which is buildup from a 90 cm EDF and a laser diode pump with central wavelength at 1480 nm.

As a result, we have obtained the amplified pulse of average output power about 180 mW near 1580 nm. Finally, the amplified pulse is compressed to 70 fs near TL pulse, after the compression by an anomalous dispersion SMF of a length about 2.9 m.

## 3.1 Overview of the EDFA Experimental Setup

### 3.1.1 Pre-compression of the Pulse

The output pulse from the NPR is broadened due to the combination of SPM and GVD in the normal (*i.e.* positive GVD) dispersion laser cavity. This came clear to us after we have measured the temporal pulse duration with the aid of a commercial intensity autocorrelator (AC). Assume the electric field of input pulse is $E(t)$, and then the electric field of pulse after the delay translation stage with time-delay $\tau$ can be written as $E(t-\tau)$. The electric field of the signal after the SHG crystal is:

$$E_{sig}^{SHG}(t,\tau) \propto E(t) \cdot E(t-\tau) \tag{3.1}$$

then, we have the intensity of the signal that is proportional to the intensity of each pulse.

$$I_{sig}^{SHG} \propto I(t) \cdot I(t-\tau) \tag{3.2}$$

However, the detectors are usually very slow that the SHG signal cannot be directly detected, so the measured autocorrelation intensity $I_{AC}(\tau)$ is the integration of the SHG signal intensity as [92]:

$$I_{AC}(\tau) \propto \int_{-\infty}^{+\infty} I(t) \cdot I(t-\tau)\, dt \tag{3.3}$$

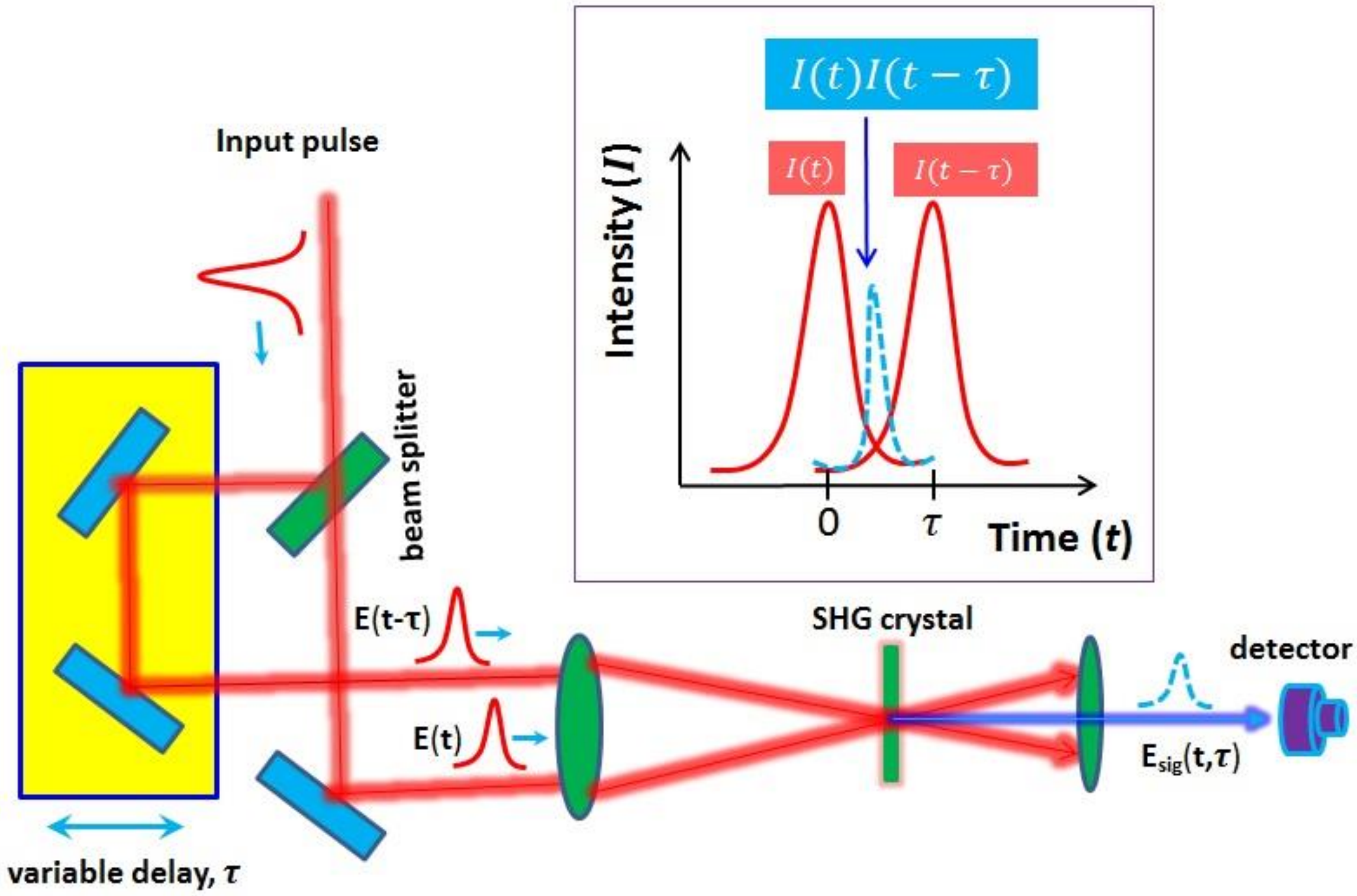

**Figure 3.1 The schematic configuration of the commercial intensity autocorrelator (AC). The input pulse will be divided into two arms and the pulses are overlapped through the second harmonic generation crystal. The SHG pulse energy versus the delay time will be measured, which yields the AC pulse trace. The inset is the overlap between the two replicas of the pulse. Note: the intensity autocorrelation is only nonzero when the pulses overlap. Figure redrawn from Ref. [92].**

The commercial intensity autocorrelator (as in Figure 3.1) is a tool to obtain the temporal information of an ultrashort pulse. While it reveals the pulse width of a pulse, it measures nothing about actual pulse intensity and phase. Therefore, a device called frequency-resolved optical gating (FROG) is needed for a more precise measurement to obtain the frequency-dependent intensity and phase of an ultrashort pulse. Nevertheless, as far as the pulse width concerned, the intensity AC is a better choice in terms of alignment and cost. Usually, the input pulse is aligned into the intensity AC from outside, in order to get the original pulse and the time-delayed pulse to be well overlapped on the SHG crystal. Of course, we need to calibrate the intensity AC every time for each experiment before we use it in order to determine the right calibration coefficient between the pulse width $T_{oscilloscope}$that is directly measured from an oscilloscope which is connected to the detector and the AC pulse width $T_{AC}$ which is linked to the real pulse width $T_{FWHM}$by Equation 3.5 and 3.6.

The pulse before the EDFA is should be compressed to near TL pulse duration since the shorter the pulse duration the higher the peak power, while the average power is kept the same. Another reason for obtaining near TL pulse before amplification is that SMF fiber pre-chirps the pulse so that nonlinearity is minimized during the amplification [93]. For wavelengths$> 1.3\mu m$, a single mode fiber can act as a dispersive delay line and compress the pulse if its length is suitable [94].

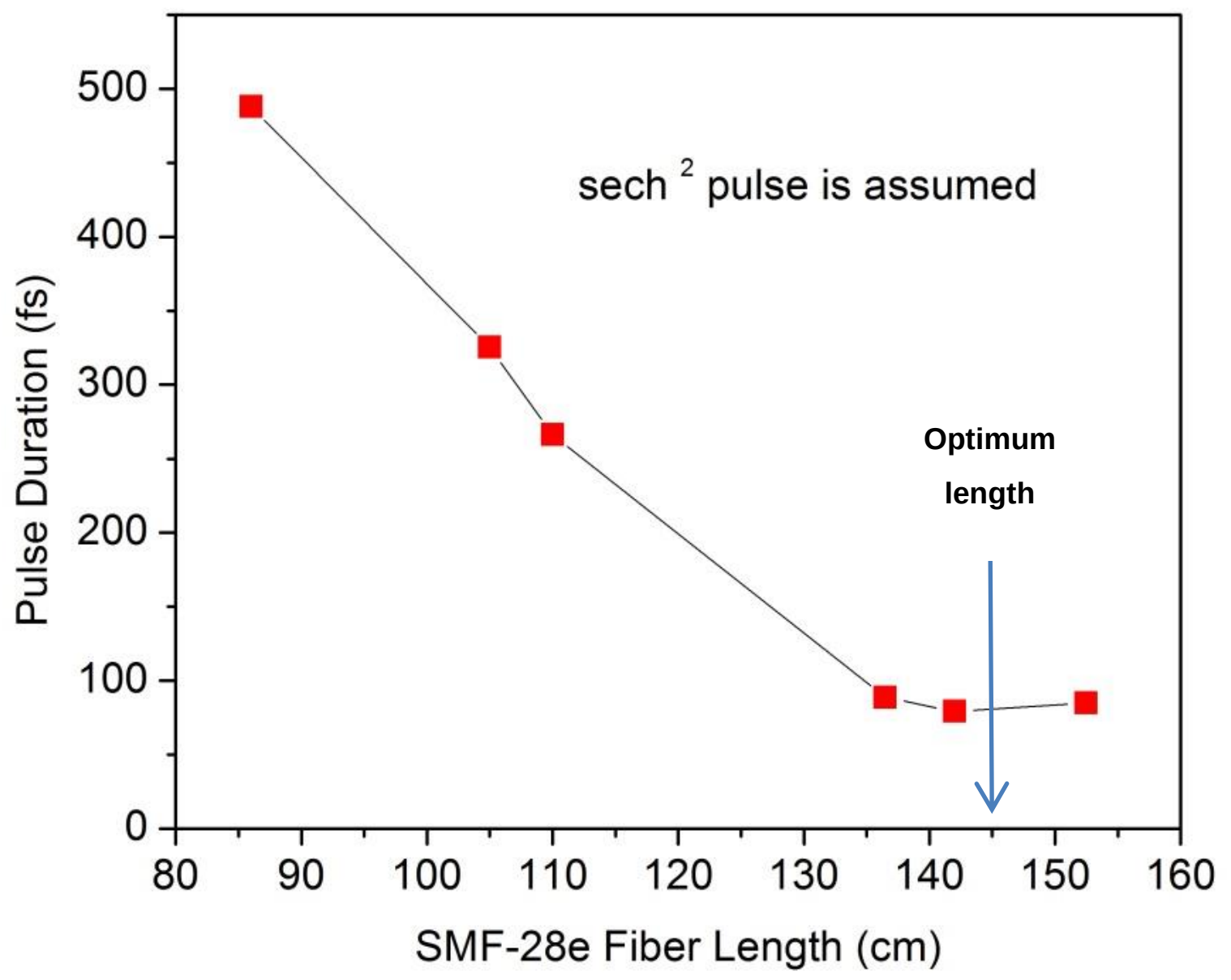

**Figure 3.2 The pulse duration versus SMF length before the EDFA.**

As can be seen from Figure 3.2, the pulse duration will decrease (*i.e.* pulse is compressed) until it reaches to the optimum length of 145 cm. However, the pulse is again broadened after we further increase the length of the SMF. At this optimum length of the SMF-28e, we have measured the pulse duration for Gaussian pulse shape and sech$^2$ pulse shape , respectively, 86 fs and 79 fs at full width half maximum (FWHM), while transform limited pulse duration is 80 fs (Gaussian pulse shape) and 57 fs (sech$^2$ pulse shape) . Theoretical pulse shape fitting for the measured pulse shape also performed to find that the pre-compressed pulse is closer to a $sech^2$ pulse shape as shown in Figure 3.3.

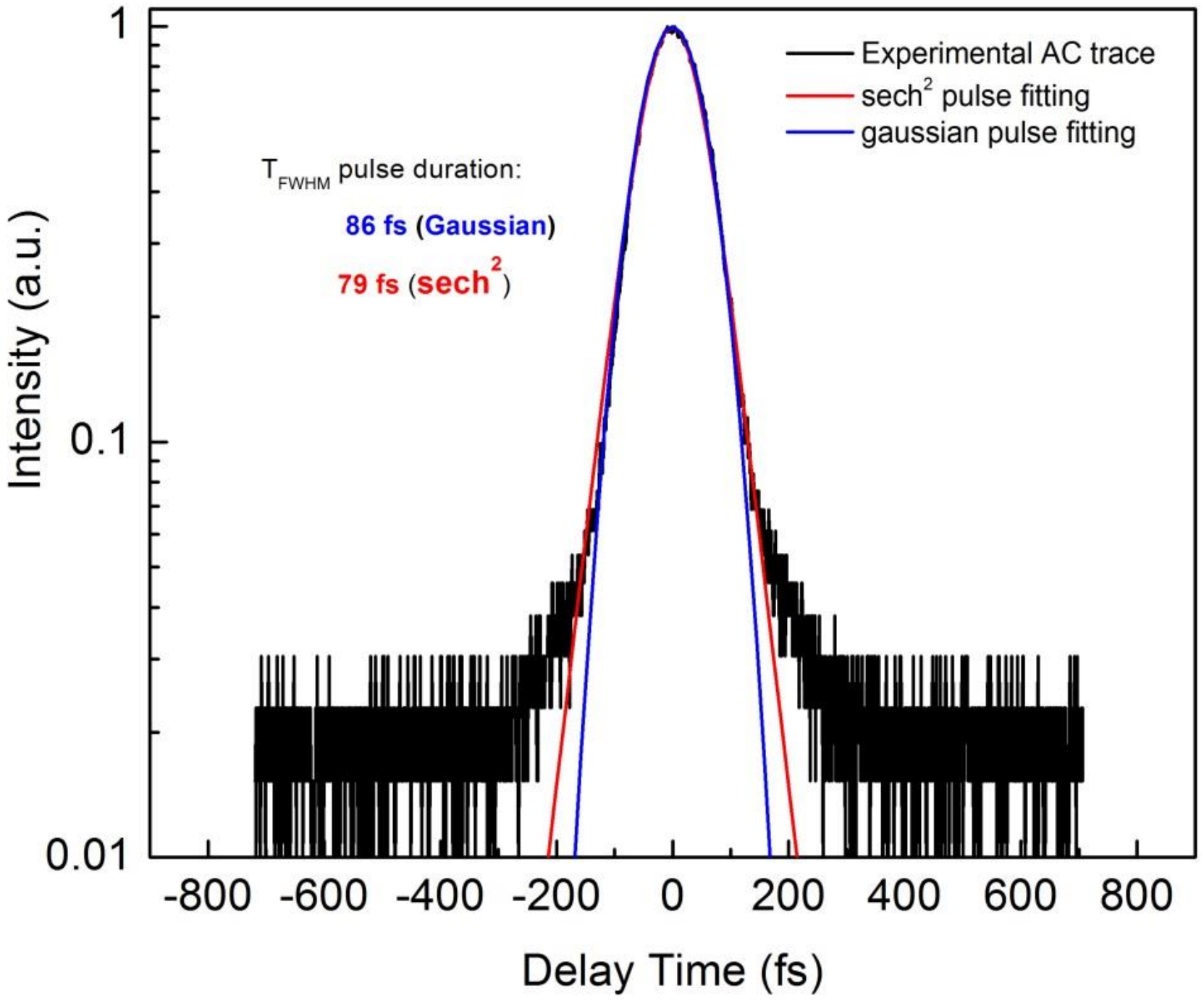

**Figure 3.3 The AC trace of the near TL pulse and theoretical fitting of Gaussian and hyperbolic-secant pulse shape. Note that he vertical axis is plotted in logarithmic scale while the horizontal axis is plotted in linear scale.**

The approximate peak power of the pulse (square pulse shape assumed) is calculated as:

$$P_{peak} = \frac{P_{ave}/f_{rep}}{T_{FWHM}} \tag{3.4}$$

The full width half maximum (FWHM) pulse duration is calculated by considering the calibration coefficient of AC and the directly measured delay time using oscilloscope.

$$T_{AC} = 5.70572\, T_{oscillosope} \tag{3.5}$$

$$T_{FWHM} = \begin{cases} 0.648\, T_{AC} \;\; (for\ a\ sech^2\ pulse\ shape) \\ 0.707\, T_{AC} \;\; (for\ a\ Gaussian\ pulse\ shape) \end{cases} \tag{3.6}$$

The intensity of an unchirped Gaussian pulse is assumed to have the form as [95]:

$$f(t) = exp\left[-\left(\frac{t}{T_0}\right)^2\right] \tag{3.7}$$

where we have the $T_0 = \frac{1}{2\sqrt{\ln 2}} T_{FWHM} = \frac{1}{1.65} T_{FWHM} \approx 0.6\, T_{FWHM}$

For the intensity of unchirped hyperbolic-secant pulse shape we have the form [95]:

$$f(t) = sech^2\left(\frac{t}{T_0}\right) \tag{3.8}$$

where we have the $T_0 = \frac{1}{\ln(1+\sqrt{2})} T_{FWHM} \approx 0.57\, T_{FWHM}$

### 3.1.2 Experimental Setup

In our experiment, we have used EDF as the gain fiber and 1480 nm laser diode pump which make up the main part of the EDFA. As we can see from Figure 2.7, EDF has strong absorption near 1550 nm that it has the potential to amplify the NPR comb pulses to a higher power. For the pumping scheme we have chosen backward pumping instead of forward pumping in which the signal and pump propagate in the same direction. In general, the forward pumping and backward pumping configurations make no difference for a small pump power that keeps the amplifier remains unsaturated. However, in the saturation regime the backward pumping configuration is better because of lower ASE [96].

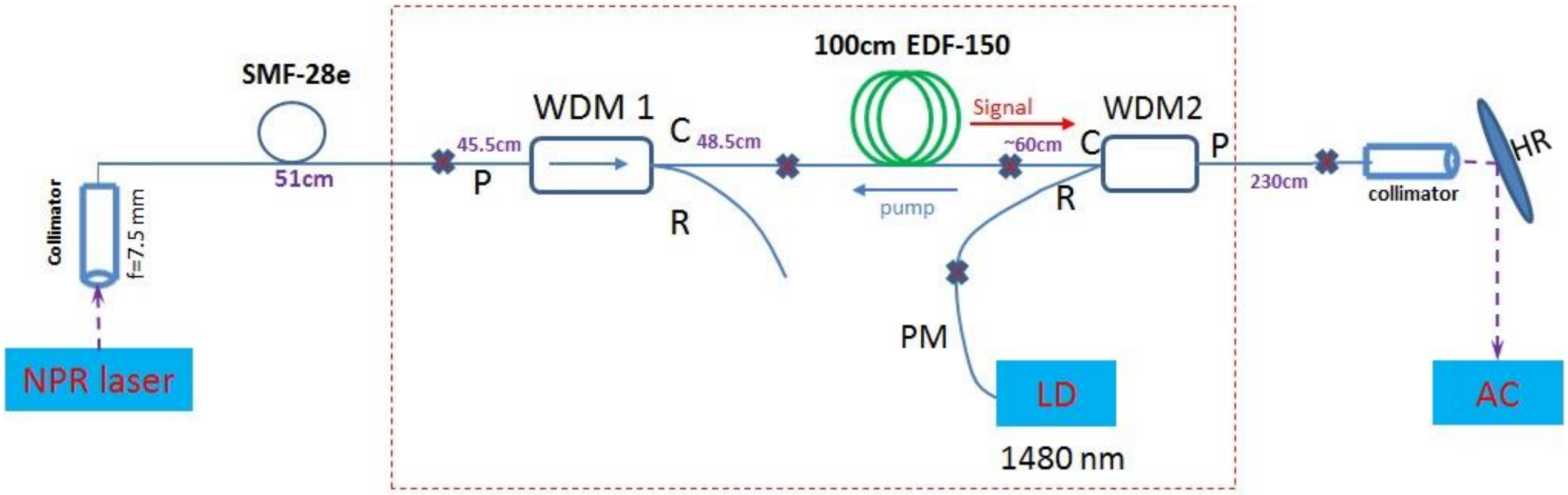

**Figure 3.4 The final schematic of EDFA experimental setup. WDM: wavelength division multiplexer, HR: high reflective mirror, LD: laser diode, AC: intensity autocorrelator, P: signal port of WDM, R: pump port of WDM, C: common port of WDM, EDF-150: erbium doped fiber, SMF-283: standard Corning single mode fiber, Collimator: aspheric lens mount connected with optical fiber connector.**

## 3.2 EDF Length Optimization for the Amplifier

In order to achieve the highest pumping efficiency with EDF, we need to choose an optimum length of gain fiber. As mentioned in Section 2.2, the optimized length of EDF is chosen to be 90 cm. This optimum length is determined by a cut-back measurement that the pump is absorbed by the gain fiber and no part of the gain fiber is left unused. In reality, we have also taken advantage of the green fluorescence, which we could observe during the experiment after turned on the EDFA pump, to determine the final optimized length of EDF [91].

Due to high absorption of the gain fiber, only short length of EDF is needed to extract most of the pump, thus minimizing the amount of nonlinear phase shift during the amplification process. It is important to choose a normal dispersion gain fiber, as this leads to SPM-dominated spectral generation whereby the resulting chirp is largely linear and compressible [97].

### 3.2.1 Group Velocity Dispersion Pulse Broadening

We can write down the dispersion length $L_D$ and the nonlinear length $L_{NL}$ as below [95]:

$$L_D = \frac{T_0^2}{|\beta_2|} \quad , \qquad L_{NL} = \frac{1}{\gamma P_0} \tag{3.9}$$

where $T_0$ is the initial half-pulse-width (the relationship between $T_0$ and $T_{FWHM}$ is given in Equation 3.7 and 3.8 ) and $P_0$ is the peak power.

When the fiber length is such that $L_D \ll L \ll L_{NL}$, the GVD effect dominates during pulse evolution and the nonlinear effect plays minor role in it. On the other hand, when $L_{NL} \ll L \ll L_D$, then the GVD effect is negligible compared to the nonlinear effect. In this case, the pulse evolution in fiber is governed by SPM which leads to spectral broadening of the pulse. As expected, when the fiber length is comparable to $L_D$ and $L_{NL}$ the dispersion effect and nonlinear effect act together as the pulse propagates along the fiber.

### 3.2.2 Backward Pumping Scheme for the Amplifier

The EDFA is built up by backward pumping configuration considering the high pump power and gain saturation of the EDF. For a 90 cm long EDF-150, we have measured the pump power and corresponding amplified power. The pumping efficiency is 32.4% as shown in Figure 3.5. This is comparable to about 41% pumping efficiency reported by Chao [93].

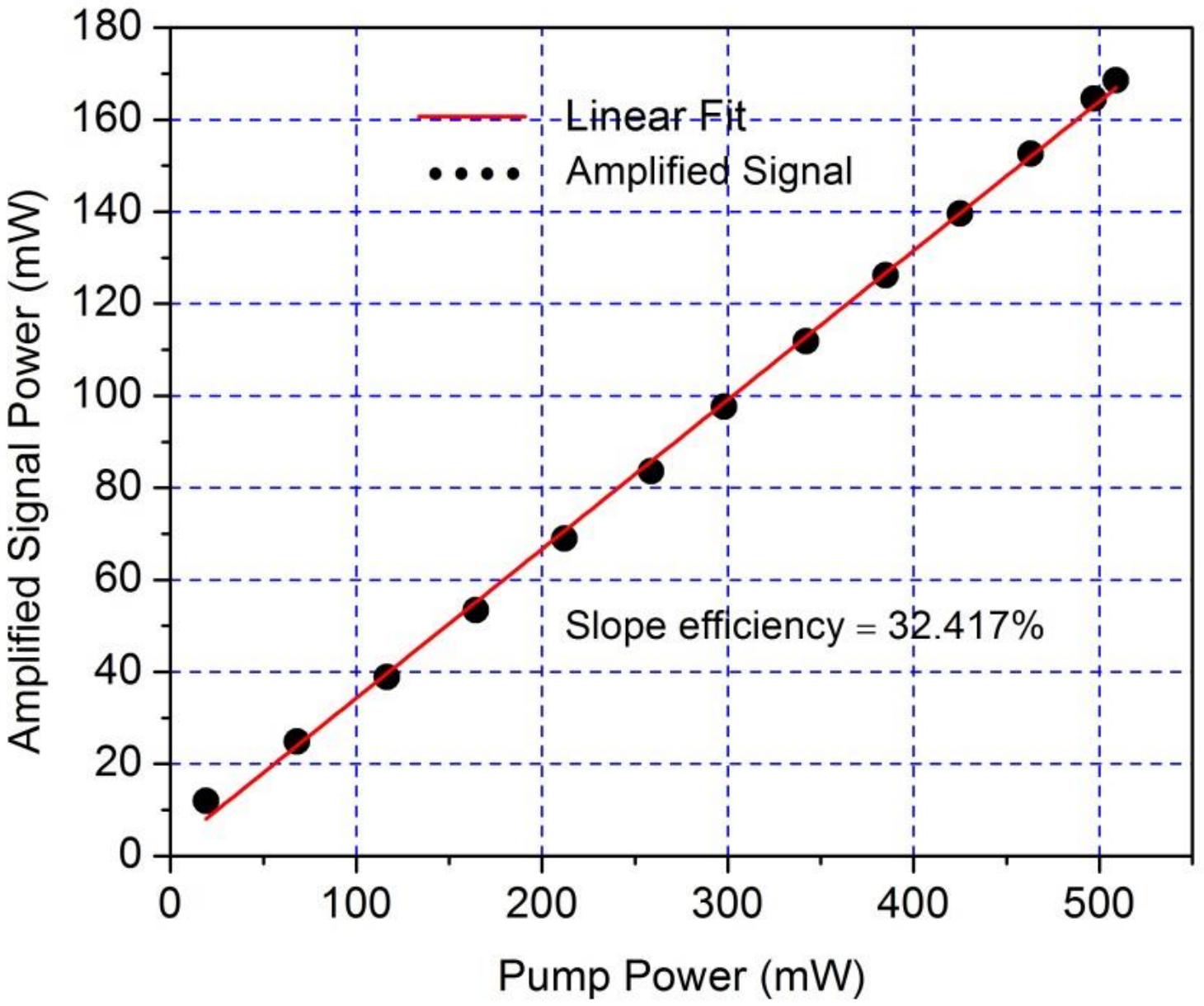

**Figure 3.5 The EDFA pump efficiency measured for pump power versus amplified output. The amplified average signal power was measured by a power meter made by Coherent.**

The difference between forward and backward pumping is essentially the existence of ASE noise in the gain fiber. Desurvire [98] reported that backward pumping configuration has less ASE noise, which makes it potentially eligible for higher pumping efficiency in terms of pump power and amplified signal.

## 3.3 EDFA Experimental Results and Explanations

We have measured the spectrum of the amplified pulse after EDFA corresponding to different output power at 1480 nm wavelength. As shown in Figure 3.6, there is SPM induced symmetric pulse broadening. Besides, the pulse peaks are red-shifted as the EDFA pump increased. In the wavelength range of normal GVD (*e.g.* in EDF), the combination of GVD and SPM makes a strong intensity pulse broaden and change shape in optical spectrum towards final rectangular pulse shape.

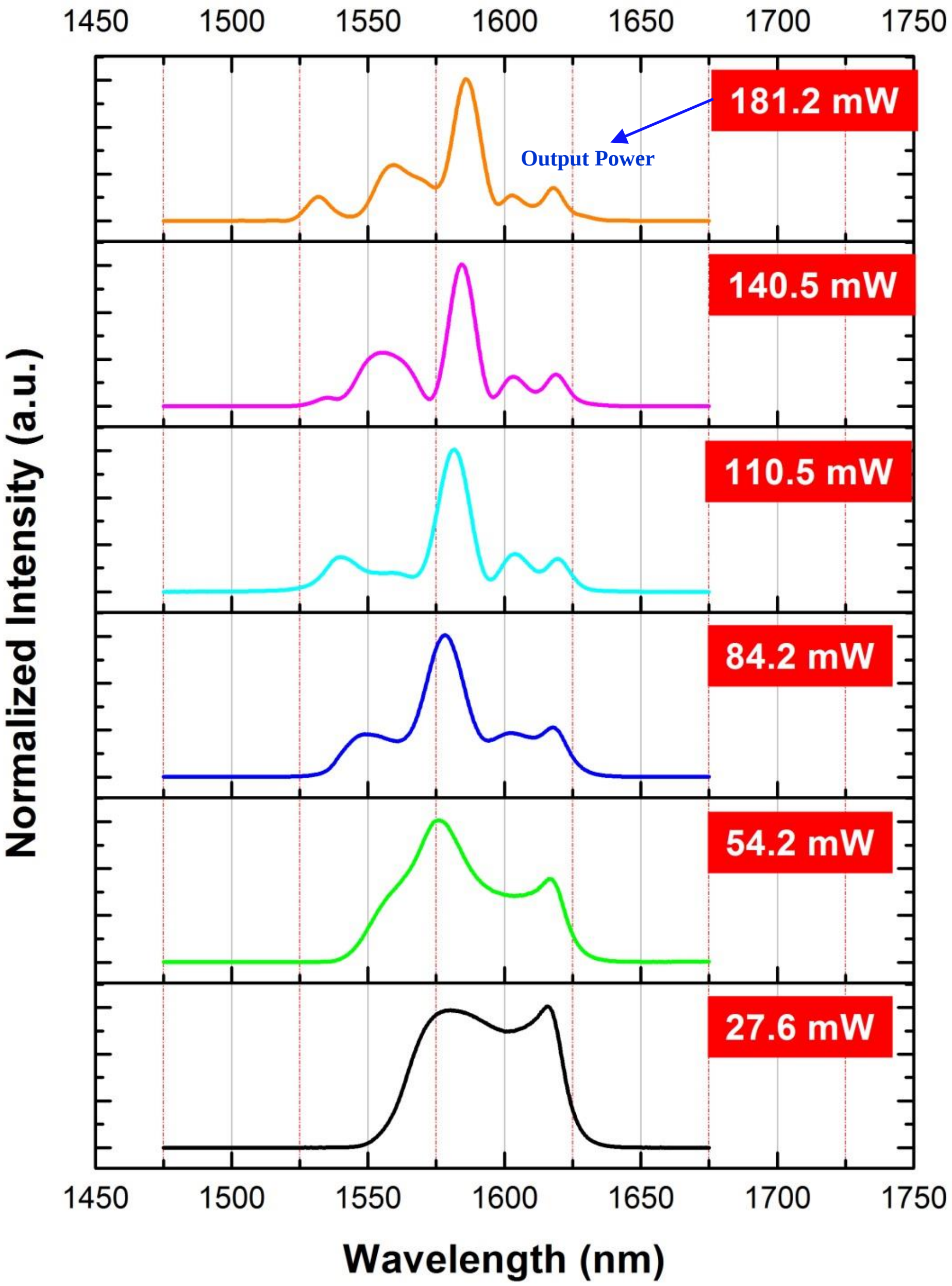

**Figure 3.6 The measured OSA spectrum of amplified pulses after EDFA corresponding to different output amplified power. As the1480 nm EDFA pump power increases (which in turn the output amplified power also increased), we can see pretty obvious frequency red-shift. Simultaneously, the pulse break up appears due to high pump power. Moreover, the central wavelength of the pulse is being shifted to longer wavelength.**

The physical mechanism behind the pulse distortion and spectra shift is SPM which is induced by temporal change in refractive index of the amplifier gain medium [94, 99]. As clearly seen from Figure 3.6, the pulses are going through symmetric pulse splitting (or wave breaking) as the pump power increased to 500 mW level. Zhong *et al.* [100]reported that there is optical wave breaking due to combined effects of SPM and GVD when the initial hyperbolic-secant pulses go through normal dispersion regime of fibers. Wave breaking is found to involve in two independent processes, including overtaking of different parts of the pulse and nonlinear generation of new frequencies during the overtaking [101].

## 3.4 Ultrashort Pulse Compression

Single mode fiber (SMF) has been used extensively in the field of fiber-based femtosecond pulse compression. In terms of experimental application such as supercontinuum (SC) generation, the pulse compression in SMF is very convenient since the SC generation fiber can be directly fusion spliced to the SMF with low splicing loss [102]. There are two different compression schemes used in the past depending on the sign of GVD in the optical fibers. In the case of normal GVD the fiber imposes nearly linear frequency chirp on the pulse, which is subsequently compressed by passing it through a grating pair. In the case of anomalous GVD, the pulse is chirped and compressed by the same fiber [103]. In the optical regime, the necessary chirp arises from SPM resulting from the intensity dependent refractive index.

### 3.4.1 Dispersion Management and SMF Length Optimization

The length of the compressor (*i.e.* the SMF-28e) is roughly estimated first by calculation according to the relationship of GVD and fiber length given below [104]:

$$\beta_2^{EDF} \cdot L^{EDF} + \beta_2^{SMF} \cdot L^{SMF} = 0 \tag{3.10}$$

where we define $\beta_2^{EDF}$ is the second-order dispersion of the gain fiber EDF-150 at 1550 nm, $L^{EDF}$ is the length of EDF-150, $\beta_2^{SMF}$ is the second-order dispersion of the anomalous dispersion fiber SMF-28e after the EDFA and $L^{SMF}$ is the length of SMF-28e.

Since the pulse before being amplified through the EDFA is nearly transform limited (*i.e.* the chirp is decreased to the minimum), therefore, we considered calculating the optimum length of SMF-28e using the Equation 3.2. Note that the optimum length here is calculated by compensating the normal dispersion that induced by the EDFA, which is to say compensating the positive chirp induced by the normal GVD of the EDF. Although, the chirp induced by the SPM during amplification is not considered, which is to be calculated through further calculation by solving the nonlinear Schrödinger equation (NLSE) of the pulse propagating through EDF. As a result, we have calculated the optimum length of SMF-28e after the EDFA is about 2.7 m. Of course, this calculation does not include the nonlinear chirp caused by SPM.

We have experimentally pursued the optimum length of SMF-28e by trying different lengths of SMF28-e and measuring the pulse duration with intensity AC (*i.e.* cut-back method). Finally, the shortest pulse is measured as 70 fs with 2.9 m length of SMF-28e.

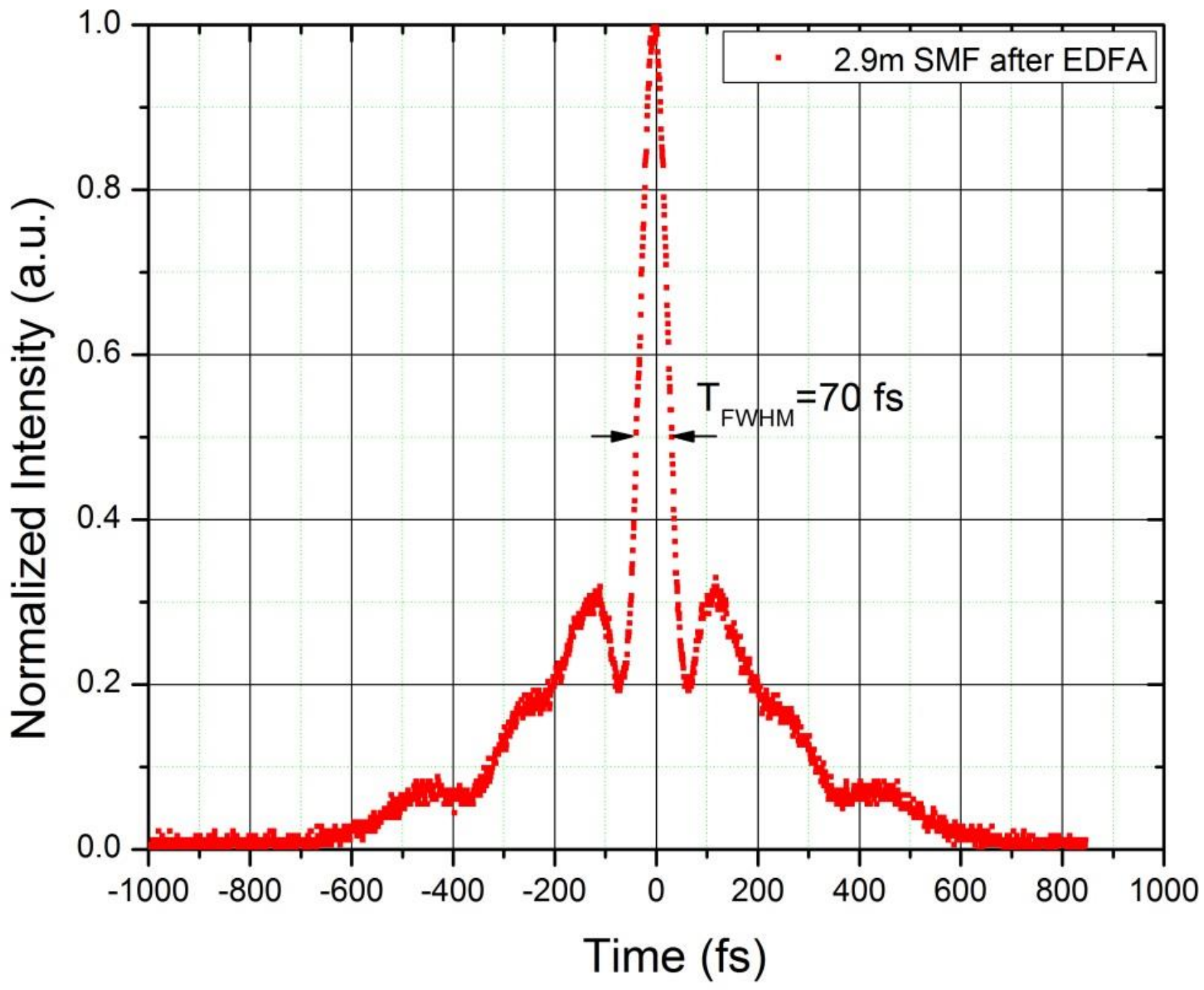

**Figure 3.7 The AC trace of the shortest near TL pulse with optimum SMF length of 2.9 m. This experimentally achieved optimum length value is very close to the theoretically calculated optimum length value of 2.7 m.**

### 3.4.2 Nonlinear Effect during Pulse Compression

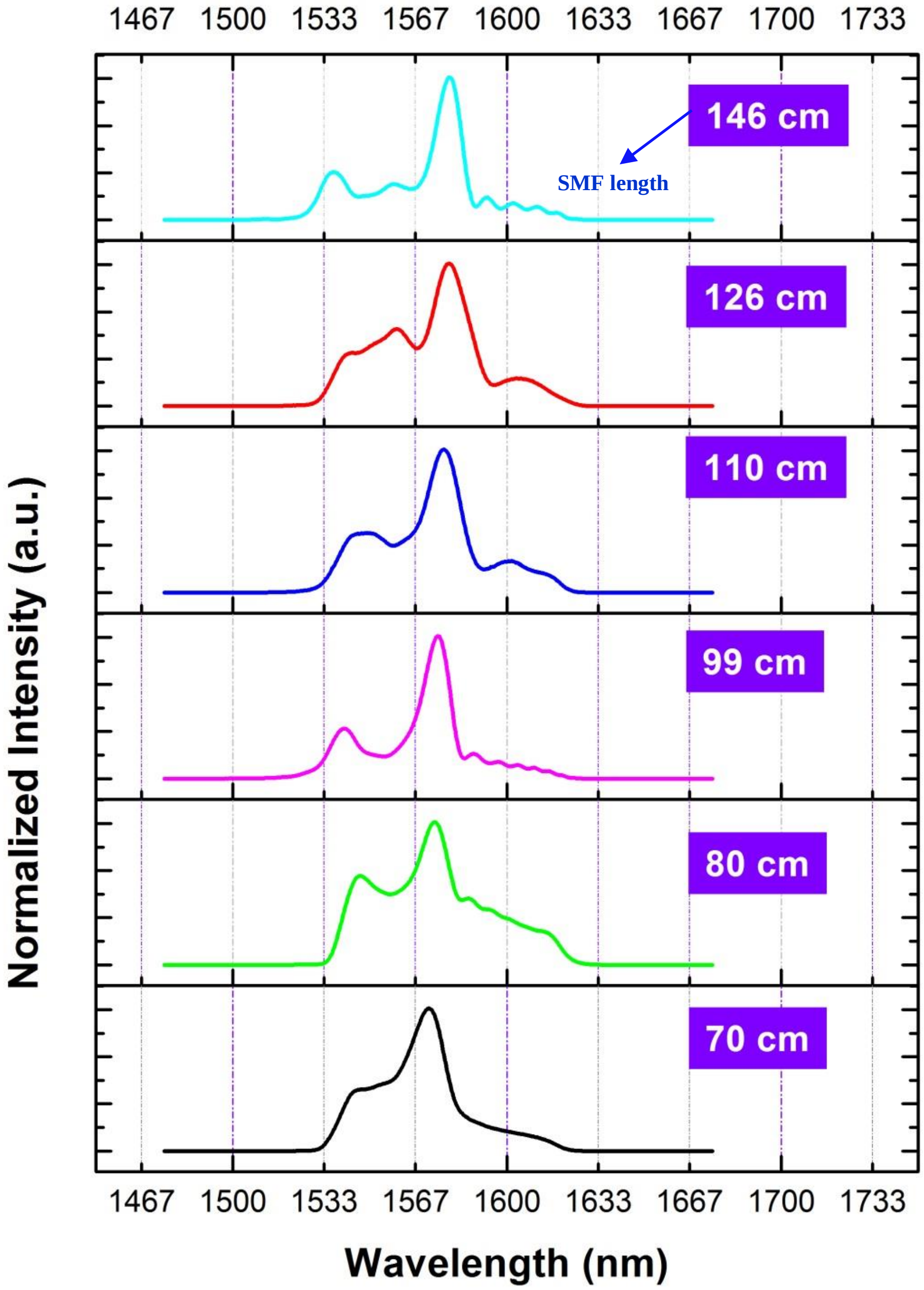

**Figure 3.8 The measured OSA spectrum for different lengths of SMF after the EDFA. The EDFA pump power is the same for all of these lengths. Vertical axis is offset position for plotting different spectrum and horizontal axis is wavelength.**

As clearly seen from Figure 3.8, the amplified pulses are red-shifted as they propagate along the SMF. In fact, the red shift can be explained by Raman scattering self-frequency shift [105]. In general, pules that are shorter than 100 fs will have enough spectral width that allows Raman gain to amplify the red components at the expense of blue components which acts as the pump [95]. During this process, the energy of the blue components is continuously transferred to red components.

Agrawal [106] argued that spectral changes induced by SPM are direct consequence of nonlinear phase shift $\emptyset_{NL}$ and the chirp induced SPM increases in magnitude with propagation distance, which is what we have experimentally observed as in Figure 3.8. This frequency shift during the pulse compression can also be described by the effect called intrapulse stimulated Raman scattering (ISRS). Moreover, the ISRS effect can improve the soliton-effect pulse compression both quantitatively and qualitatively [107].

As we know, the pulses tend to be affected by more and more nonlinear effects (*e.g.* SPM or XPM) as it propagates along the fiber. Figure 3.8shows the pulse shape evolution along the SMF. It can be seen that pulse breaking have been the main phenomenon during this process. Although, it is generally believed that SPM is associated with spectral broadening, it is not always the case if the pulse is initially chirped. That is to say, the sign of the initial chirp of the pulse is the key element when it comes to whether SPM will compress or broaden the spectrum. It is believed that for positively chirped or transform-limited pulse, the SPM will narrow the spectrum in the anomalously dispersive nonlinear media (such as SMF) [108].

When considering the high intensity pulse propagation in the SMF, the combined effect of SPM and GVD lead to a phenomenon called wave breaking (or pulse splitting) [109]. On the leading edge of the pulse the SPM leads to red shift; while the trailing edge of the pulse experiences a corresponding blue shift.

However, it is worthwhile noticing that when the pulses are undergoing asymmetric wave breaking as in Figure 3.6,  there is cross-phase modulation (XPM) induced optical wave breaking [103] which is due to crosstalk between the signal and pump light in the EDFA. In spite of the identical nature of the underlying physical mechanism, optical wave breaking exhibits different qualitatively features in XPM case compared to the SPM case. Agrawal reported that XPM wave breaking usually occurs when weak femtosecond pulse is launched with an intense pump pulses and optimizing the initial delay between the two.

In conclusion, we have observed wave breaking both in Figure 3.6 and Figure 3.8. But, there is one very obvious distinction between the two cases, that is, symmetric and asymmetric wave breaking. In addition, the distinction between the wave-breaking in these two cases is due to the XPM effect between pump and signal pulses in the amplifier as in Figure 3.6, while there is no pump pulse (thus, no XPM effect )in case of pulse compression as in Figure 3.8.

### 3.4.3 Pulse Duration Measurement with Autocorrelation

This is the final result demonstrating that we have achieved the shortest pulse duration of 70 fs.

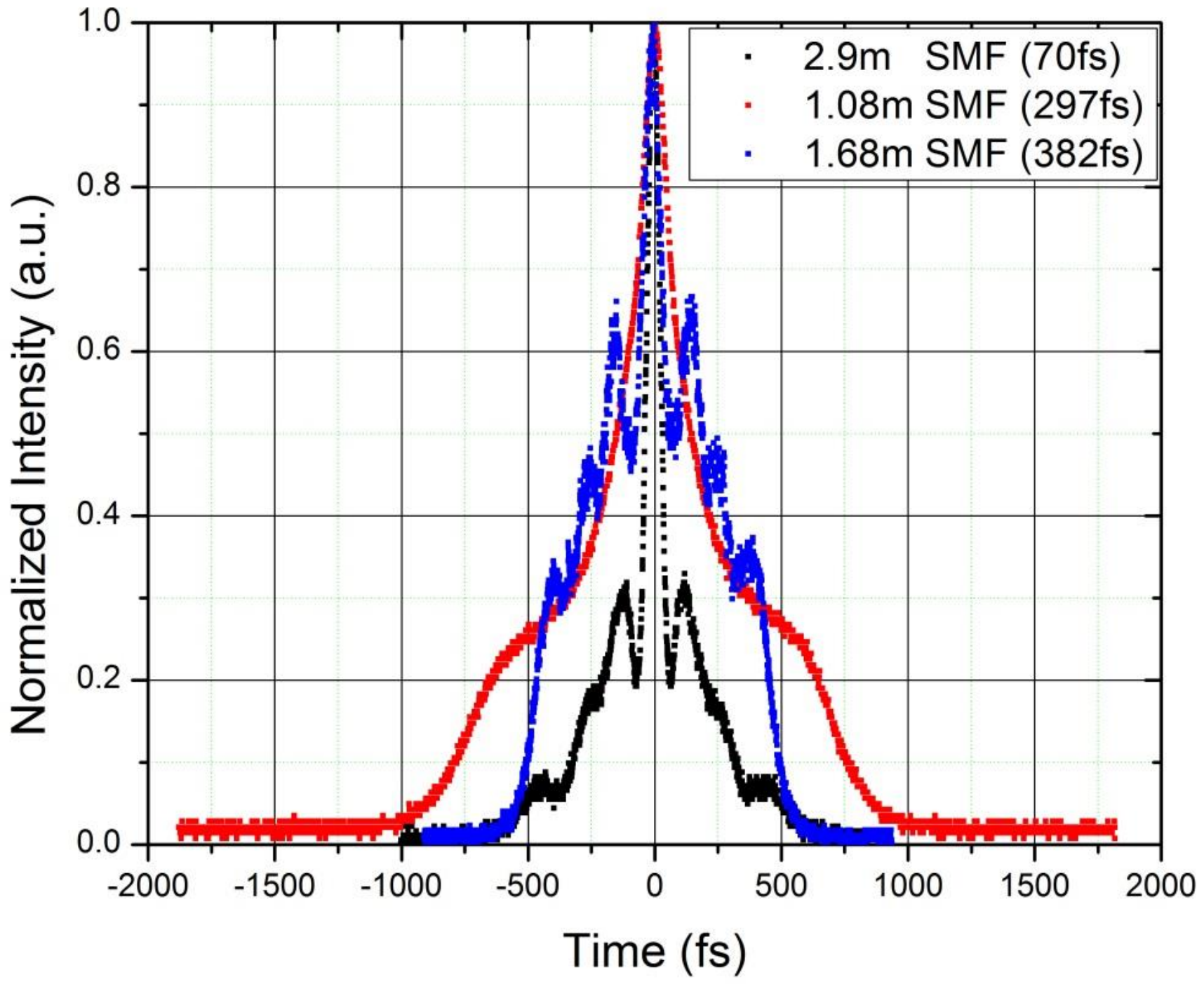

**Figure 3.9 Measured AC pulse trace for three different SMF-28e length after the EDFA. The pulses are amplified by the EDFA to an average power of 180 mW, while they were broadened in the time domain due to nonlinearity and group velocity dispersion.**

From the result as shown in Figure 3.9, we can see that the SPM induced chirp in the amplified pulse is compensated by the GVD of the anomalous group velocity dispersion fiber. Nevertheless, we can still see large pedestal in the autocorrelation trace of the post-compressed pulse, which is due to the uncompensated nonlinear chirp by fiber compression [110].In fact, when considering the pulse compression of high intensity pulses we need to take the third order dispersion $\beta_3$ and stimulated Raman scattering (SRS) into account [111].

As a result, the wing structure pulse shape in Figure 3.9 can be explained using this model in [111], which is the combined effect of $\beta_3$ and SRS. In addition, as the fiber length increases the corresponding fiber loss will increase so that there is trade-off between loss-induced degradation and higher-order nonlinearity induced degradation. This balance between loss and nonlinearity effect gives rise to an initial pulse-width dependence of the in soliton-effect compression and an optimum initial pulse width that lead to maximum pulse compression for a fixed soliton order. Given the initial pulse width $T_0$ and peak power $P_0$, we can calculate the soliton order from $N = \sqrt{\frac{L_D}{L_{NL}}}$ for constant second-order dispersion $\beta_2$ and nonlinear coefficient $\gamma$ of SMF-28e. The higher the soliton order the greater the compression ratio tend to get. In fact, there is near linear relation between soliton order N and the self-compression factor $F_{SC}$ as mentioned in [112]

$$F_{SC} = \frac{T_0}{T_{SC}} \approx \sqrt{2}N \tag{3.11}$$

where $T_0$ is the initial pulse width and $T_{SC}$is the self-compression width.

## 3.5 Summary

In this chapter, we reported the experimental results and theoretical discussion of ultrashort pulse amplification and compression. First of all, we have demonstrated the pre-compression of the output pulse from the NPR laser comb to a hyperbolic-secant pulse with pulse duration of 79 fs. Then we presented the experimental setup configuration for EDFA and post-compression of the amplified pulse via anomalous dispersion fiber, that is, SMF-28e.The optimum length of EDF as 1 m is chosen by cut-back method. Moreover, we have discussed the backward pumping configuration in this EDFA setup and have reviewed the advantage of backward pumping configuration over forward pumping configuration in the case of high intensity pump power.

Next, we have reported the high output power amplification experimental result and provided corresponding theoretical explanation for the observed phenomenon of dispersion and nonlinearity, as well as fiber loss induced effects.

We also put forward the experimental results of post-compression of amplified pulses and discussed the related theoretical background. The optimum length of the SMF for post-compression is not only calculated theoretically but also determined experimentally using the cut-back method. We have noted that the so called optical wave breaking effect occurs during both pulse amplification and pulse post-compression. However, the wave breaking is induced by the combined effect of GVD and XPM during the pulse amplification and that the wave breaking in the optical spectrum is symmetric. While, the wave breaking is induced by the combined effect of GVD and SPM during pulse compression and the wave breaking is asymmetric. It is also known to us that the frequency shift in the pulse spectrum is mainly due to the SRS.

So far we have managed to obtain a 70 fs ultrashort pulse with average power of about 180 mW at wavelength near 1550 nm (*i.e.* the peak power is about 28kW). In the next chapter, we will talk about the supercontinuum generation in highly nonlinear fibers using this high intensity ultrashort pulse that we reported in this chapter.

# Chapter 4 - Supercontinuum Generation in HNLF

In this chapter, we will discuss supercontinuum (SC) broadband generation using the high power ultrashort pulse as described in Chapter 3 - as the light source. The amplified then post-compressed pulse provides 70 fs sech$^2$ like pulse shape centered at about 1575 nm with repetition frequency near 89 MHz. The highly nonlinear fiber (HNLF) is spliced to the end of the post-compressor SMF-28e. We have used the cut-back method to get optimum length of HNLF, in order to achieve the broadest SC generation spectrum that ranges from 1000 nm to 2100 nm.

## 4.1 Overview of the Octave-spanning SC generation

### 4.1.1 Supercontinuum Generation Principle

Supercontinuum generation from femtosecond pulses in nonlinear fibers has attracted applications in the field of frequency metrology. Previous studies showed that there are several physical phenomena involved in the SC generation process, like, self-phase modulation, intrapulse Raman scattering (IPRS), cross-phase modulation (XPM), modulation instability, and dispersive wave (DW) generation (which also is called non-solitonic radiation) [113]. In general, there are three major physical processes that responsible for the generation of SC, such as self-phase modulation (SPM), stimulated Raman scattering (SRS) and self-steepening (SS) [93]. SPM effect causes spectral broadening when the initial pulse is positively chirped or transform limited, while it causes spectral compression when initial pulse is negatively chirped. So we can say that the effect of SPM depends on the sign of initial pulse chirp [114]. Erbium doped femtosecond fiber laser pumped SC generation , that spanned from visible to infrared, has the potential to be used in the field of spectroscopy and frequency metrology [115]. In general, there are two major types of HNLFs, such as, small core step-index silica fibers and photonic crystal fibers (PCF). In our experiment, however, we only used the step-index HNLF. Nevertheless, for the sake of visual comparison, the geometric structure and index profile are given for both types of HNLFs as shown in Figure 4.1.

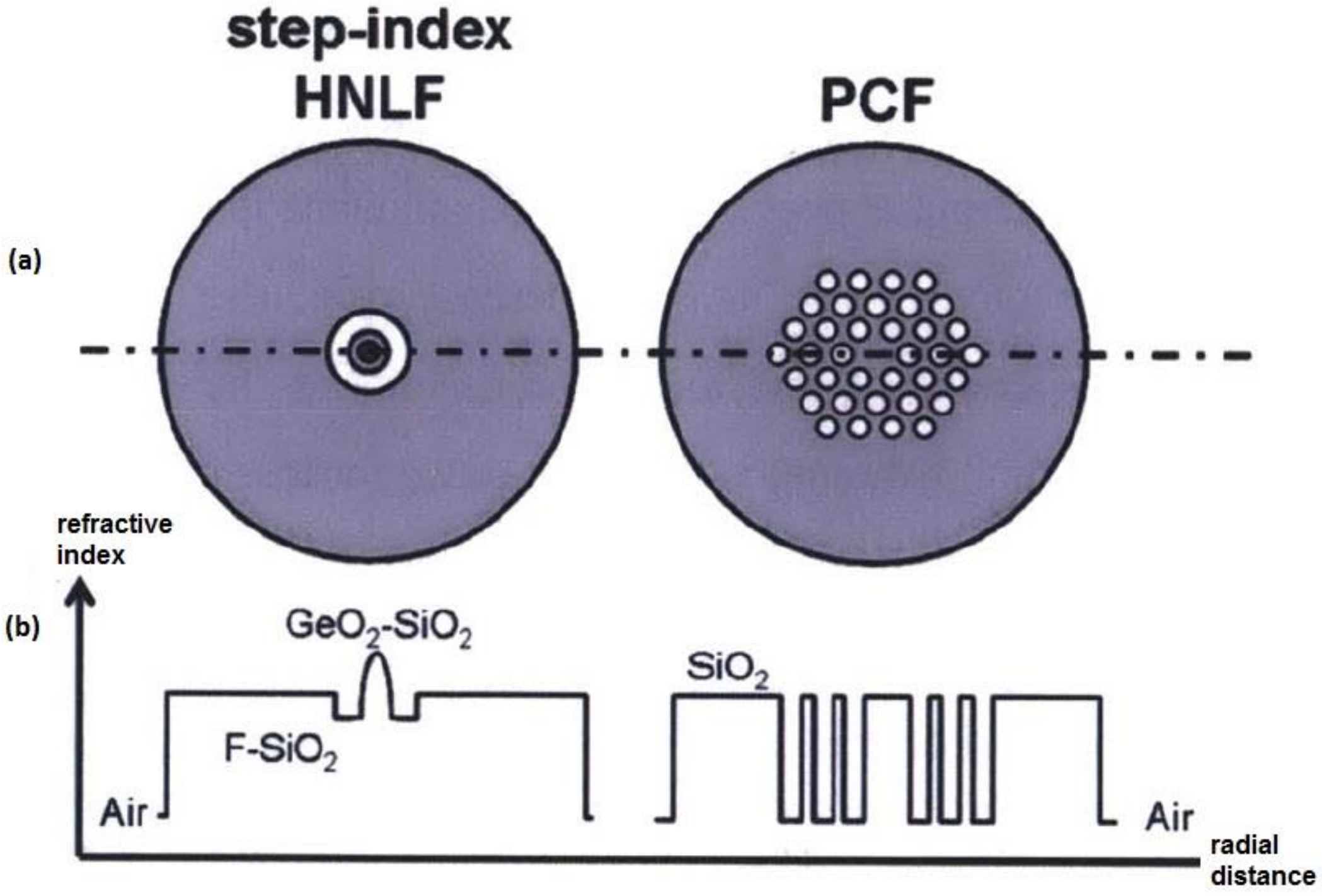

**Figure 4.1 The simple schematic structure (a) and refractive index profile (b) of two major types of HNLFs. Step-index HNLFs are more easily compatible with standard SMF, while the PCF design has more flexibility on controlling nonlinearity and dispersion. Figure reproduced from Ref. [93].**

In previous experiments, a type of ultraviolet (UV) exposed [116]HNLF has been used to obtain SC generation spectra from 1.0 $\mu m$ to 2.2 $\mu m$ [91]. The short-wavelength edge of SC spectrum usually shifts to shorter wavelength [116] after the HNLF exposed to UV which changes the dispersion of HNLF led by the change of refractive index.

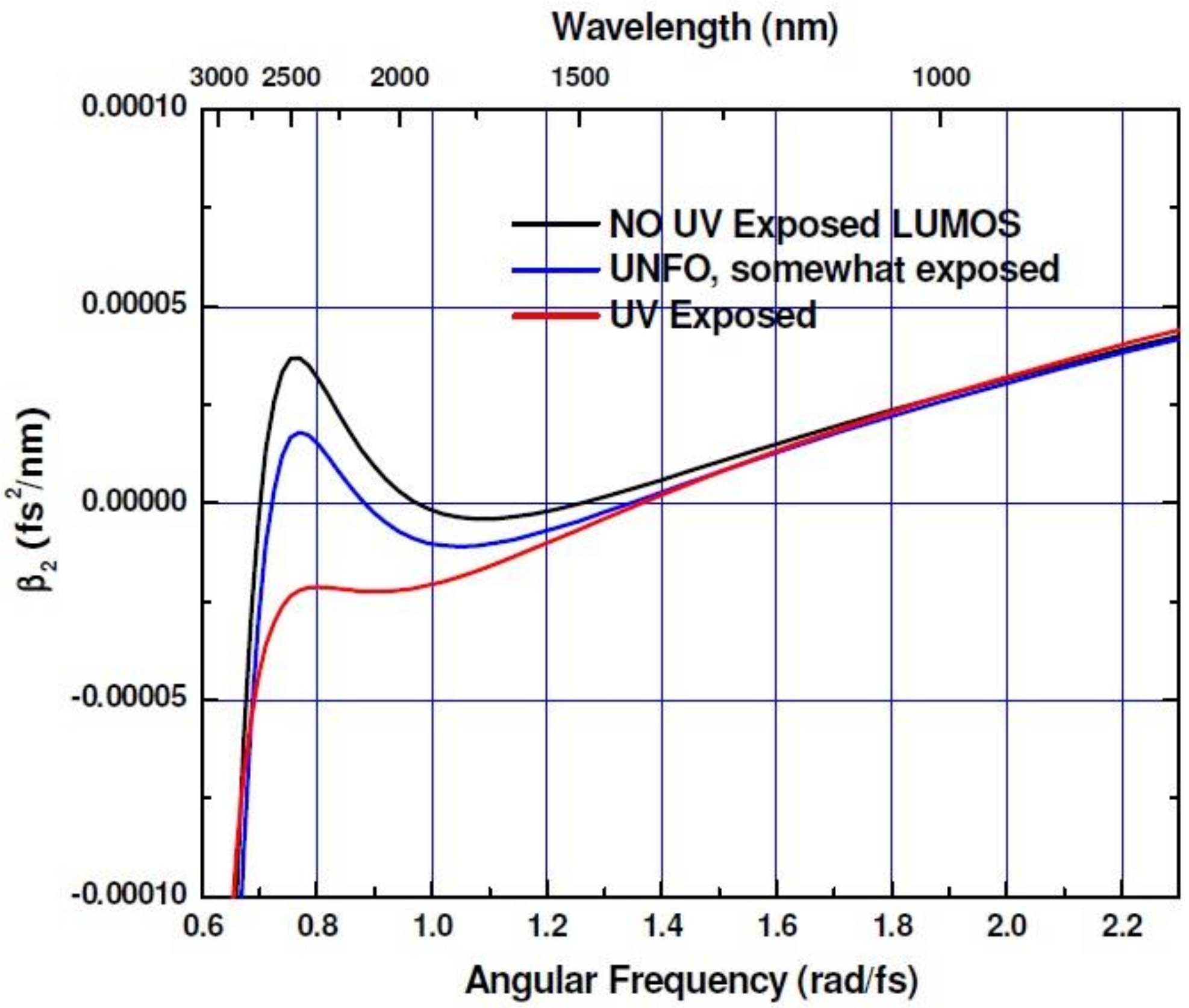

**Figure 4.2 The calculated GVD of three different HNLF that used in our lab. Figure reproduced from Ref. [91].**

In general, the nonlinearity of HNLF is about 10 times higher than standard SMF because of small effective area, although its nonlinearity is significantly smaller than what can be obtained with a small-core microstructure fiber [117]. The nonlinear coefficient of a fiber is:

$$\gamma = \frac{2\pi n_2}{A_{eff}\lambda} \tag{4.1}$$

where $n_2$ is the intensity dependent refractive index of the fiber, $A_{eff}$ is the effective mode-confinement area of the fiber.

### 4.1.2 Supercontinuum Generation Experimental Setup

In our experiment, we used a type of germanosilicate HNLF to generate SC from 1 $\mu m$ to 2.1 $\mu m$. This HNLF has zero dispersion parameter $D$ (i.e. $D \approx 0.0 \pm 1.0\, ps/nm/km$) at wavelength of near 1550 nm, dispersion slope of 0.019 $ps/nm^2\ /km$, mode field diameter (MFD) of 3.9 $\mu m$, nonlinear coefficient of 20 $W^{-1}km^{-1}$ and cut-off wavelength at 1300 nm. As shown in Figure 4.3, the HNLF is fusion spliced to the end of post-compressing SMF-28e and the splicing loss is less than 0.2 dB. The femtosecond pulse from $Er^{3+}$doped fiber laser comb has gone through the pre-chirp SMF-28e, erbium doped fiber amplifier (EDFA) and the post-chirp SMF-28e, in sequence. Thus, we have a high power ultrashort pulse as the pump source of the SC generation in HNLF. Finally, we can measure the SC generation spectrum using an OSA at end of the HNLF with FC/APC fiber connectors. The PC is installed on near edge the SMF right before the HNLF, so that we can adjust the polarization of pulse entering the HNLF. This is to further enhance the SC generation spectrum for our purpose.

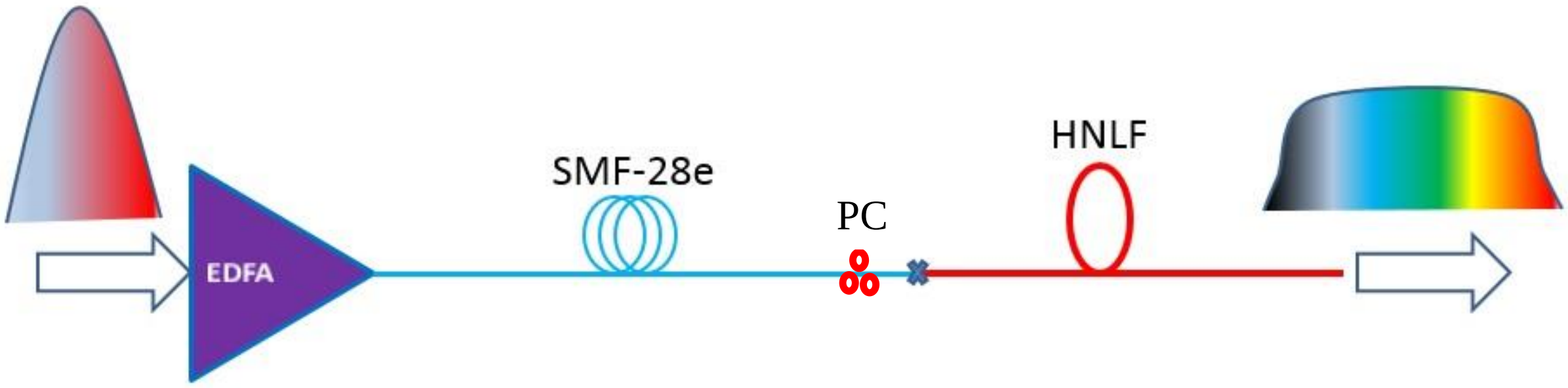

**Figure 4.3 The schematic configuration of SC generation experimental setup. EDFA: erbium doped fiber amplifier, HNLF: highly nonlinear fiber, SMF-28e: standard Corning single mode fiber, PC: fiber based (inline) polarization controller.**

## 4.2 HNLF Length Optimization

In order to determine the optimum length of HNLF to generate the broadest octave spanning SC, we have measured SC spectrum for different length of HNLF. As shown in Figure 4.4, we have measured the corresponding SC spectrum much earlier than we have determined the optimum length of pre-compressor SMF for near transform limited (TL) short pulse and built up the EFDA to amplify the light before the HLNF. We only have measured the SC spectrum for short wavelength component since the maximum measuring scale of OSA is 1750 nm. Thus we could see the shortest end of the SC spectrum wavelength, while we could not see the longest end of the SC spectrum wavelength but to estimate from symmetric pulse broadening due to SPM in negative dispersion ($\beta_2 < 0$) HNLF. Nevertheless, the spectrum is not broad enough at short-wavelength edge to get SC components around 1000 nm, most probably due to the fact that the pulse going into HNLF is not compressed to near TL pulse duration and not amplified to high peak power. In case of 100 cm HNLF, we have used a Vytran (FFS-2000) splicer which produces a splice with very small loss, in contrast with splicing made by the electric arc fusion splicer (Ericsson FSU-995) splicer in case of 95 cm and 218 cm HNLF. Thus, the difference in spectral shape of the SC among these three different lengths of HNLF in Figure 4.4 probably ca be attributed to the different input power into the HNLFs. The SPM effect is dominant during the process of SC generation at short-wavelength edge in 100 cm HNLF (red), while non-solitonic Raman radiation (*i.e.* dispersive wave generation) is responsible for the short-wavelength edge of SC in case of 95 cm (black) and 218 cm (blue) length of HNLFs.

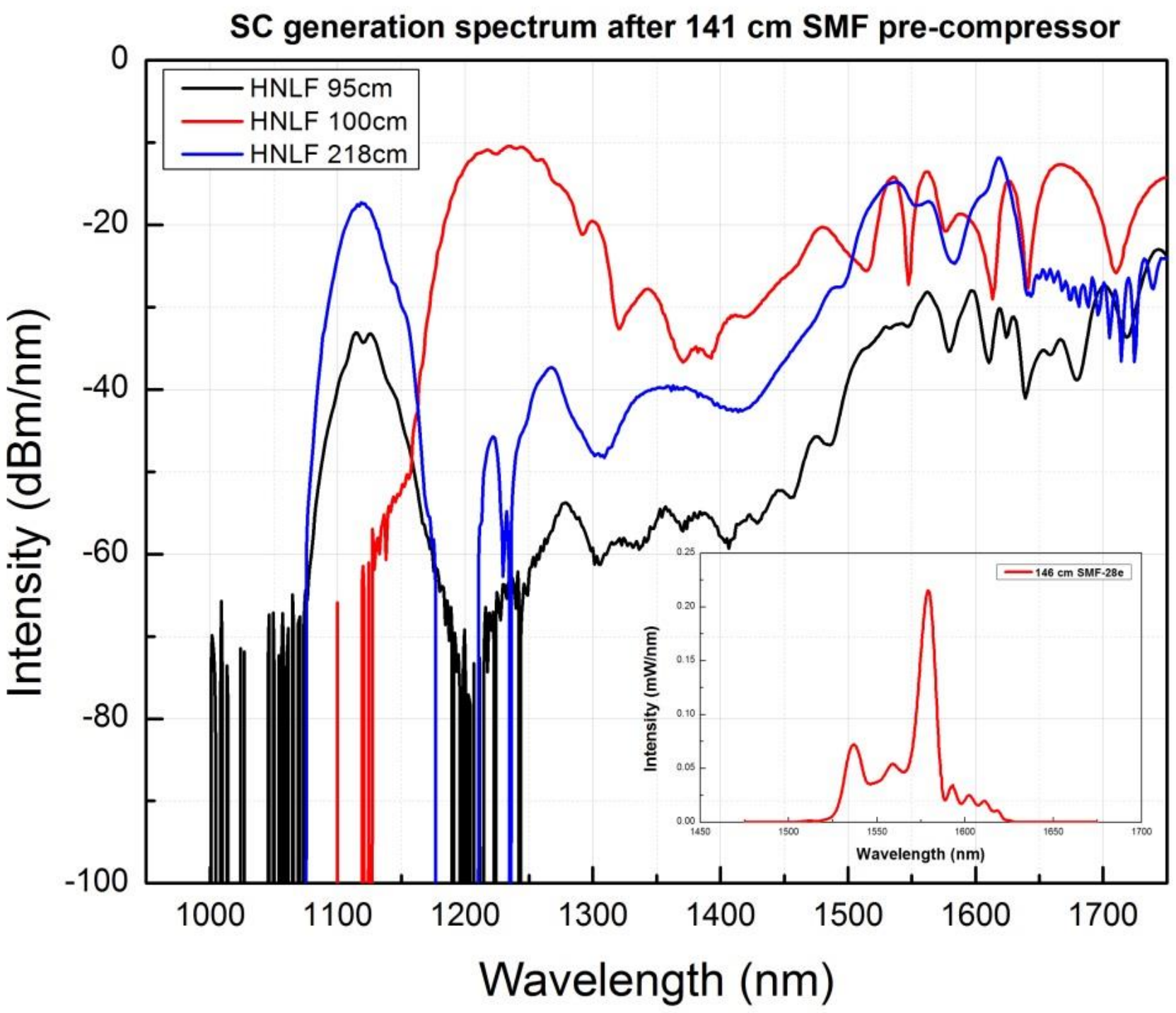

**Figure 4.4 The SC spectrum measured by optical spectrum analyzer (OSA) for three different length of HNLF while the pre-compressor SMF length is about 141 cm. This SC spectrum was taken even before we inserted the EDFA between the NPRL comb and HNLF. The average output power after HNLF for these three lengths is 6 mW~10 mW at 1575 nm. The inset is the spectrum of compressed pulse for 146 cm SMF before the HNLF.**

After we have designed the EDFA to amplify the NPRL comb pulse, we then compressed the pulse to near TL pulse duration. The FWHM pulse duration of the input pulse into the HNLF is 70 fs (sech$^2$ pulse assumed), average power is about 180 mW and the peak power is estimated at about 28 kW. Then we measured the SC generation spectrum corresponding to the optimized length 2.7 m of HNLF as in Figure 4.5, using the cut-back measurement method. However, we have done further optimization on the length of HNLF concerning the SC stability for the application of self-referencing and beat frequency detection. The final optimum length of HNLF is about 1m.

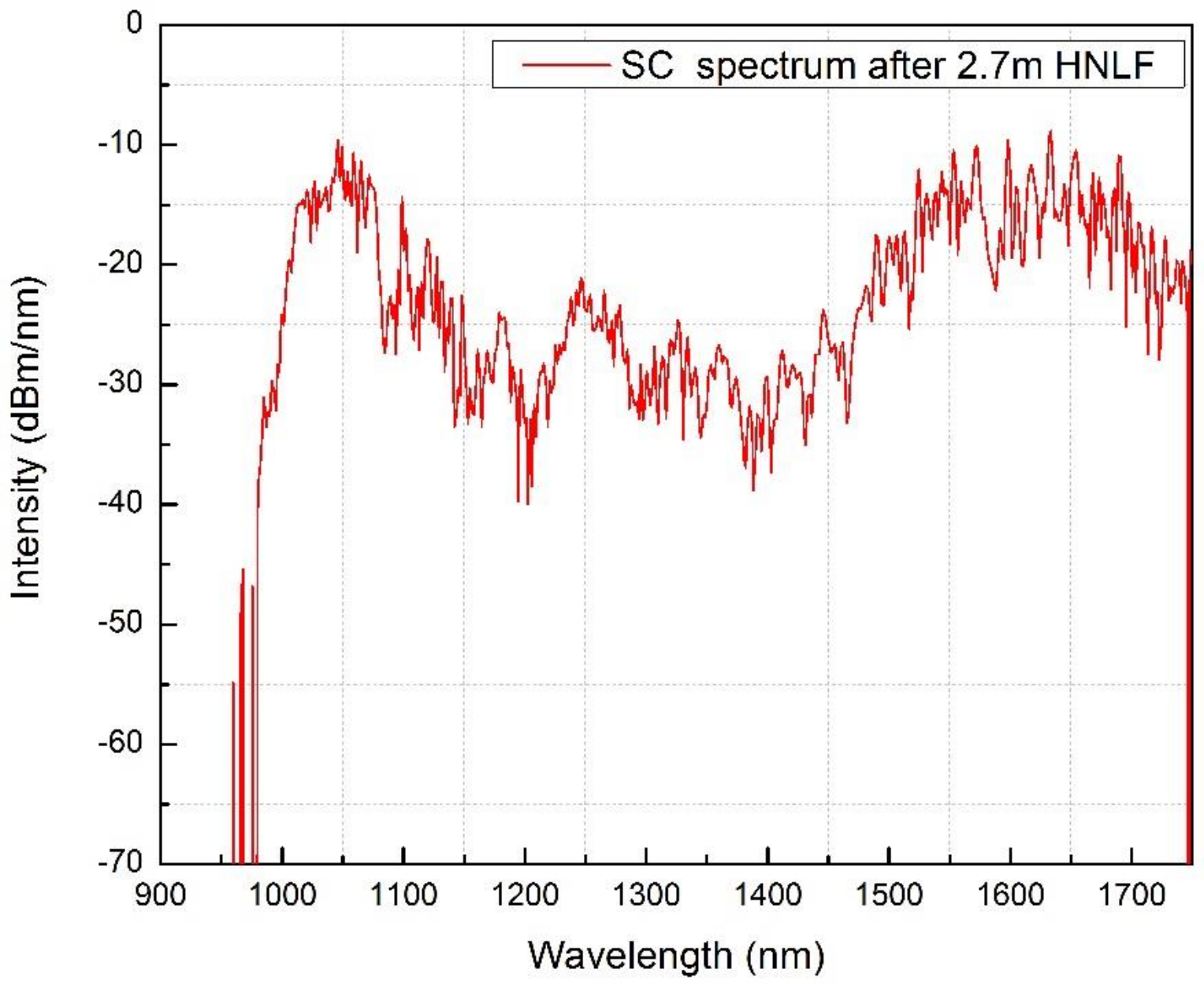

**Figure 4.5 The SC spectra for 2.7m HNLF after we applied the amplified and compressed pulse. This spectrum is measured in logarithmic scale by OSA, and the maximum measured wavelength is 1750 nm. The average power of the input pulse before HNLF is about 180 mW.**

## 4.3 SC Experimental Result and Discussion

As shown in Figure 4.6, we have measured the SC spectrum using optimal length about 1 m of HNLF which is determined from the HNLF length optimization process in previous section. Our goal was to generate octave-spanning SC generation corresponding to 1025 nm at short-wavelength edge and 2050 nm at long-wavelength edge, which are the two wavelengths that we use for self-referencing then for optical heterodyne beating between the 1025 nm and second harmonic generation (SHG) signal of 2050 nm, for the detection of carrier envelope offset (CEO) frequency. The short-wavelength edge (as can be seen from (a) and (b)) was observed from the SC spectrum that measured by OSA, while a monochromator was needed to measure the SC spectrum for long wavelengths beyond 1750 nm (as can be seen from (c) and (d)).

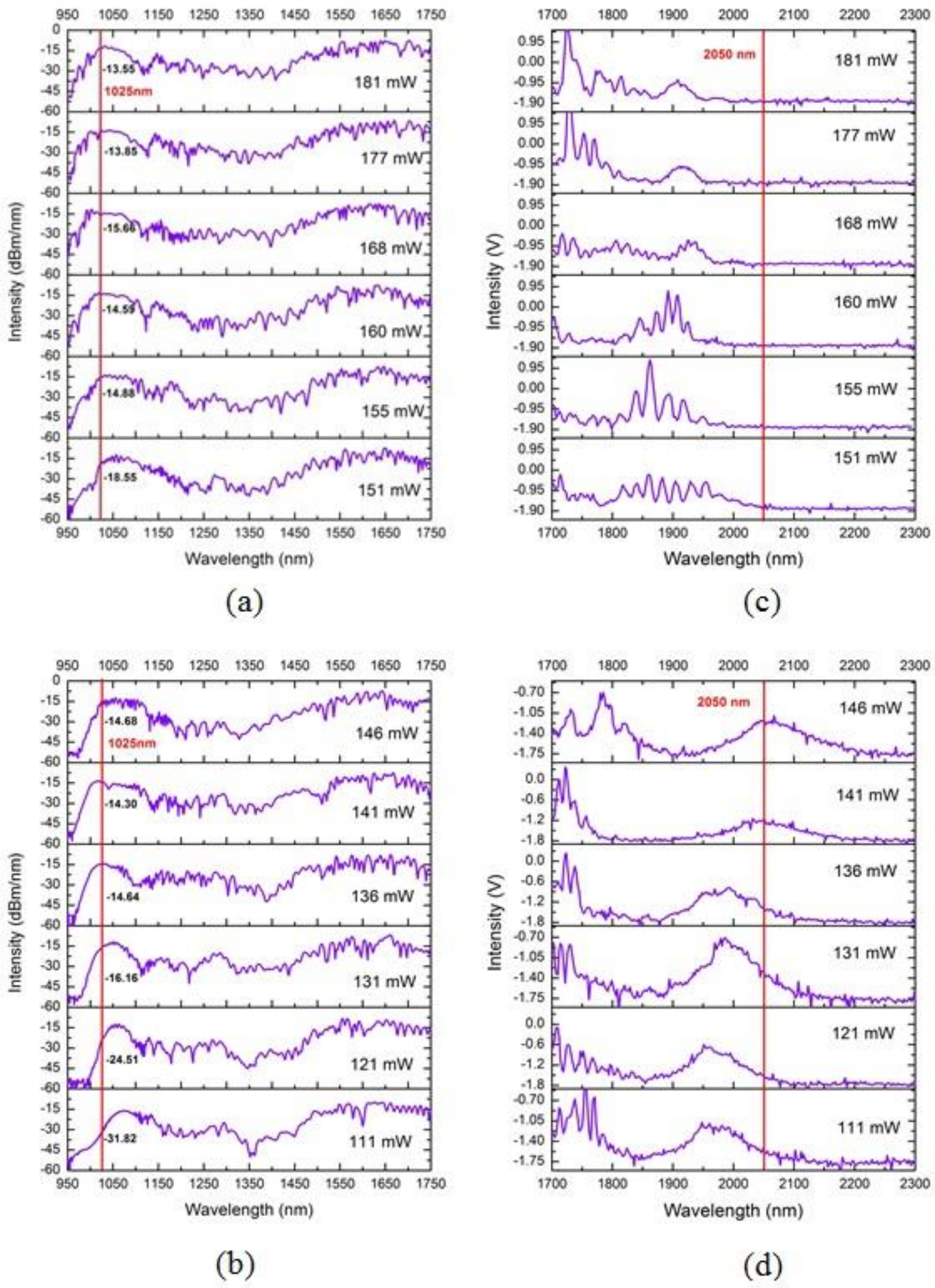

**Figure 4.6 The SC generation spectrum for different input power (*i.e.* the amplified power after EDFA). (a) and (b) are measured using OSA, (c) and (d) are measured using monochromator. As can be seen in (b) and (d), we get the broadest SC generation spectrum when the input power is about 146 mW. The two red lines are at 1025 nm and 2050 nm, which are respectively corresponding to $2f$ and $f$ for the self-referencing. The optimum length of the HNLF is about 1 m.**

We have measured the SC generation spectrum corresponding to different amplified power before the HNLF (which was realized by changing the EDFA pump), in order to find the maximum intensity for 1025 nm and 2050 nm. Note that the EDFA pump power corresponding to 111 mW and 181 mW amplified power is, respectively, 314 mW and 509 mW at 1480 nm. In Figure 4.6, (a) and (b) show that the 1025 nm component is significantly increased when the input pulse average power is increased from 111 mW  to 146 mW (as in (b)) , while there is no obvious increase when the input pulse average power keep increasing continuously up to 181 mW. However, from (c) and (d) we can see that the long wavelength component 2050 nm is at its maximum when the input pulse average power is about 146mW, but the intensity of 2050 nm suddenly dropped when the input pulse average power keep increasing up to 181 mW. Islam *et al.* [118] argued that the spectrum broadening during SC generation is due to modulation instability (MI) and soliton self-frequency shift (SSFS) effects. Actually, the evolution of long-wavelength edge of the SC spectrum due to Raman scattering as in Figure 4.6 (d) for different pump power is pretty close to what they observed.

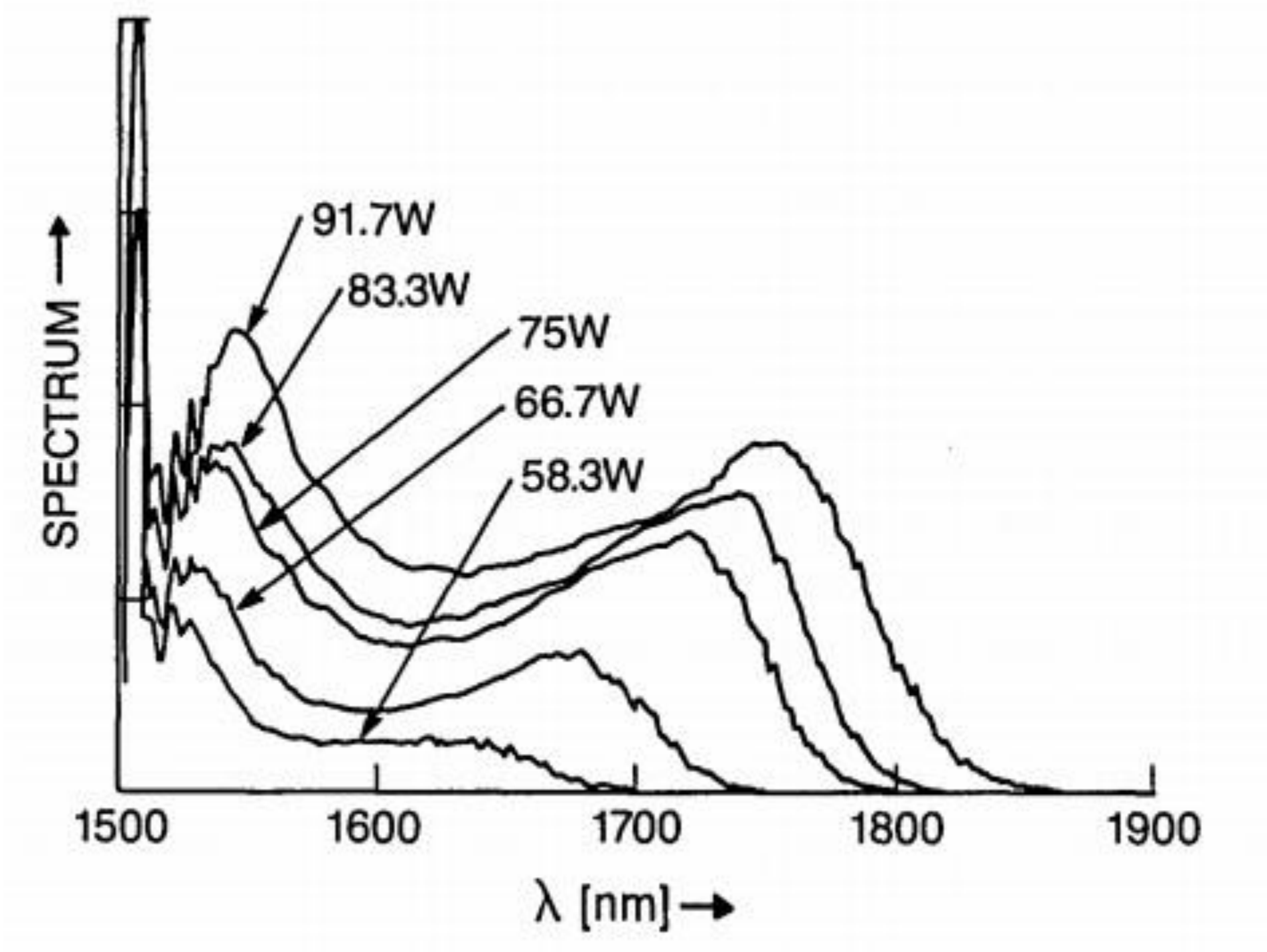

**Figure 4.7 The SC spectra as a function of peak power (*i.e.* different average power). Figure reproduced from Ref. [118].**

In previous work, Stumpf *et al.* [119]reported that the long-wavelength edge spectral broadening is attributed to Raman scattering self-frequency shift, while the short-wavelength edge spectral broadening is contributed by non-solitonic Raman radiation (*i.e.* dispersive wave generation). The dispersive wave is generated such that the main pulse component loses its energy due to high nonlinearity in HNLF [120]. Basically, the initial fundamental soliton propagation in the HNLF is associated with the generation of dispersive wave spectral components via resonant energy transfer across zero dispersion wavelength (ZDW). As the soliton continuously propagate along the HNLF, the SC generation keeps being shifted to broader spectrum at the long-wavelength edge through Raman scattering self-frequency shift. On the other hand, the generated Raman soliton and dispersive waves can couple via XPM, which in turn leads to generate more additional frequency components that contribute to make the overall SC spectrum even broader [121].

In general, the nonlinearity induced SPM effect and dispersion induced GVD are responsible for the pulse broadening during SC generation process. At the beginning, right after the SC pump (*i.e.* the amplified and compressed femtosecond laser) pulse launched into the HNLF, the SPM effect broadened the spectrum because of high peak power. After that, the long-wavelength edge reaches the anomalous dispersion wavelength beyond the ZDW regime, so that the Raman scattering self-frequency shift further extended the SC spectrum to longer wavelength continuously and the dispersion wave was generated in the short-wavelength edge when phase matching condition was met [122].

As can be seen from (a) and (b), the SC spectrum has maintained high degree of cross-coherence as the pump power (average input power into the HNLF) due to short ultra-short pulse duration of the input pulse [123]. In fact, there is coherence degradation near 1560 nm where the ASE spectrum of the EDFA becomes relatively strong [124, 125].It is also worth mentioning that to generate broader SC in even shorter length of HNLF, we may take advantage of UV exposing technique that continuously broaden the SC spectrum more than 100 nm by changing the refractive index of the doped core of the HNLF [126].

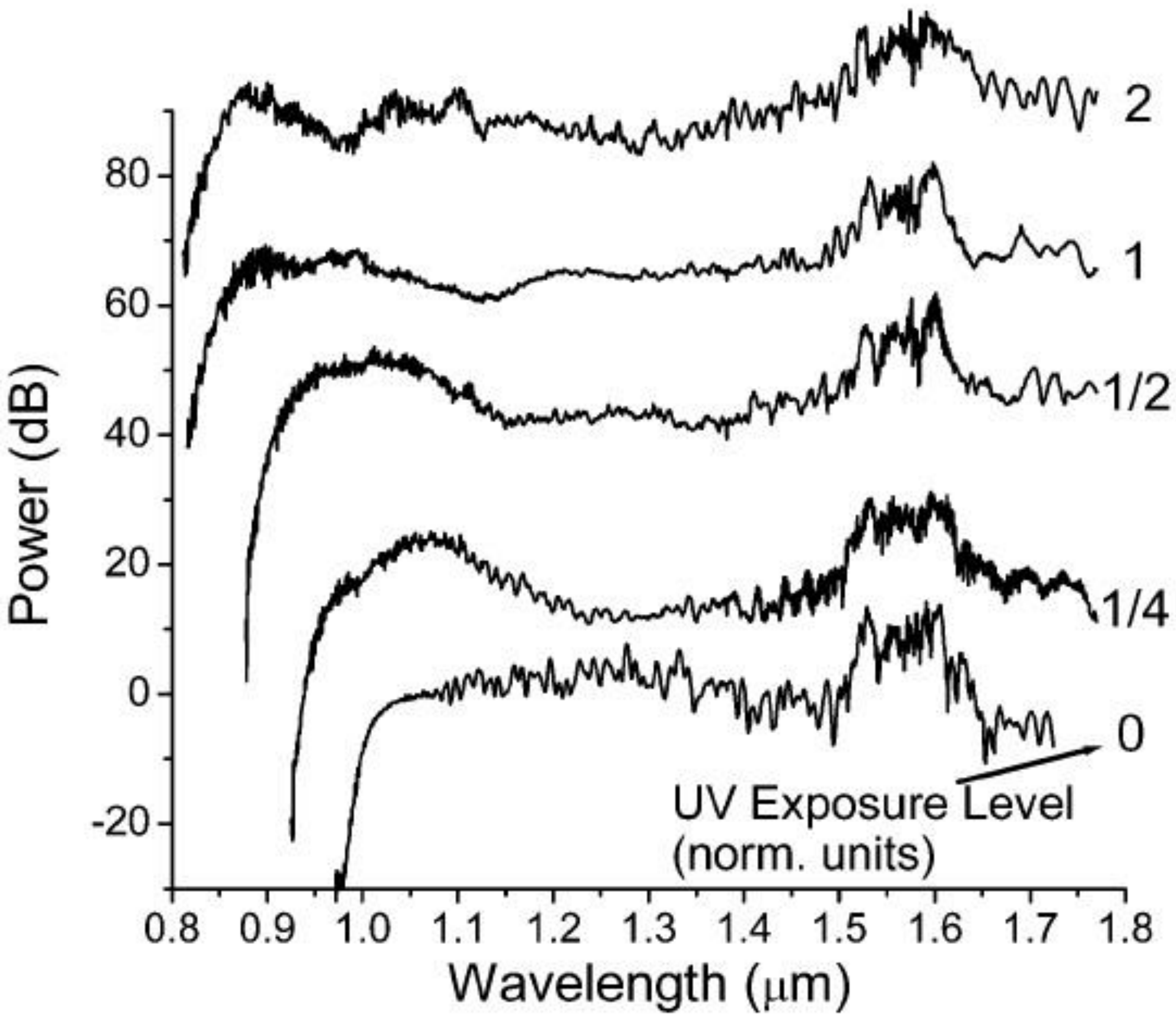

**Figure 4.8 The SC spectrum corresponding to different UV exposure level. Figure reproduced from Ref. [126].**

Another very important mechanism of SC generation is the polarization of the incident pulse. As in Figure 4.3, we have used the PC to control the polarization of pulse entering the HNLF, which effects the magnitude and spectral width of SC [127]. The magnitude of the incident pulse depends on the polarization of the incident pulse and increases as it evolves from circular to linear polarization.

Nevertheless, we have determined the optimum input pulse average power before the HNLF to guarantee the broadest SC generation spectrum both for short-wavelength edge and long-wavelength edge, as about 146 mW. At this point, the output average power of SC generation is about 102 mW at 1575 nm corresponding to the input of 146 mW at 1575 nm.

## 4.4 Summary

In this chapter, SC generation in a HNLF pumped by high power ultrashort femtosecond pulse from an $Er^{3+}$doped fiber laser comb is investigated. To generate ultrabroad SC spectrum, we have used different length of HNLF, which in turn, led us to the optimum length of HNLF. In addition, we have obtained different intensities of short-wavelength component (*e.g.* 1025 nm) and long-wavelength component (*e.g.* 2050 nm), by controlling the EDFA pump power. Eventually, we have optimized the HNLF length to about 1 m. Thus, we have presented the SC spectra ranging from near 1000 nm to 2100 nm with optimal EDFA pump power of 411 mW at 1480 nm, which corresponding to amplified power of 146 mW at 1575nm before the HNLF.

Moreover, we have presented the theoretical reasoning and explanations for the observed experimental SC generation results. So far, we have short-wavelength component near 1025 nm with intensity of about -14 dBm/nm and substantial amount of corresponding long-wavelength component near 2050 nm which will be used to obtain second harmonic generation (SHG) in periodically polled lithium niobate (PPLN) crystal. In the next chapter, we are going to investigate SHG of long-wavelength portion of supercontinuum generation (SCG) and the detection of optical heterodyne beat (*i.e.* the carrier envelop off-set frequency,$f_0$ or $f_{ceo}$) between short-wavelength portion of SCG and SHG from the PPLN crystal.

# Chapter 5 - Phase Stabilization of the Comb using $f$ to $2f$ Interferometer

In this chapter, we will present the experimental results and theoretical discussions on the attempt to phase stabilize the NPR mode-locked laser frequency comb using free-space $f$ to $2f$ interferometer. First, we put forward the $f$ to $2f$ interferometer schematic of experimental setup and brief overview of related previous researches. In addition, we need to compensate the time delay between the short wavelength portion (*e.g.* 1025nm) and long wavelength portion (*e.g.* 2050 nm) through the about 1 m long HNLF, as well as, the time delay due to different lengths of free-space light-path and optical devices that transmitting light (*e.g.* periodically polled lithium niobate).Thus, the SHG light and fundamental light from the two arms of the $f$ to $2f$ interferometer can overlap very well spatially and temporally. Then, the overlapped beams from the two arms can be coupled into the SMF patchcord that will be connected to the photodiode (PD) which converts the light signal into electric signal, after which we will be able to detect beat note with the electric signal analyzer (ESA).

## 5.1 Overview of Phase Stabilization and the $f$ to $2f$ Interferometer

Phase stabilization of femtosecond laser combs is very important for the application of frequency metrology [57, 128], spectroscopy [129, 130] and etc. The phase stabilization of mode-locked femtosecond lasers, such as Ti:Sapphire lasers and rare earthed doped fiber lasers, is achieved by detecting and controlling the carrier envelope offset(CEO) phase in $f$ to $2f$ interferometer [131]. In general,  we control the rate of change in CEO phase which corresponds to the controlling the rate of change in the CEO frequency in the frequency domain [132].Thus, we can obtain optical frequency combs that generate stable frequency lines that are uniformly separated by cavity repletion rate $f_{rep}$ and extremely high power per comb line [133].The development of phase stabilized frequency combs was not yet experimentally achieved until 1999 [54], even though the theoretical basis for the frequency comb and carrier envelope phase (CEP) was already came to known in the 1970s [134, 135].Without knowing the beat note (*i.e.* CEO frequency) $f_0$, the absolute frequency measurement could not be achieved. In a ring cavity ultrashort pulse laser comb, the comb teeth are separated by cavity round trip pulse period $T_{rep} = \frac{1}{f_{rep}} = \frac{L}{v_g}$ , where L is the ring cavity length and $v_g = c/n_g = c/(n + \omega dn/d\omega)$ is the group velocity of the pulse envelope for $n = n(\omega)$ which is the refractive index as a function of angular frequency [136].Then, the $n^{th}$ tooth of the comb is given by $f_n = nf_{rep} + f_0$, where *n* is integer that is on the order of $10^6$.

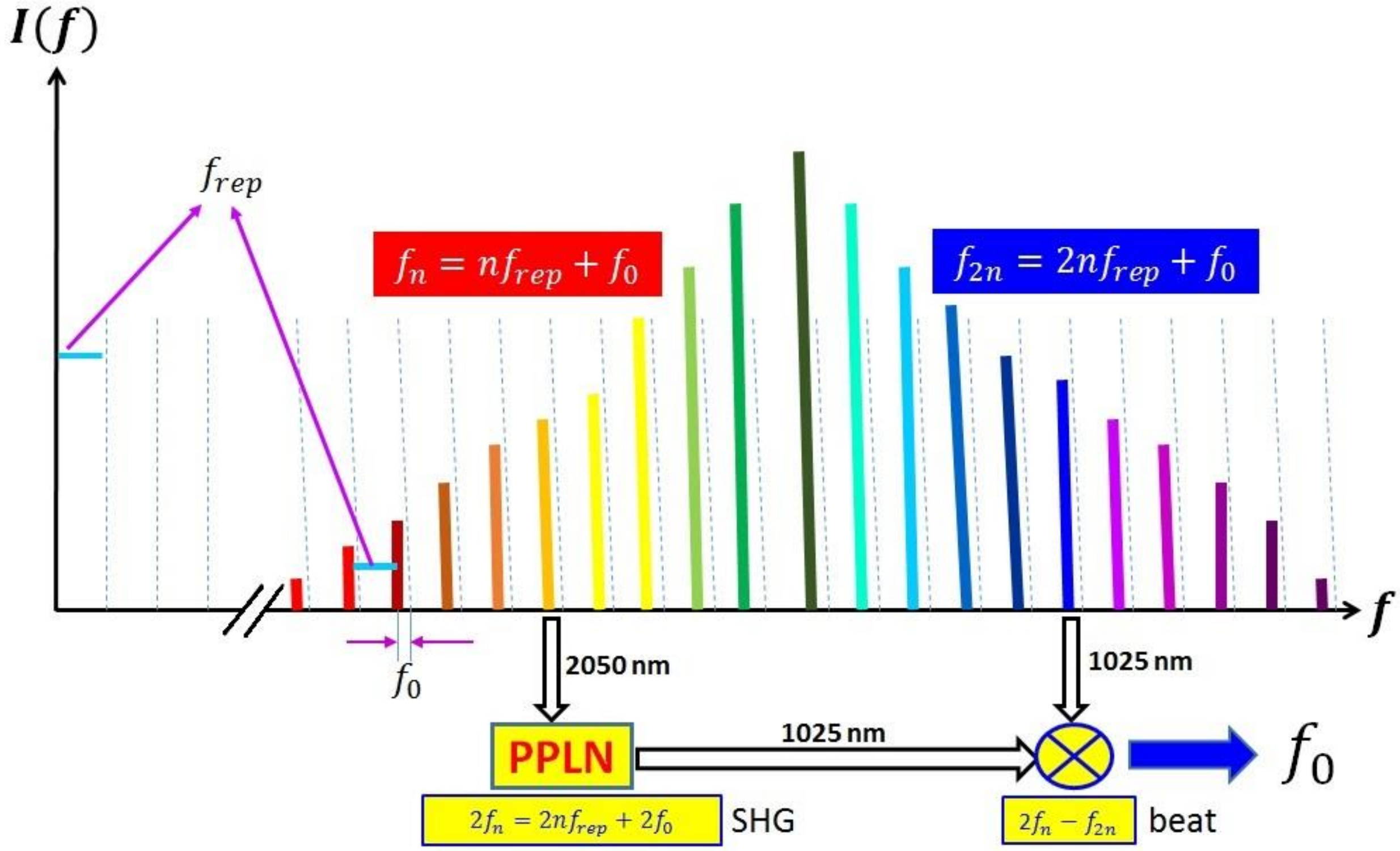

**Figure 5.1 The broadband SC is divided into two arms in f-2f interferometer. Low-frequency component is frequency-doubled through PPLN crystal and then mixed with high-frequency component (i.e. optical heterodyne beat between $2f_n$ and $f_{2n}$).The $f_0$ is detected using a fast 125 MHz photodetector. The entire comb (solid) is offset from integer multiples (dotted) of $f_{rep}$ by an offset frequency $f_0 = \frac{\Delta\phi_{ceo}}{2\pi} f_{rep}$, where the CEO phase is the pulse to pulse phase shift. It is constant in a phase-locked stable laser, while it changes in a nondeterministic manner in unstable lasers.**

It is relatively simple to phase lock the repetition frequency $f_{rep}$ to a RF source, but it is difficult to detect $f_0$ [43]. We can detect the repetition frequency directly after the NPR oscillator with a 10% fiber splitter output. The $f_{rep}$ can be controlled by changing the cavity length, which is done via the piezoelectric transducer (PZT) which is placed under one of the fiber collimator in the free-space NPR scheme (as shown in Figure 2.1). When it comes to beat note, the $f_0$ can then be detected through the heterodyne beat frequency between the high-frequency portion and the doubled low-frequency portion of the SC in an $f$ to $2f$ interferometer. The $f_0$ can be detected by changing the pump power of the NPR oscillator, which is done by a servo-controlled [137] feedback loop electronics that is connected to the oscillator pump laser diode. Finally, the comb systems is called phase stabilized, if the $f_0$ is phase locked to a stable RF oscillator by feedback to the laser [43].

The schematic of experimental setup for the $f$ to $2f$ interferometer in our research is shown in Figure 5.2. The SC is collimated through an aspheric lens with a focal length 4.5 mm. Then the broadband SC spectrum is divided into two arms with short-wavelength (*i.e.* high-frequency) portion goes to the fundamental arm and the long-wavelength (*i.e.* low-frequency) portion goes to the SHG arm of the $f$ to $2f$ interferometer. More specifically, the light of wavelength shorter than 1750 nm is reflected to the fundamental arm, while the light of wavelength longer than 1850 nm transmitted to SHG arm. The delay arm can insert negative or positive time delay on the pulses according to the time delay that has been induced by the HNLF on the short-wavelength and long-wavelength portion of the SC broadband. In our case, we have inserted positive time delay on the pulses through delay arm, since the shorter wavelength travels faster than longer wavelength in the HNLF. We then will be able to measure the beat note between fundamental arm light and SHG arm light, which is coupled into a short length of SMF patchcord through the PBS.

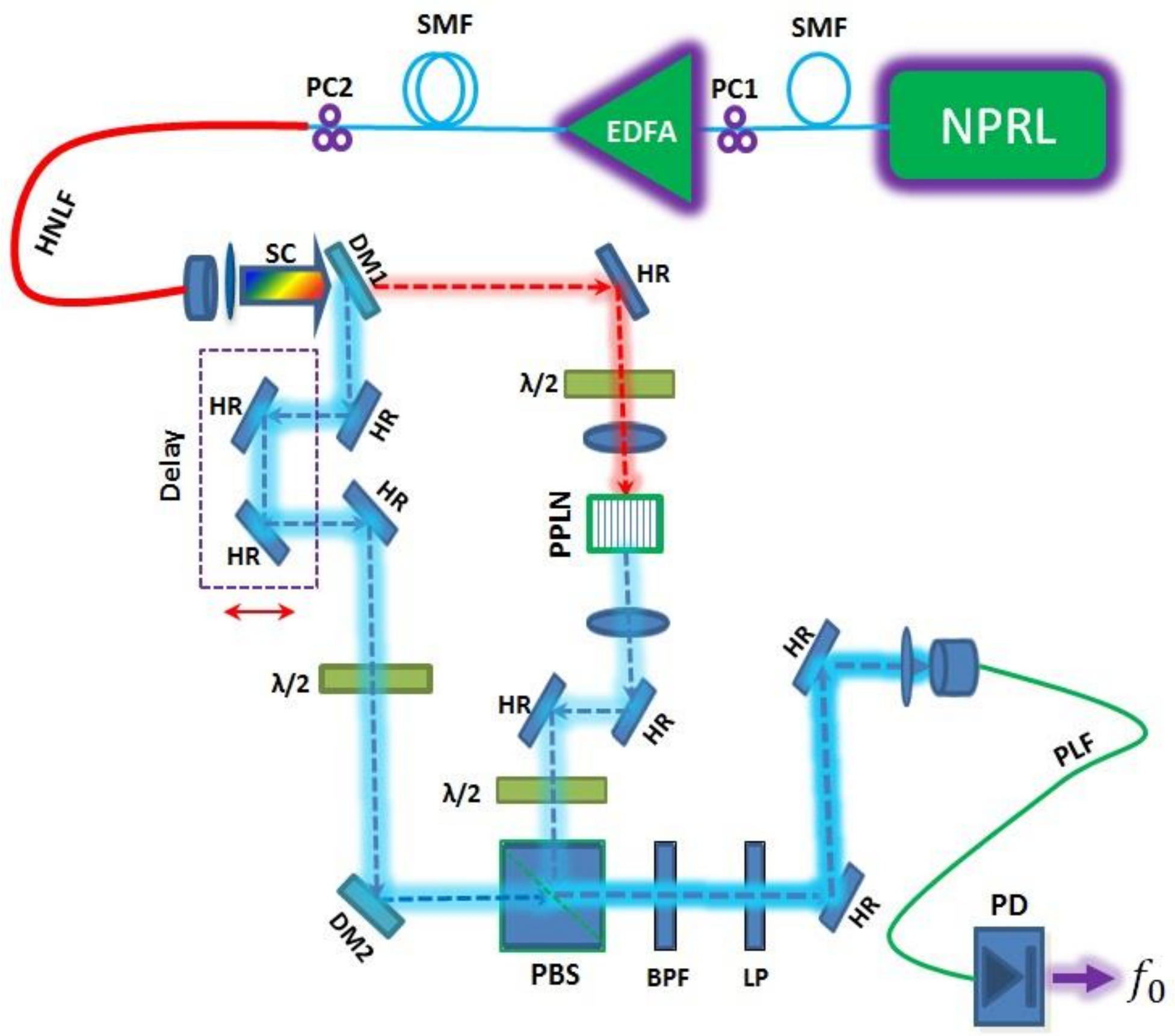

**Figure 5.2 The schematic of $f$ to $2f$ self-referenced interferometer used for carrier-envelope offset frequency (*i.e.* $f_{ceo}$ or $f_0$) detection. NPRL: NPR laser comb, PC: inline polarization controller, SC: supercontinuum, DM: dichroic mirror, HR: high-reflective mirror, PPLN: periodically poled lithium niobate, PBS: polarization beam splitter, BPF: band-pass filter, LP: linear polarizer, PLF: patch-cord lead fiber, PD: photodiode. All the other abbreviations are already defined in previous sections.**

## 5.2 Time-delay Measurement

The delay time between the different wavelengths of SC has been measured so that we could obtain best temporal overlapping of SHG arm light and fundamental arm light. The delay arm is designed such that we can adjust the length of beam path by which we can adjust the amount of delay to be inserted on the pulses passing through this delay-arm, by changing the position of the two HR that are placed on a micrometer adjustable stage. Moreover, we have measured the optical pulse fringe of the coupled pulse from the two arms of the $f$ to $2f$ interferometer, in order to precisely find position of the delay stage that insert enough delay so that the time delay between the short-wavelength 1025nm and SHG of long-wavelength (*i.e.* 1025 nm) is compensated to be zero. At this point, the lights from two arms of the $f$ to $2f$ interferometer may be best overlapped temporally and we can adjust the spatial overlapping by moving the reflective mirrors HR, which may enable us to detect stable and low noise beat note RF frequency through ESA.

### 5.2.1 Time Delay of Pulses in HNLF

The time delay in HNLF is due to chromatic dispersion corresponding to different wavelength components in SC spectrum. The first order dispersion is $\beta = \frac{n(\omega)}{c}\omega$ , then:

$$\Delta t = 10^9[\beta_1(\omega_1) - \beta_1(\omega_2)] \text{ and } \beta_1(\omega) = \frac{\partial\beta}{\partial\omega} = \frac{1}{v_g} \tag{5.1}$$

where $\Delta t$ is the time-delay between high and low frequency components of the SC and the unit of $\Delta t$ is fs/m; $\omega_1 = \frac{2\pi c}{\lambda_1}$ and $\omega_2 = \frac{2\pi c}{\lambda_2}$ are, respectively, stands for high-frequency(*i.e.* short-wavelength such as 1025 nm ) component and low-frequency (*i.e.* long-wavelength such as 2050 nm).

Basically, different wavelength corresponds to different group velocity,$v_g$, which leads to the case that short-wavelength portion comes out of the HNLF prior to the long-wavelength portion. That means we need to insert positive time delay on the short-wavelength components of SC output after the HNLF.

### 5.2.2 Time Delay of Pulses in Free-space $f$ to $2f$ Interferometer

In order to obtain best temporal overlap between SHG and fundamental arm light, extra delay will be inserted on the short-wavelength portion through the delay-stage in the fundamental arm of $f$ to $2f$ interferometer. The fringe spectrum as in Figure 5.3 is measured using OSA after both are coupled into a short length of MMF since it makes the coupling process much easier and time-saving. When the fringe is first observed, it shows many interference oscillation spectra as in Figure 5.3from (a) to (d). Next, we need to adjust the position of time-delay translational stage, until we get the spectrum in Figure 5.3 (e) which is corresponding to zero-delay between the two pulses. Eventually, the two pulses which are combined via PBS have zero delay (or near the zero-delay point) in time domain [138, 139].

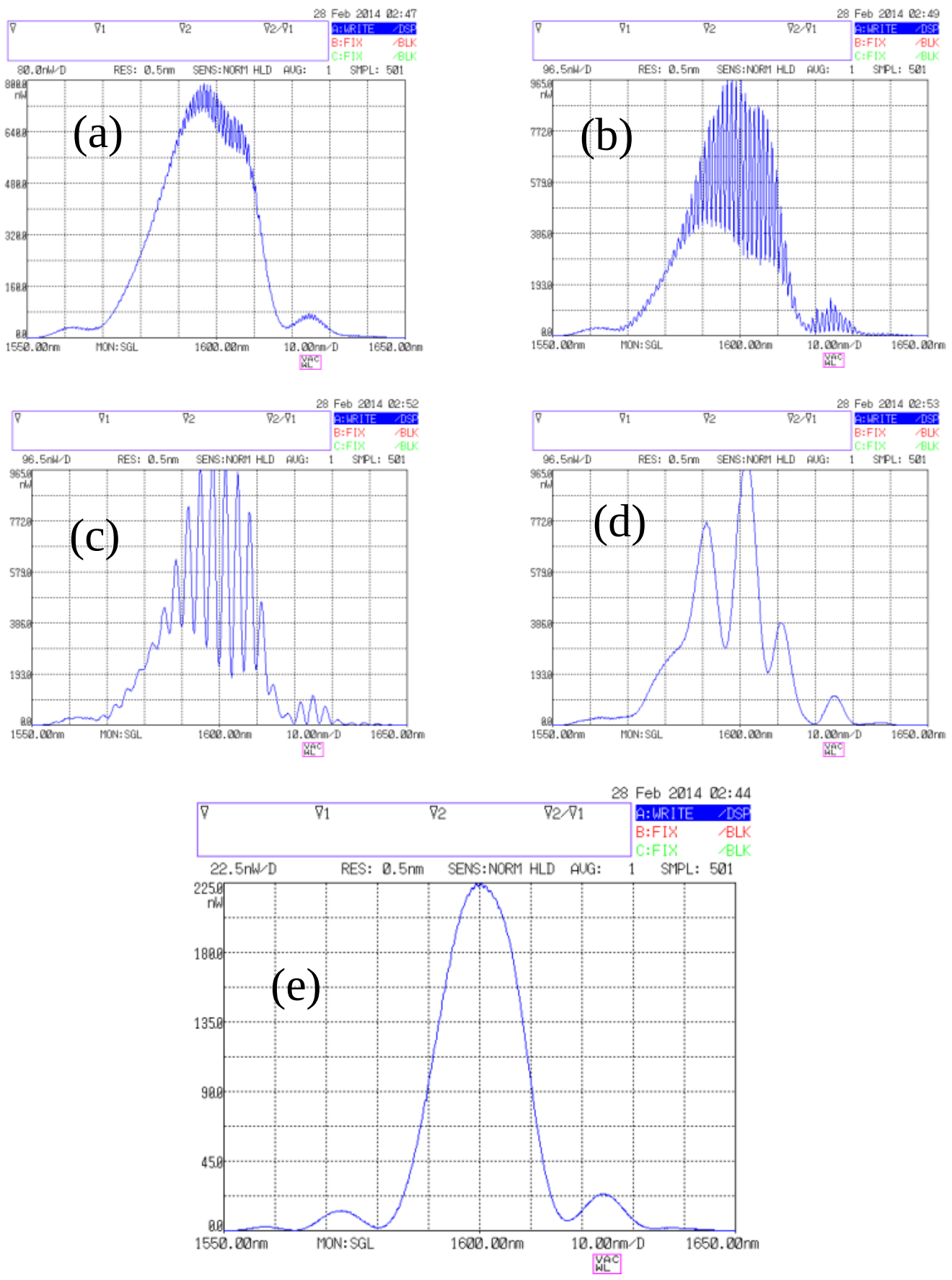

**Figure 5.3 The measured two pulse fringes for time-delay compensation in the $f$ to $2f$ self-referencing interferometer. From (a) to (e), the time-delay difference between the two arm pulses is decreasing. At the delay stage position of fringe spectra (e), the time-delay is zero.**

## 5.3 Quasi-phase Matching and Frequency Doubling

### 5.3.1 Overview of Quasi-phase Matching

Quasi-phase-matching (QPM) [140, 141] is a technique to match the phase [142] of nonlinear optical interactions in nonlinear medium in which the relative phase in the medium is periodically changed by the use of electric fields applied through lithographically defined electrodes [143].When light propagates through a linear-optical media the electric field accumulates a phase due to $e^{-ik(\omega)z}$,where the $k(\omega)$ is the propagation constant (also called wave-vector amplitude)and z is the propagation distance. In a nonlinear-optical medium, the phase is modified by nonlinear polarization [144]. The phase accumulated by a spectral component in linear-medium is different from the one in nonlinear-medium. Thus, we call the phase difference between these two processes as phase-mismatch. For fundamental harmonic (FH) light of $\omega_1$ and second harmonic (SH) light of $\omega_2$ , then the phase difference is $\Delta k = k_2 - 2k_1 = \frac{4\pi}{\lambda}[n(\omega_2) - n(\omega_1)]$. Due to dispersion, the phase difference is not zero, *i.e.* $\Delta k \neq 0$ , unless the medium is specially treated such as QPM nonlinear crystals or birefringence phase matching (BPM) crystals. In a QPM material, the nonlinear coefficient is modulated with a period twice the coherent length of the interaction to compensate the accumulated phase mismatch [145]. Boyd calculated [146] the phase-mismatch as $\Delta k = 3.2/l$, which gives $\Delta k$ =32 in our case for the PPLN length of 10 mm.

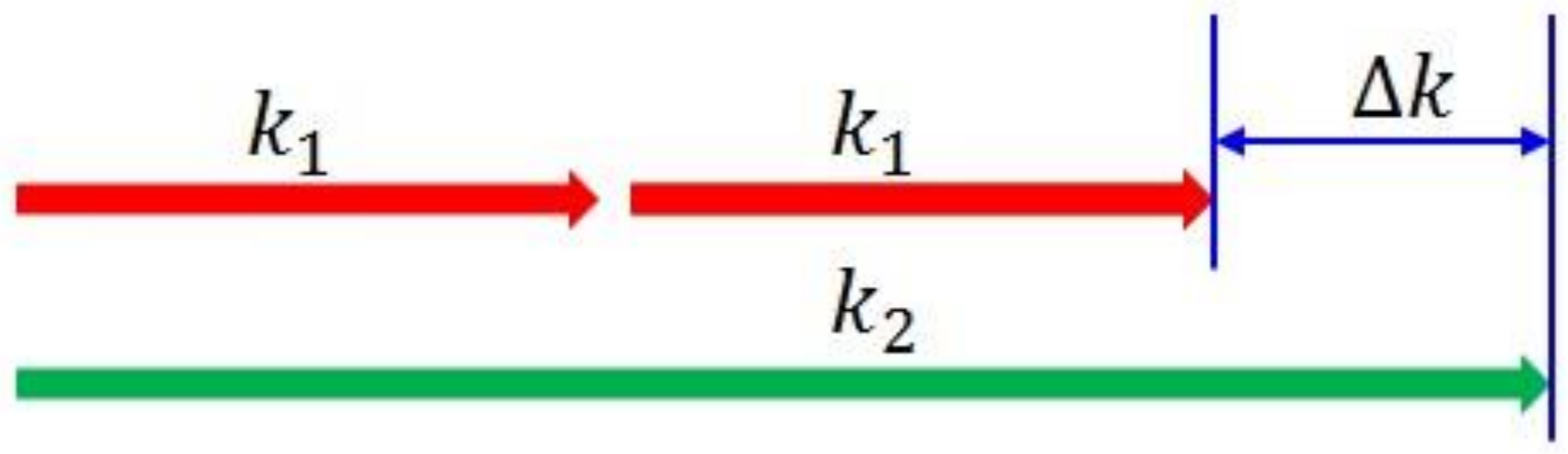

**Figure 5.4 The schematic of phase-mismatch in a nonlinear medium. Because of dispersion, the wavenumber of SH is more than two times larger than of FH, i.e. $k_2 > 2k_1$, where $k_1 = k(\omega_1)$ and $k_2 = k(\omega_2)$. And this can be phase-matched by QPM or BPM.**

There are two major ways to increase the efficiency of the nonlinear interactions (*i.e.* efficiency of SHG in our case), including QPM and birefringence -phase-matching (BPM). The latter is perfect phase-matching (*i.e.* $\Delta k = 0$ ) process, while QPM is imperfect (*i.e.* $\Delta k$ is not exactly equals to zero). In the case of QPM, the phase-mismatch condition becomes $\Delta k = 2k_1 - k_2 - k_g$ , where the grating vector $k_g = 2\pi/\Lambda_g$ and $\Lambda_g$ is the grating period (or QPM period as in Figure 5.6). As seen in Figure 5.5, the QPM can increase the efficiency of SHG and it can be achieved using non-birefringent materials [147]. Easily derived from the nonlinear wave equation, the intensity of SHG can be written as [148]

$$I(\omega_2) = \frac{2\mu_0^3 c^3 \omega_1^2 d_{eff}^2 l^2 I(\omega_1)^2}{n(\omega_1)^2 n(\omega_2)} \left(\frac{sin(\Delta kl/2)}{\Delta kl/2}\right)^2 \tag{5.2}$$

where $\mu_0$ is the permeability of free space, $d_{eff} = \frac{2}{\pi} d_{33}$ is the effective second-order nonlinear coefficient of PPLN and other parameters are already defined.

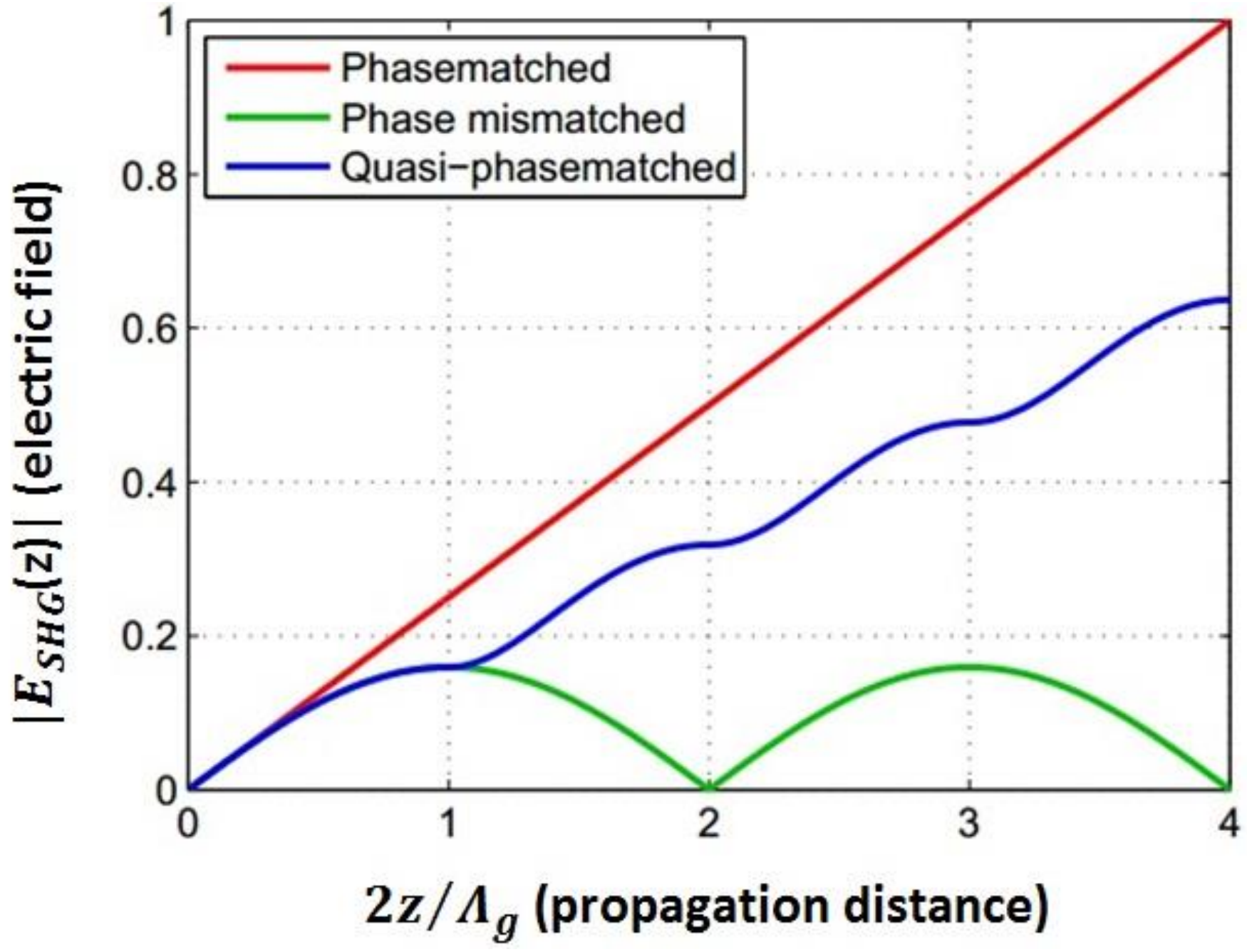

**Figure 5.5 The evolution of the SHG electric field with propagating distance in the PPLN crystal. In case of phase-matched (red), *i.e.* the BPM: linear growth of electric field and quadratic growth of intensity for the SHG. Figure reproduced from Ref. [147].**

In the past, QPM has been achieved using different nonlinear materials, such as, $LiNbO_3$, $LiTaO_3$, GaAs, CdTe and quartz [147, 149]. Among those, the periodically poled lithium niobate (PPLN) ferroelectric crystal is a very attractive candidate for QPM, due to a mature fabrication technology, large effective nonlinearity ($d_{eff}$ ~16.5 pm/V), wide acceptance range of input wavelength (which is from 0.35$\mu m$ to 4$\mu m$ [150]), wider temperature acceptance and readily availability in interaction lengths up to 50 mm [151].

### 5.3.2 Quasi-phase Matching in PPLN

In our experiment, we have used a Covesion [152]made PPLN crystal (size of 10 mm × 10 mm × 0.5 mm corresponding to length, width and thickness) for the purpose of QPM second harmonic generation (SHG). It has total seven periodically poled gratings (*i.e.* poling periods), including 29.5 $\mu m$, 30.0 $\mu m$, 30.5 $\mu m$, 31.0 $\mu m$, 31.5 $\mu m$, 32.0 $\mu m$ and 32.5 $\mu m$. Both of the input and output facets of the PPLN crystal are AR coated to less than 1.0% reflectivity at 1000-1150nm/2000-2300nm. Each poling period is 0.5 mm wide and separated by 0.2 mm regions of unpoled material. The heating temperature is up to 200$^{o}$C. It is worth noting that the photorefractive [153] effect is easier to avoid when heating temperature is relatively higher.

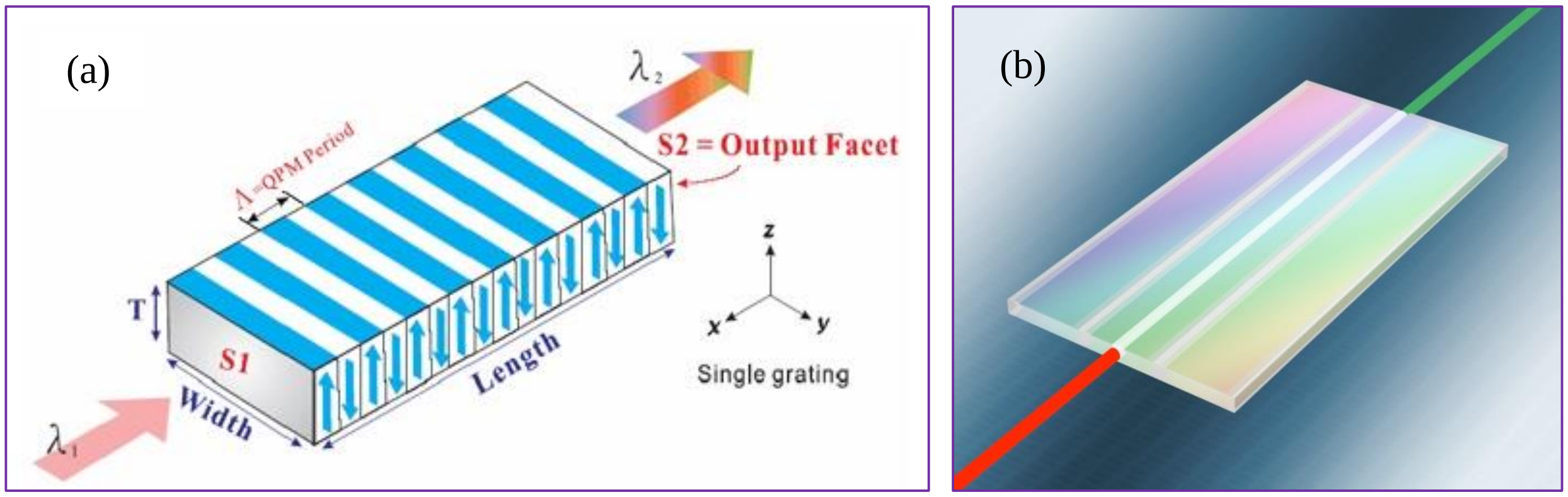

**Figure 5.6 The schematic structure of single-grating PPLN crystal (a), and the graphic model of a triple-grating PPLN crystal (b). The dielectric polarization of the crystal is periodically changed inversely, and the polarization direction of the input and output light is in the thickness direction (*i.e.* on the z-axis)of the PPLN. Figure reproduced from Ref. [154, 155].**

### 5.3.3 SNLO Calculation of Quasi-phase Matching

We have used the SNLO [156] software, in order to theoretically calculate the right poling period corresponding to the 2060 nm pump wavelength (*i.e.* the input wavelength $\lambda_1$).As is shown in Figure 5.7, the corresponding poling period to obtain SHG at 1030 nm is 30.0 $\mu m$.

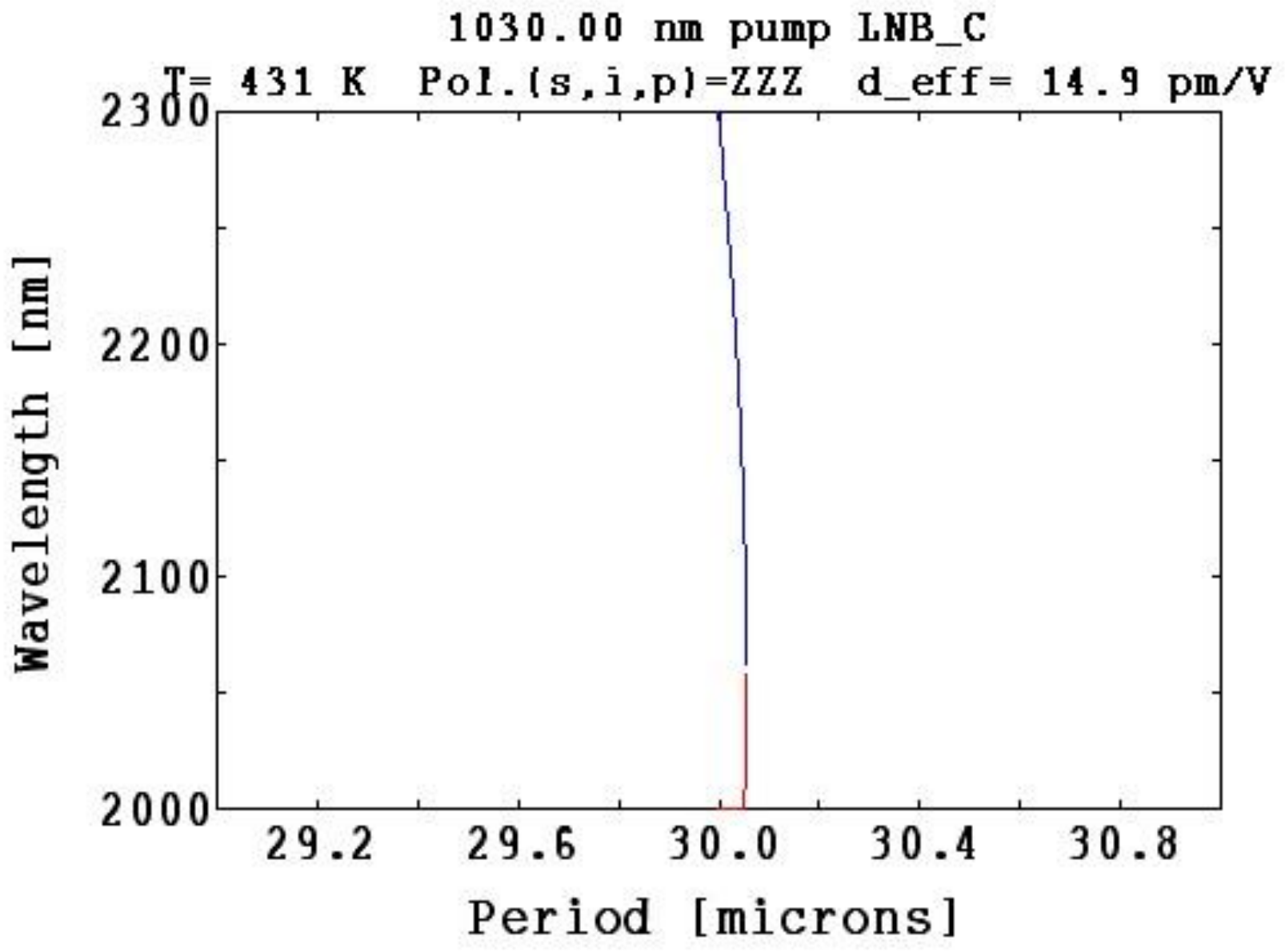

**Figure 5.7 The SNLO calculation result of the poling period for input wavelength at 1030nm.The poling period of 30 $\mu m$ is calculated, which is close to the experimental result.**

## 5.4 SHG Experimental Results

### 5.4.1 Review of Second Harmonic Generation in PPLN

Second harmonic generation (SHG) was first observed using ruby laser in crystalline quartz by Franken *et al.* in 1961 [157]. In the existence of high intensity electric-field, the nonlinear polarization of the medium is written by

$$P = \epsilon_0 \cdot (\chi^{(1)} \cdot E + \chi^{(2)} \cdot E^2 + \chi^{(3)} \cdot E^3 + \cdots) \qquad (5.3)$$

where $\epsilon_0$ and $\chi$ are, respectively, the vacuum permittivity and susceptibility among which the $\chi^{(1)}$is the linear susceptibility term, $\chi^{(2)}$is the second order susceptibility (which results in the SHG) and the rest are due to higher order nonlinear effects. Physically, the SHG process can be seen as an exchange of photons between the different frequency components of the electric field.

The first experimental result of SHG in lithium niobate was reported by Boyd *et al.* in 1964. According to Boyd-Kleinman theory [146], the SHG can be described as a single quantum-mechanical process in which two photons having a frequency of ω are destroyed while a single photon of frequency 2ω is created. Among all, the conversion efficiency performance of the nonlinear crystal to be used for SHG experiment is of vital importance. Therefore, we have chosen the PPLN crystal to obtain SHG, since there is no walk-off phenomenon in PPLN as in other common crystals such as $\beta$-barium borate(BBO) [158-160].

### 5.4.2 The Measured Experimental SHG Spectrum

The QPM SHG nonlinear crystal, *i.e.* PPLN, is placed on a three axes translational stage, where x-direction controls the poling period that light goes through in PPLN, y-direction controls the vertical position of the PPLN and the z-direction controls the horizontal position of the PPLN in the light propagating direction. Initially, the PPLN crystal center is placed at the focusing point of the Gaussian pulse, which is focused by a 30 mm achromatic lens (or lens1 for short).In the x-direction, it is placed at poling period of 30 $\mu m$ which is predicted by SNLO in Section 5.3.3. Then we can let the fundamental light go through the PPLN from the center of the input facet, as we gradually lower the PPLN in the y-direction from an initial higher position that exactly block the fundamental light on the metal that holds the PPLN.

We have measured the SHG spectrum using the OSA, after the SHG arm of the $f$ to $2f$ interferometer is coupled into a multi-mode-fiber (MMF). Our goal was to obtain SHG as high as -30dBm/nm in the SMF. To reach this point, we have taken advantage of the fiber based PC (as shown in Figure 5.2) to adjust the polarization of the fundamental light as well as the bandwidth of the SC spectrum that injected into the $f$ to $2f$ interferometer. Moreover, the HWP right before the PPLN focusing lens and the PC are used in group to improve the SHG spectrum. In addition, the heating temperature of the PPLN is vital to tune the SHG central wavelength to the wanted range of wavelength (*i.e.* ~1030 nm) due to the working wavelength of the optical devices and the limitations of the SC bandwidth. Finally, we found that at heating temperature of 160 $^{o}$C the central wavelength of the SHG is near 1030 nm as shown in Figure 5.8. As this result shows, in (a) the SHG central wavelength red-shifts as the poling period of PPLN increases, which is consistent with the given data sheet from the factory. On the other hand, in (b) the SHG central wavelength also red-shifts as the heating temperature increase. In the meantime, however, the intensity of the SHG gradually decreases proportional to the heating temperature and poling period of the PPLN crystal.

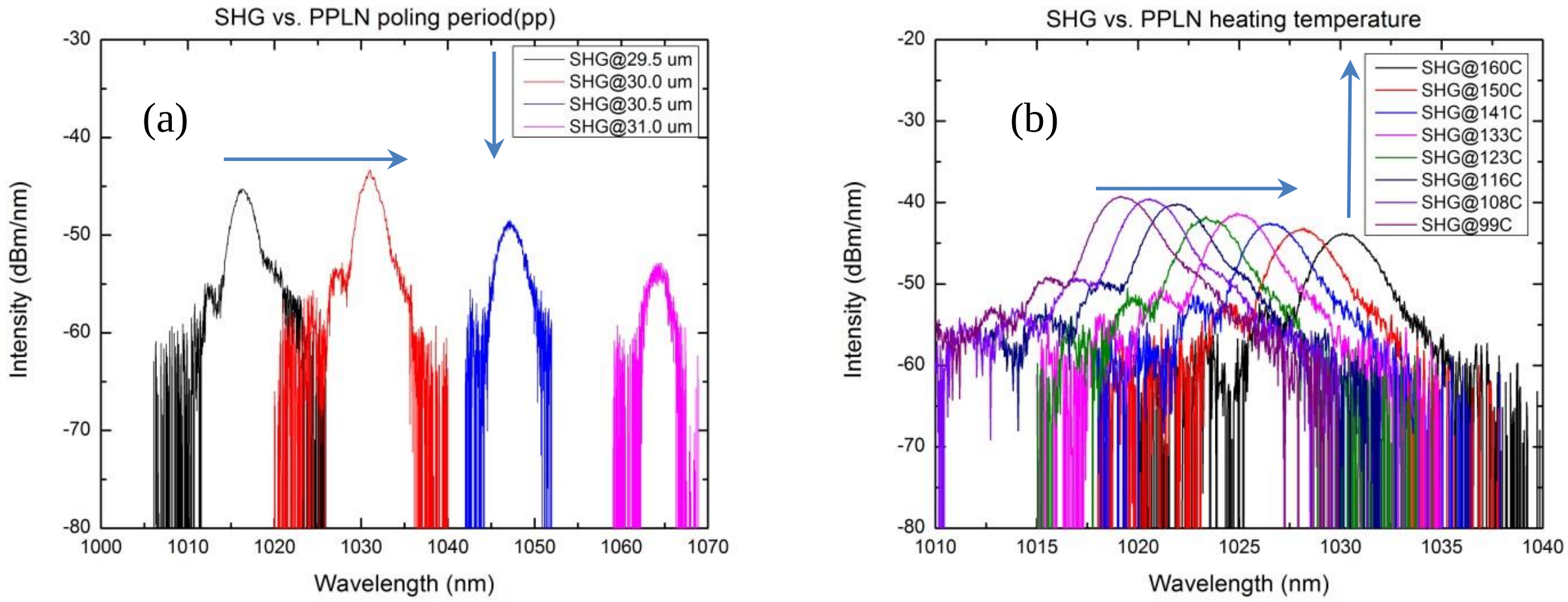

**Figure 5.8 The experimental result of SHG. (a) is the SHG spectra corresponding to different poling period of the PPLN crystal . (b) is the SHG spectra corresponding to different heating temperature of the PPLN crystal. By changing the poling period or the heating temperature, we were able to tune the SHG central wavelength. The PPLN used in this experiment is Covesion SHG7-0.5-10.**

### 5.4.3 Difficulties and Existing Problems in Obtaining SHG

There are several practical issues in terms of obtaining stable and high intensity SHG, including the repeated free-space alignment of the beam in the $f$ to $2f$ interferometer, low coupling efficiency of the beam into SMF, determining the optimum PPLN heating temperature and etc. Especially, the process of adjusting the PPLN position and realigning the beam over and over again is very time consuming. It is hard to see the light on the input facet of the PPLN, therefore it is not very obvious to see if the light is going through at the middle of the PPLN, which is of vital importance to ensure high power density in the PPLN to initiate high nonlinear interactions.

It is also worth mentioning that we did not conduct any theoretical simulation of the SC generation and SHG process, which would have been very helpful to determine the optimum condition for the best experimental result.

## 5.5 Summary

In this chapter, we have presented the experimental results of our attempt towards the CEO phase stabilization of the NPR mode-locked laser comb system using a self-referenced $f$ to $2f$ interferometer. Due to chromatic dispersion, there is time delay between long-wavelength and short-wavelength portion of the SC spectrum. To avoid that, we have implemented a time-delay translation stage in the $f$ to $2f$ interferometer. We also have mentioned the theoretical background and reasoning of QPM SHG in a PPLN nonlinear optical crystal. The experimental result of SHG and the theoretically predicted poling period for optimum QPM are in good agreement. Lastly, the experimental result of SHG has presented and briefly explained. Indeed, the SHG result is not very satisfying since we were not able observe the beat note between the SHG light and the short-wavelength edge of the SC spectrum. Finally we have discussed the existing issues and potential problems in terms of unsuccessful try for beat note.

# Chapter 6 - Conclusions and Outlook

## 6.1 Thesis Summary

In this work, we have put forward a hybrid design of free-space NPR mode-locking $Er^{3+}$ doped fiber laser cavity towards the application of optical frequency comb. The effective length of the cavity is shortened because of the free-space section in the laser cavity. Moreover, this free-space NPR mode-locking scheme enables the laser oscillator to extract high average output power of about 30 mW at 1575 nm with repletion rate of about 89 MHz. In addition, this NPR laser has output pulse spectral bandwidth of about 50 nm. On the other hand, the NPR laser operates at stretched pulse mode-locked state due to its net normal cavity dispersion. Thus, it makes the NPR laser a very stable and robust tool to build up potential frequency comb system.

Further pulse amplification and compression was needed to generate broadband supercontinuum (SC) generation, even though the output average power from the oscillator is about 10 times higher than traditional all-fiber ring cavity oscillators. The pulse is first pre-chirped through single mode fiber (SMF) and amplified via erbium doped fiber amplifier (EDFA) to an output average power of 181 mW at 1575 nm. Then, it is post-chirped, by dispersion compensation through a longer length of SMF, to compress the pulse to near transform limited (TL) pulse duration. As a result, we have achieved an ultrashort pulse (USP) of 70 fs full width at half maximum (FWHM) pulse duration and about 28 kW peak power.

Then, we have obtained SC generation broadband from 1000 nm to 2100 nm by injecting the above ultrashort high power pulse into a 1 m long dispersion flattened highly nonlinear fiber (HNLF). The output average power of SC generation is about 140 mW at 1575 nm. Moreover, the SC spectrum is optimized by a pair of inline polarization controller (PC) and half waveplate (HWP).

Finally, the SC generation is injected into the free-space self-referenced $f$ to $2f$ interferometer to detect the beat note frequency, namely, the carrier envelope offset (CEO) frequency ($f_0$). In this setup, the SC is divided into two arms through a dichroic mirror, including delay-arm (which contains wavelength components <1800 nm) and SHG-arm (which contains wavelength components >1800 nm). The second harmonic generation (SHG) is obtained through quasi-phase-matching (QPM) crystal of periodically poled lithium niobate (PPLN). Meanwhile, the delay-arm has been used to obtain zero-time-delay between the short-wavelength edge of SC generation and the SHG of long-wavelength edge of SC generation. Thus, we are one step closer to achieve a phase-stabilized self-referencing NPR mode-locked laser comb system, although further experiment is needed in the future for CEO-phase-locking stabilization of the system.

## 6.2 Future Work

The octave spanning NPR mode-locked frequency comb can be stabilized in terms of the repetition rate $f_{rep}$ and the beat note $f_0$ [39, 55, 161-165] .The CEO frequency $f_0$ is measured by a fast-photodetector (PD) using self-referencing technique and can be stabilized to a RF synthesizer. Then, the error signal will be fed-back using servo electronics to the pump power (which is directly controlled by laser diode pump current) of the oscillator. While, the repetition frequency $f_{rep}$ can be measured directly from the oscillator using a fast-photodiode then with RF counter. Then, it will be also stabilized to a RF synthesizer and fed-back using servo to the piezoelectric transducer (PZT) that controls the oscillator cavity length [166].Once the $f_{rep}$ and $f_0$ are stabilized, the frequency comb can be used as an 'optical ruler' for unknown frequency measurement. In the future, the phase-stabilized NPR mode-locked laser based optical frequency comb can be used in our laboratory (LUMOS/UNFO) for absolute frequency measurements. Given the repetition frequency $f_{rep}$ and CEO frequency $f_0$, we can write down the unknown frequency $f_u$ in terms of $f_{rep}$, $f_0$ and the mode number of the comb $n_c$ as [39, 162, 167]:

$$f_u = f_{n_c} \pm f_b = f_0 + n_c f_{rep} \pm f_b \tag{6.1}$$

where the $f_b$ is the beat note between the unknown laser and the frequency comb. The sign ambiguity of $f_b$ arises because it is not known that whether the individual comb line is at higher or lower frequency than of the laser frequency $f_u$. One can determine the sign in front of $f_b$ by adding or subtracting a known offset to the frequency of the laser and observe the deviation of $f_b$.

# References

1. Hänsch, T.W., *Nobel Lecture: Passion for precision.* Reviews of Modern Physics, 2006. **78**(4): p. 1297-1309.
2. Hall, J.L., *Nobel Lecture: Defining and measuring optical frequencies.* Reviews of Modern Physics, 2006. **78**(4): p. 1279-1295.
3. Moulton, P.F., *Sceptroscopic and laser characteristics of Ti:$Al_2O_3$.* J. Opt. soc. Am. B, 1986. **3**(1): p. 125-133.
4. Spence, D.E., P.N. Kean, and W. Sibbett, *60-fsec pulse generation from a self-mode-locked Ti:sapphire laser.* Optics Letters, 1991. **16**(1): p. 42-44.
5. Pinto, J.F., et al., *Improved Ti: Sapphire Laser Performance with New High Figure of Merit Crystals.* IEEE Journal of Quantum Electronics, 1994. **30**(11): p. 2612-2616.
6. Ell, R., U. Morgner, and F.X. Kartner, *Generation of 5-fs pulses and octave-spanning spectra directly froma Ti:Sapphire laser.* Optics Letters, 2001. **26**(6): p. 373-375.
7. Fortier, T.M., D.J. Jones, and S.T. Cundiff, *Phase stabilization of an octave-spanning Ti:sapphire laser.* Optics Letters, 2003. **28**(22): p. 2198-2200.
8. Lyon, M. and S.D. Bergeson, *Precision spectroscopy using a partially stabilized frequency comb.* Applied Optics, 2014. **53**(23): p. 5163-5168.
9. Nelson, L.E., et al., *Ultrashort-pulse fiber ring laser.* Appl. Phys. B, 1997. **65**: p. 277-294.
10. Newbury, N.R. and W.C. Swann, *Low-noise fiber-laser frequency combs (Invited).* J. Opt. Soc. Am. B, 2007. **24**(8): p. 1756-1700.
11. Deng, D., et al., *55-fs pulse generation without wave-breaking from an all-fiber Erbium-doped ring laser.* Optics Express, 2009. **17**(6): p. 4284-4288.
12. Byun, H., et al., *Compact, stable 1 GHz femtosecond Er-doped fiber lasers.* Applied Optics, 2010. **49**(29): p. 5577-5582.
13. Diddams, S.A., *The evolving optical frequency comb [Invited]*. J. Opt. Soc. Am. B, 2010. **27**(11): p. B51-B62.
14. Richardson, D.J., J. Nilsson, and W.A. Clarkson, *High power fiber lasers: current status and future perspectives [Invited]*. J. Opt. Soc. Am. B, 2010. **27**(11): p. B63-B92.
15. Deng, D., et al., *70-femtosecond Gaussian pulse generation in a dispersion-managed erbium-doped fibre laser.* Journal of Modern Optics, 2011. **58**(7): p. 625-630.
16. Kim, S., et al., *Hybrid mode-locked Er-doped fiber femtosecond oscillator with 156 mW output power.* Optics Express, 2012. **20**(14): p. 15054-15060.
17. Sibbett, W., A.A. Lagatsky, and C.T.A. Brown, *The development and application of femtosecond laser systems.* Optics Express, 2012. **20**(7): p. 6989-7001.
18. Yang, T., et al., *A Compact 500 MHz Femtosecond All-Fiber Ring Laser.* Applied Physics Express 2013. **6**.
19. Li, X., W. Zhou, and J. Chen, *41.9 fs hybridly mode-locked Er-doped fiber laser at 212 MHz repetition rate.* Optics Letters, 2014. **39**(6): p. 1553-1556.
20. Xia, H., et al., *Nanosecond pulse generation in a graphene mode-locked erbium-doped fiber laser.* Optics Communication, 2014. **330**: p. 147-150.
21. Sawai, S., et al., *Demonstration of a Ti:sapphire mode-locked laser pumped directly with a green diode laser.* Applied Physics Express, 2014. **7**.
22. Roth, P.W., D. Burns, and A.J. Kemp, *Power scaling of a directly diode-laser-pumped Ti:sapphire laser.* Optics Express, 2012. **20**(18): p. 20629-20634.

23. Roth, P.W., et al., *Direct diode-laser pumping of a mode-locked Ti:sapphire laser*. Optics Letters, 2011. **36**(2): p. 304-306.
24. Roth, P.W., et al., *Directly diode-laser-pumped Ti:sapphire laser*. Optics Letters, 2009. **34**(21): p. 3334-3336.
25. Durfee, C.G., et al., *Direct diode-pumped Kerr-lens mode-locked Ti:sapphire laser*. Optics Express, 2012. **20**(13): p. 13677-13683.
26. Wilken, T., et al., *A frequency comb and precision spectroscopy experiment in space*, in *CLEO*. 2013: San Jose, California, USA.
27. Lee, J., et al., *Testing of a femtosecond pulse laser in outer space*. Scientific Reports, 2014. **4**(5134).
28. Sinclair, L.C., et al., *Operation of an optically coherent frequency comb outside the metrology lab*. Optics Express, 2014. **22**(6): p. 6996-7006.
29. Eckstein, J.N., A.I. Ferguson, and T.W. Hänsch, *High-Resolution Two-Photon Spectroscopy with Picosecond Light Pulses*. Phys. Rev. Lett., 1978. **40**(13): p. 847-850.
30. Telle, H.R., D. Meschede, and T.W. Hänsch, *Realization of a new concept for visible frequency division: phase locking of harmonic and sum frequencies*. Optics Letters, 1990. **15**(10): p. 532-534.
31. Spence, D.E., P.N. Kean, and W. Sibbett. *Sub-100fs Pulse Generation from a Self-Modelocked Titanium:Sapphire Laser*. in *Conference on Lasers and Electro-optics*. 1990. Washington.
32. Rauschenberger, J., et al., *Control of the frequency comb from a modelocked Erbium-doped fiber laser*. Optics Express, 2002. **10**(24): p. 1404-1410.
33. Tamura, K., et al., *77-fs pulse generation from a stretched-pulse mode-locked all-fiber ring laser*. Optics Letters, 1993. **18**(13): p. 1080-1082.
34. Kafka, J.D., T. Baer, and D.W. Hall, *Mode-locked erbium-doped fiber laser with soliton pulse shaping*. Optics Letters, 1989. **14**(22): p. 1269-1271.
35. Hargrove, L.E., R.L. Fork, and M.A. Pollack, *Locking of He-Ne Laser Induced by Synchronous Intracavity Modulation*. Appl. Phys. Lett., 1964. **5**(1): p. 4-5.
36. Keller, U., *Recent developments in compact ultrafast lasers*. Nature, 2003. **424**: p. 831-838.
37. Ippen, E.P., C.V. Shank, and A. Deines, *Passive mode locking of the cw dye laser*. Appl. Phys. Lett., 1972. **21**(8): p. 348-350.
38. http://www.rle.mit.edu/ultrafast/research/ultrashort-pulse-laser-technology/.
39. Jones, D.J., et al., *Carrier-Envelope Phase Control of Femtosecond Mode-Locked Lasers and Direct Optical Frequency Synthesis*. Science, 2000. **288**: p. 635-639.
40. Keller, U., *Ultrafast solid state laser oscillators: a success story for the last 20 years with no end in sight*. Appl. Phys. B, 2010. **100**: p. 15-28.
41. http://www.nobelprize.org/nobel_prizes/physics/laureates/2005/advanced-physicsprize2005.pdf.
42. Nicholson, J.W., *Advances in Femtosecond Fiber Lasers*, in *Optical Fiber Communication Conference and Exposition and The National Fiber Optic Engineers Conference*. 2007: Anaheim, California.
43. Washburn, B.R., S.A. Diddams, and N.R. Newburry, *Phase-locked, erbium-fiber-laser-based frequency comb in the near infrared*. Optics Letters, 2004. **29**(3): p. 250-252.

44. Adler, F., K. Moutzouris, and A. Leitenstorfer, *Phase-locked two-branch erbium-doped fiber laser system for long-term precision measurementsof optical frequencies.* Optics Express, 2004. **12**(24): p. 5872-5880.
45. Spaulding, K.M., et al., *Nonlinear dynamics of mode-locking optical fiber ring lasers.* J. Opt. Soc. Am. B, 2002. **19**(5): p. 1045-1054.
46. Zhong, Y.H., Z.X. Zhang, and X.Y. Tao, *Passively ModeLocked Fiber Laser Based on Nonlinear Optical Loop Mirror with Semiconductor Optical Amplifier.* Laser Phys. Lett., 2010. **20**(8): p. 1756-1759.
47. Keller, U., et al., *Semiconductor Saturable Absorber Mirrors (SESAM's) for Femtosecond to Nanosecond Pulse Generation in Solid-State Lasers.* IEEE Journal of Selected Topics in Quantum Electronics, 1996. **2**(3): p. 435-453.
48. Set, S.Y., et al., *Laser Mode Locking Using a Saturable Absorber Incorporating Carbon Nanotubes.* Journal of Light Wave Technology, 2004. **22**(1): p. 51-56.
49. Martinez, A. and Z. Sun, *Nanotube and graphene saturable absorbers for fibre lasers.* Nature Photonics, 2013. **7**.
50. Baek, I.H., et al., *Efficient Mode-Locking of Sub-70-fs Ti:Sapphire Laser by Graphene Saturable Absorber.* Applied Physics Express 5 (The Japan Society of Applied Physics), 2012.
51. Chen, J., et al., *High repetetion rate, low jitter, low intensity noise, fundamentally mode-locked 167 fs soliton Er-fiber laser.* Optics Letters, 2007. **32**(11): p. 1566-1568.
52. Holman, K.W., *Distribution of an Ultrastable Frequency Reference Using Optical Frequency Combs*, in *Department of Physics*. 2005, University of Colorado. p. 151.
53. Reichert, J., et al., *Phase Coherent Vacuum-Ultraviolet to Radio Frequency Comparison with a Mode-Locked Laser.* Phys. Rev. Lett., 2000. **84**(15): p. 3232-3235.
54. Udem, T., et al., *Accurate measurement of large optical frequency differences with a mode-locked laser.* Optics Letters, 1999. **24**(13): p. 881-883.
55. Holzwarth, R., T. udem, and T.W. Hänsch, *Optical Frequency Synthesizer for Precision Spectroscopy.* Phys. Rev. Lett., 2000. **85**(11): p. 2264-2267.
56. Diddams, S.A., et al., *An Optical Clock Based oon a Single Trapped* $^{199}$*Hg+ Ion* Science, 2001. **293**: p. 825-828.
57. Udem, T., R. Holzwarth, and T.W. Hänsch, *Optical Frequency Metrology.* Nature, 2002. **416**: p. 233-237.
58. Diddams, S.A., et al., *Standards of Time and Frequency at the Outset of the 21st Century.* Science, 2004. **306**: p. 1318-1324.
59. Joo, K.-N. and S.-W. Kim, *Absolute distance measurement by dispersive interferometry using a femtosecond pulse laser.* Optics Express, 2006. **14**(13): p. 5954-5960.
60. Hyun, S., et al., *Absolute length measurement with the frequency comb of a femtosecond laser.* Meas. Sci. Technol., 2009. **20**: p. 1-6.
61. Mehta, N., et al., *Compressive multi-heterodyne optical spectroscopy.* Optics Express, 2012. **20**(27): p. 28363-28372.
62. Steinmetz, T., et al., *Laser Frequency Combs for Astronomical Observations.* Science, 2008. **321**: p. 1335-1337.
63. http://www.menlosystems.com/.
64. Hänsch, T.W., et al., *Precision spectroscopy of hydrogen and femtosecond laser frequency combs.* Phil. Trans. R. Soc. A, 2009. **363**: p. 2155-2163.

65. Kozma, I.Z., et al., *Metrology Under Control: Fiber Technology Shoots Frequency Combs to Outer Space*. Optik&Photonik, 2007(4): p. 22-27.
66. Nielsen, C.K., *Mode Locked Fiber Lasers: Theoretical and Experimental Developments*, in *Department of Physics and Astronomy*. 2006, University of Aarhus: Denmark. p. 146.
67. http://www.rp-photonics.com/mode_locked_fiber_lasers.html.
68. K.Tamura, et al., *Technique for obtaining high-energy ultrashort pulses from an additive-pulse mode-locked erbium-doped fiber ring laser*. Optics Letters, 1994. **19**(1): p. 46-48.
69. Haus, H.A., *Mode-locking of Lasers*. IEEE, 2000. **6**(6): p. 1173-1185.
70. Ortac, B., et al., *Salf-starting passively mode-locked chirped pulse fiber laser*. Optics Express, 2007. **15**(25): p. 16794-16799.
71. Mastas, V.J., et al., *Characterization of a self-starting, passively mode-locked fiber ring laser that exploits nonlinear polarization evolution*. Optics Letters, 1993. **18**(5): p. 358-360.
72. Fermann, M.E. and I. Hartl, *Ultrafast Fiber Laser Technology*. IEEE, 2009. **15**(1): p. 191-206.
73. Tamura, K.R., *Additive Pulse Mode-locked Erbium-Doped Fiber Lasers*, in *Department of Eelectrical Engineering and Computer Science*. 1994, MIT. p. 169.
74. Washburn, B.R., R.W. Fox, and N.R. Newbury, *Fiber-laser-based frequency comb with a tunable repetition rate*. Optics Express, 2004. **12**(20): p. 4999-5004.
75. Tamura, K., et al., *Unidirectional ring resonators for self-starting passively mode-locked lasers*. Optics Letters, 1993. **18**(3): p. 220-222.
76. Tropf, W.J., M.E. Thomas, and T.J. Harris, *Optical and Physical Properties of Materials*.
77. http://www.schott.com/advanced_optics/us/abbe_datasheets/schott_datasheet_all_us.pdf.
78. Agrawal, G.P., *Fiber Optic Communication Systems*. Third ed. 2002, Rochester, NY: John Wiley& Sons, Inc.
79. http://people.seas.harvard.edu/~jones/ap216/lectures/ls_2/ls2_u5/ls2_unit_5.html
80. http://www.thorlabs.com/NewGroupPage9.cfm?ObjectGroup_ID=1504
81. Tang, D.Y., et al., *Mechanism of multisoliton formation and soliton energy quantization in passively mode-locked fiber lasers*. Physical Review A, 2005. **72**.
82. Seigman, A.E., *LASERS*. 1986, Mill Valley, CA: University Science Books.
83. Antoncini, C., *Ultrashort Laser Pulses*. The University of Reading.
84. http://course.ee.ust.hk/elec342/notes/Lecture%205_waveguiding%20in%20optical%20fibers.pdf.
85. Bagwell, P.F., *Mode Competition in Gas and Semiconductor Lasers*. Physics Optics, 2008.
86. Smith, A.V. and J.J. Smith, *Mode competition in high power fiber amplifiers*. Optics Express, 2011. **19**(12): p. 11319-11329.
87. Feng, X., H.-y. Tam, and P.K. Wai, *Stable and uniform multiwavelength erbium doped fiber laser using nonlinear polarization rotation*. Optics Express, 2006. **14**(18): p. 8205-8210.
88. Namiki, S., et al., *Relaxation oscillation behavior in polarization additive pulse mode-locked fiber ring lasers*. Appl. Phys. Lett., 1996. **69**(26): p. 3969-3971.
89. Lin, Y.-H. and G.-R. Lin, *Kelly sideband variation and self four-wave-mixing in femtosecond fiber soliton laser mode-locked by multiple exfoliated graphite nano-particles*. Laser Phys. Lett. , 2013. **10**: p. 12.

90. Dennis, M.L. and I.N. Duling, *Experimantal Study of Sideband Generation in Femtosecond Fiber Lasers.* IEEE Journal of Quantum Electronics, 1994. **30**(6): p. 1469-1477.
91. Lim, J., *All-fiber frequency comb employing a single walled carbon nanotube saturable absorber for optical frequency metrology in near infrared*, in *Department of Physics*. 2011, Kansas State University: Manhattan, KS. p. 201.
92. [http://www.swampoptics.com/PDFs/tutorials_autocorrelation.pdf](http://www.swampoptics.com/PDFs/tutorials_autocorrelation.pdf).
93. Chao, D., *Self-referenced 1.5 um Fiber Frequency Combs at GHz Repetition Rates*, in *Department of Electrical Engineering and Computer Science* 2012, MIT: March 20. p. 142.
94. Agrawal, G.P. and N.A. Olsson, *Amplification and compression of weak picosecond optical pulses by using semiconductor -laser amplifiers.* Optics Letters, 1989. **14**(10): p. 500-502.
95. Agrawal, G.P., *Nonlinear Fiber Optics*. First ed. Quantum electronics- Principles and Application. 1989, New York: Academic Press.
96. Agrawal, G.P., *Application of Nonlinear Fiber Optics*. Second ed. 2008, NewYork: Academic Press. 508.
97. Tomlinson, W.J., R.H. Stolen, and C.V. Shank, *Compression of optical pulses chirped by self-phase modulation in fibers.* J. Opt. Soc. Am. B, 1984. **1**(2): p. 139-149.
98. Desurvire, E., *Analysis of Gain Difference Between Forward- and Backward-Pumped in Erbium-Doped Fiber Amplifiers the Saturation Regime* IEEE Photonics Technology Letters, 1992. **4**(7): p. 711-714.
99. Olsson, N.A. and G.P. Agrawal, *Spectral shift and distortion due to selfphase modulation of picosecond pulses in 1.5 um optical amplifiers.* Appl. Phys. Lett., 1989. **55**(1): p. 13-15.
100. Zhong, X., et al., *Evolution of hyperbolic-secant optical pulses towards wave breaking in quintic nonlinear fibers.* Optics and Laser Technology, 2011. **44**: p. 669-674.
101. Anderson, D., et al., *Wave breaking in nonlinear-optical fibers.* J. Opt. Soc. Am. B, 1992. **9**(8): p. 1358-1361.
102. Nicholson, J.W., et al., *High power, single mode, all-fiber source od femtosecond pulses at 1550 nm and its use in supercontinuum generation.* Optics Express, 2004. **12**(13): p. 3025-3034.
103. Agrawal, G.P., *Optical wave breaking and pulse compression due to cross-phase modulation in optical fibers* Optics Letters, 1989. **14**(2): p. 137-139.
104. Nakatsuka, H. and D. Grischkowsky, *Recompression of optical pulses broadened by passage through optical fibers.* Optics Letters, 1981. **6**(1): p. 13-15.
105. Stolen, R.H., et al., *Raman response function of silica-core fibers.* J. Opt. Soc. Am. B, 1989. **6**(6): p. 1159-1166.
106. Agrawal, G.P., *Nonlinear fiber optics: its history and recent progress[Invited]* J. Opt. Soc. Am. B, 2011. **28**(12): p. A1-A10.
107. Agrawal, G.P., *Effect of intrapulse stimulated Raman scattering on soliton-effect pulse compression in optical fibers.* Optics Letters, 1990. **15**(4): p. 224-226.
108. Liu, J., et al., *Spectrum reshaping and pulse self-compression in normally dispersive media with negeatively chirped femtosecond pulses.* Optics Express, 2006. **14**(2): p. 979-987.

109. Tomlinson, W.J., R.H. Stolen, and A.M. Johnson, *Optical wave breaking of pulses in nonlinear optical fibers.* Optics Letters, 1985. **10**(9): p. 457-459.
110. Peng, J., et al., *All-fiber ultrashort similariton generation, amplification, and compression at telecommmunication band.* J. Opt. Soc. Am. B, 2012. **29**(9): p. 2270-2274.
111. Chan, K.C. and H.F. Liu, *Effect of third-order dispersion on soliton-effect pulse compression.* Optics Letters, 1994. **19**(1): p. 49-51.
112. Chen, C.-M. and P.L. Kelley, *Nonlinear pulse compression in optical fibers: scaling laws and numerical analysis.* J. Opt.Soc. Am. B, 2002. **19**(9): p. 1961-1967.
113. Roy, S., et al., *Dynamics of Raman soliton during supercontinuum generation near the zero-dispersion wavelength of optical fibers* Optics Express, 2011. **19**(11): p. 10443-10455.
114. Washburn, B.R., J.A. Buck, and S.E. Ralph, *Transform-limited spectral compression due to self-phase modulation in fibers.* Optics Letters, 2000. **25**(7): p. 445-447.
115. Nicholson, J.W., et al., *Visible continuum generation using a femtosecond erbium-doped fiber laser and a silica nonlinear fiber.* Optics Letters, 2008. **33**(1): p. 28-30.
116. Nicholson, J.W., et al., *Supercontinuum generation in ultraviolet-irradiated fibers.* Optics Letters, 2004. **29**(20): p. 2363-2365.
117. Nicholson, J.W., et al., *All-fiber, octave-spanning supercontinuum.* Optics Letters, 2003. **28**(8): p. 643-645.
118. Islam, M.N., et al., *Broad bandwidths from frequency-shifting solitons in fibers.* Optics Letters, 1989. **14**(7): p. 370-372.
119. Stumpf, M.C., et al., *Self-referencable frequency comb from a 170-fs, 1.5-um solid-state laser oscillator.* Appl. Phys. B, 2010. **99**: p. 401-408.
120. Takushima, Y., *High average power, depolarized supercontinuum generation using a 1.55um ASE noise source.* Optics Express, 2005. **13**(15): p. 5871-5876.
121. Genty, G., S. Coen, and J.M. Dudley, *Fiber supercontinuum sources (Invited).* J. Opt. Soc. Am. B, 2007. **24**(8): p. 1771-1785.
122. Gao, W., et al., *All-fiber broadband supercontinuum source with high efficiency in a step-index high nonlinear silica fiber.* Applied Optics, 2012. **51**(8): p. 1071-1075.
123. Nicholson, J.W. and M.F. Yan, *Cross-coherence measurements of supercontinua generated in highly-nonlinear, dispersion shifted fiber at 1550 nm.* Optics Express, 2004. **12**(4): p. 679-688.
124. Kim, S., et al., *Coherent Supercontinuum generation using Er-doped fiber laser of hybrid mode-locking.* Optics Letters, 2014. **39**(10): p. 2986-2989.
125. Nicholson, J.W., A.D. Yablon, and M.F. Yan, *Coherence of supercontinua generated by ultrashort pulses compressed in optical fibers.* Optics Letters, 2008. **33**(18): p. 2038-2040.
126. Westbrook, P.S., et al., *Improved Supercontinuum Generation Through UV Processing of Highly Nonlinear Fibers.* Journal of Light Wave Technology, 2005. **23**(1): p. 13-18.
127. Srivastava, A. and D. Goswami, *Control of supercontinuum generation with polarization of incident laser pulses.* Appl. Phys. B, 2003. **77**: p. 325-328.
128. Ye, J., H. Schnatz, and L.W. Hollberg, *Optical Frequency Combs: From Frequency Metrology to Optical Phase Control.* IEEE Journal of Selected Topics in Quantum Electronics, 2003. **9**(4): p. 1041-1058.

129. Rauschenberger, J., *Phase-stabilized Ultrashort Laser Systems for Spectroscopy*, in *Dpertment of Physics*. 2007, Ludwig-Maximilians University: Munchen. p. 135.
130. Murphy, M.T., et al., *Laser frequency comb techniques for precise astronomical spctroscopy*. Mon. Not. R. Astron. Soc., 2012. **422**: p. 761-771.
131. Wei, Z., et al., *Measurement and Control of Carrier-Envelope Phase in Femtosecond Ti:sapphire Laser* Advances in Solid-State Lasers: Development and Applications, ed. M. Grishin. 2010: Intech, Croatia. 630.
132. Kobayashi, Y., et al., *Relative Carrier-Envelope-Offset Phase Control Between Independent Femtosecond Light Sources*. IEEE Journal of Selected Topics in Quantum Electronics, 2003. **9**(4): p. 1011-1017.
133. Klenner, A., et al., *Phase-stabilization of the carrier-envelope-offset frequency of a SESAM modelocked thin disk laser*. Optics Express, 2013. **21**(21): p. 24770-24780.
134. Teets, R., J. Eckstein, and T.W. Hänsch, *Coherent Two-Photon Excitation by Multiple Light Pulses*. Phys. Rew. Lett. , 1976. **38**(14): p. 760-763.
135. Bernfeld, S.B., *Stabilization of a Femtosecond Laser Frequency Comb*, in *Department of Physics and Astronomy*. 2009, Oberlin College.
136. Ye, J. and S.T. Cundiff, *Femtosecond Optical Frequency Comb: Principle, Operation, and Applications* 2004: Springer.
137. Tillman, K.A., et al., *Significant Carrier Envelope Offset Frequency Linewidth Narrowing in a Prism-based Cr:forsterite Frequency Comb*, in *Conference of Lasers and Electro-Optics/Quantum Electronics and Laser Science* 2008: San Jose,CA.
138. Oksenhendler, T., *Self-referenced spectral interferometry theory*. 2012.
139. Erskine, D.J. and J. Edelstein, *High- Resolution Broadband Spectral Interferometry*, in *Conf. on Astronomical Instrumentation*. 2002: Hawaii.
140. Paul, A., et al., *Quasi-phase-matched generation of coherent extreme-ultraviolet light.* Nature, 2003. **421**: p. 51-54.
141. Hum, D.S. and M.M. Fejer, *Quasi-phasematching*. Comptes Rendus Physique, 2007. **8**: p. 180-198.
142. Rundquist, A., et al., *Phase-Matched Generation of Coherent Soft X-rays*. Science, 1998. **280**: p. 1412-1415.
143. Miller, G.D., et al., *42%-efficient single-pass cw second-harmonic generation in periodically poled lithium niobate*. Optics Letters, 1997. **22**(24): p. 1834-1836.
144. Kleinman, D.A., *Nonlinear Dielectric Polarization in Optical Media*. Phys. Rev., 1962. **126**.
145. Myers, L.E., et al., *Quasi-phase-matched 1.064-um -pumped optical parametric oscillator in buk periodically poled* $LiNbO_3$. Optics Letters, 1995. **20**(1): p. 52-54.
146. Boyd, G.D. and D.A. Kleinman, *Parametric Interaction of Focused Gaussian Light Beams*. J. Appl. Phys., 1968. **39**: p. 3597-3639.
147. Phillips, C.R., *Broadband Optical Sources based on Highly Nonlinear Quasi-Phasematched Interactions*, in *Department of Electrical Engineering*. 2012, Stanford University. p. 286.
148. http://www.ist-brighter.eu/tuto11/CONF2/Cambridge_Petersen.pdf.
149. Fejer, M.M., et al., *Quasi-Phase-Matched Second Harmonic Generation: Tuning na dTolerances*. IEEE Journal of Quantum Electronics, 1992. **28**(11): p. 2631-2653.
150. Gliko, O.A. and I.I. Naumova, *Nonlinear Optical Properties of PPLN (Periodically Poled LiNbO3)*. Journal of the Korean Physical Society, 1998. **32**: p. S464-S467.

151. Guo, S., et al., *Invesitgation of optical inhomogeneity of MgO:PPLN crystal for frequency doubling of 1560 nm laser*. Optics Communication, 2014. **326**: p. 114-120.
152. www.covesion.com/assets/downloads/SHG7-0.5.pdf.
153. Taya, M., M.C. Bashaw, and M.M. Fejer, *Photorefractive effects in periodically poled ferroelectrics*. Optics Letters, 1996. **21**(12): p. 857-859.
154. http://www.shanghai-optics.com/products/ppln/.
155. http://www.bblaser.com/bbl_item/027hcp.html.
156. http://www.as-photonics.com/snlo.
157. Franken, P.A., et al., *Generation of Optical Harmonics*. Phys. Rev. Lett., 1961. **7**(4): p. 118-119.
158. Wang, H., *The Study of Femtosecond Harmonic Generation in Bulk Crystals with Large walkoff*. 2003, Purdue University. p. 111.
159. Han, W., et al., *Phase matching limitation of high-efficiency second-harmonic generation in both phase- and group-velocity-matched structures*. Optik, 2008. **119**: p. 122-126.
160. Abu-Safe, H.H., *Difference frequency mixing of strongly focused Gaussian beams in periodically poled $LiNbO_3$*. Applied Physics Letters, 2005. **86**: p. 231105-1~231105-3.
161. H.R. Telle, et al., *Carrier-envelope offset phase control: A novel concept for absolute optical frequency measurement and ultrashort pulse generation.* Appl. Phys. B, 1999. **69**: p. 327-332.
162. Corwin, K.L., et al., *Absolute-frequency measurements with a stabilized near-infrared optical frequency comb from a Cr:forsterite laser*. Optics Letters, 2004. **29**(4): p. 397-399.
163. Amy-Klein, A., et al., *Absolute frequency measurement in the 28-THz spectral region with a femtosecond laser comb and a long-distance optical link to a primary standard.* Appl. Phys. B, 2004. **78**: p. 25-30.
164. Cundiff, S.T. and J. Ye, *Colloquium: Femtosecond optical frequency combs.* Reviews of Modern Physics, 2003. **75**: p. 325-342.
165. Reichert, J., et al., *Measuring the frequency of light with mode-locked lasers* Optics Communication, 1999. **172**: p. 59-68.
166. Lim, J., et al., *A phase-stabilized carbon nanotube fiber laser frequency comb* Optics Express, 2009. **17**(16): p. 14115-14120.
167. Thapa, R., *Cr: Forsterite Laser Frequency Comb Stabilization and Development of Portable Frequency References Inside a Hollow Optical Fiber* in *Department of Physics*. 2008, Kansas State University. p. 141.

# Appendix A - List of Abbreviations

| | |
|---|---|
| **a.u.** | Arbitrary unit |
| **AC** | Intensity autocorrelation |
| **AOM** | Acousto-optic modulator |
| **APM** | Additive pulse mode-locking |
| **ASE** | Amplified spontaneous emission |
| **BPF** | Band-pass filter |
| **BPM** | Birefringence phase matching |
| **CEO** | Carrier envelope offset |
| **CEP** | Carrier envelope phase |
| **CW** | Continuous wavelength |
| **dB** | Decibel |
| **EDF** | Erbium doped fiber |
| **EDFA** | Erbium doped fiber amplifier |
| **EOM** | Electro-optic modulator |
| **ESA** | Electric spectrum analyzer |
| **FC/APC** | Angled-face fiber connector |
| **FC/PC** | Flat-face fiber connector |
| **FH** | Fundamental harmonics |
| **FROG** | Frequency resolved optical gating |
| **FWHM** | Full width half maximum |
| **FWM** | Four wave mixing |
| **GPS** | General positioning system |
| **GVD** | Group velocity dispersion |
| **HNLF** | Highly nonlinear fiber |
| **HR** | High-reflective mirror |
| **HWP** | Half-wave plate |
| **IPRS** | Intra-pulse Raman scattering |
| **ISO** | Polarization-sensitive isolator |
| **KLM** | Kerr lens mode-locking |
| **MFD** | Mode field diameter |
| **MI** | Modulation instability |
| **MMF** | Multi-mode fiber |
| **NLSE** | Nonlinear Schrödinger equation |
| **NPR** | Nonlinear polarization rotation |
| **NPRL** | Nonlinear polarization rotation laser |
| **OSA** | Optical spectrum analyzer |
| **PBC** | Polarization beam combiner |
| **PBS** | Polarization beam splitter |
| **PC** | Polarization controller |
| **PCF** | Photonic crystal fiber |
| **PD** | Photodiode/Photodetector |
| **PM** | Polarization maintaining |
| **PPLN** | Periodically polled lithium niobate |
| **PZT** | Piezoelectric transducer |
| **QPM** | Quasi-phase matching |
| **QWP** | Quarter-wave plate |
| **RF** | Radio frequency |
| **SC** | Supercontinuum |
| **SESAM** | Semiconductor saturable absorber mirror |
| **SHG** | Second harmonic generation |
| **SMF** | Single mode fiber |
| **SPM** | Self-phase modulation |
| **SRS** | Stimulated Raman scattering |
| **SSFS** | Soliton self-frequency shift |
| **TL** | Transform limited |
| **USP** | Ultra-short pulse |
| **WDM** | Wavelength division multiplexer |
| **XPM** | Cross-phase modulation |
| **ZDW** | Zero-dispersion wavelength |